\documentclass[useAMS,usenatbib]{mn2e}
\usepackage{graphicx}
\newcommand{\mdot}{M$_{\odot}$~}

\newcommand{\kmsec}{km s$^{-1}$}

\usepackage{booktabs,lipsum}
\usepackage{natbib}
\usepackage{rotating}
\usepackage{pdflscape}
\usepackage{color}
\pdfoutput=1

\def \oiii {[O{\small~III}]}

\def \sii {[S{\small~II}]}
\def \nii {[N{\small~II}]}
\def \ha {H$\alpha$}
\def \hb {H$\beta$}
\def \lam {$\lambda$}

\title[IFU observations of luminous type II AGN - I.]{IFU observations of luminous type II AGN - I. Evidence for ubiquitous winds}
\author[R. McElroy et al.] 
{\parbox[t]{\textwidth}{Rebecca McElroy$^{1, 2}$\thanks{E-mail: rmcelroy@physics.usyd.edu.au}, Scott M. Croom$^{1, 2}$, Michael Pracy$^{1}$, Rob Sharp$^{2, 3}$, I-Ting Ho$^{4}$, and Anne M. Medling$^{3}$}
\\ \\
$^{1}$Sydney Institute for Astronomy, School of Physics, University of Sydney, NSW 2006, Australia\\
$^{2}$ARC Centre of Excellence for All-sky Astrophysics (CAASTRO)\\
$^{3}$Research School of Astronomy \& Astrophysics, The Australian National University, Cotter Road, Weston Creek, ACT 2611, Australia\\
$^{4}$Institute for Astronomy, University of Hawaii, 2680 Woodlawn Drive, Honolulu, HI 96822, USA\\}

\begin{document}

\date{Draft as of \today}

\input{journal_short.def}

\pagerange{\pageref{firstpage}--\pageref{lastpage}} \pubyear{2002}

\maketitle

\label{firstpage}

\begin{abstract}

We present observations of 17 luminous ($\log(L_{\rm{\oiii}}/L_\odot)>8.7$)
local ($z<0.11$) type II AGN.
Our aim is to investigate the
prevalence and nature of AGN driven outflows in these galaxies by
combining kinematic and ionization diagnostic information.  
We use non-parametric methods (e.g.\ $W_{80}$, the width containing
80 percent of the line flux) to assess the line widths in the central regions of our
targets.  The maximum values of $W_{80}$ in each galaxy are in the range
$400-1600$ \kmsec, with a mean of 790 $\pm$ 90 \kmsec.  Such high velocities are
strongly suggestive that these AGN are driving ionized outflows.
Multi-Gaussian fitting is used to decompose the velocity structure
in our galaxies.  14/17 of our targets require 3 separate kinematic
components in the ionized gas in their central regions.  The broadest
components of these fits have $\rm{FWHM} = 530-2520$ \kmsec, with a mean value
of 920 $\pm$ 50 \kmsec.  By simultaneously fitting both the  \hb/\oiii\ and
\ha/\nii\ complexes we construct ionization diagnostic diagrams for
each component.  13/17 of our galaxies show a
significant ($>95$ \%) correlation between the \nii/\ha\ ratio
and the velocity dispersion of the gas.  Such a
correlation is the natural consequence of a contribution to the
ionization from shock excitation and we argue that this demonstrates
that the outflows from these AGN are directly impacting the
surrounding ISM within the galaxies.

\end{abstract}


\section{Introduction}

The importance of AGN feedback in galaxy evolution is evident from the observed correlations between the mass of galaxies' supermassive black holes (SMBH) and their stellar bulges \citep[e.g.][]{Tremaine2002}. This correlation suggests a co-dependence between the evolution of galaxies and their SMBHs, leading to the belief that the same process may limit the further growth of both systems. AGN-driven outflows are often invoked as this mechanism. The growth of SMBHs occurs through accretion during active phases. The stellar bulge grows as gas inside the galaxy collapses to form stars, or through accretion of previous generations of stars via mergers. Outflows can remove the gas required for both processes. 

Simulations have shown that feedback processes are necessary to reproduce realistic galaxies \citep[e.g.][]{Hopkins2006}, and observations of massive galaxies require a phenomenon such as radio-mode feedback to explain the steep cut-off in their luminosity function \citep{Croton2006, Bower2006}, though in this paper we will focus on radiatively efficient quasar-mode feedback. It is believed that this cut-off is due to outflows removing or heating gas in these massive galaxies. This inhibits star formation and further AGN activity leading to a lower galaxy luminosity than otherwise predicted, and limits the upper mass the galaxy can  attain \citep[e.g.][]{Murray2005, Scannapieco2004}. Outflows likely cause the observed enrichment of the intergalactic medium with metals generated in star formation  \citep[e.g.][]{Cowie1995}. This has been shown through analysis of quasar absorption spectra, and suggests that at least some of the gas driven by galactic winds can escape galaxies.

The exact role of AGN feedback is not well constrained. It is thought that the majority of AGN hosts possess outflows \citep{Ganguly2008}, but that those in higher luminosity AGN are more powerful \citep{Veilleux2013}. While limited insight into outflows within galaxies may be provided by single spectrum surveys such as SDSS \citep[see][]{Barrows2013}, spatial information is required to fully understand the dynamics at play within these galaxies. Integral field spectroscopy (IFS) is the perfect tool for this as it provides spatially resolved spectra across the 2D projection of galaxies, meaning that conditions in different regions of the galaxy may be explored. Several small IFU surveys of AGN and starburst galaxies have been successful in identifying outflows. Studies such as \cite{SharpJBH2010} concentrated on very nearby AGN and starburst galaxies known to have galactic winds. From this they found that starburst winds tend to have shock ionization, and AGN generally excite AGN ionised winds. These imply different time-scales as due to a starburst's shorter lifetime, by the time the wind is visible many of the OB stars will have already gone supernova, and the radiation driving the wind will be reduced significantly. We now know that winds are not particularly rare; \cite{Ganguly2008} showed that they are present in at least 60\% of AGN in their study of single absorption spectra of QSOs, and as such broader surveys of previously unstudied objects are being pursued.

In their study of 11 luminous, obscured, radio quiet quasars from the \cite{Reyes2008} catalogue with an average $z \sim 0.55$, \cite{Liu2013b} found that outflows were common using non-parametric measures of line width and asymmetry of only the \oiii5007 emission line. They found very extended broad emission throughout their sample with the width containing the central 80\% of flux, W$_{80}$, ranging from 500-1800 km s$^{-1}$. This sample has a mean log(L\oiii) luminosity of 43.2 erg s$^{-1}$. Similarly, \cite{Harrison2014} used the same technique on both the \oiii5007 and \oiii4959 emission lines to show that outflows were prevalent in their sample of 16 luminous type II AGN. Their AGN were required to be at z \textless\ 0.2 (with a mean z = 0.1365, and a mean log(L\oiii) of 42.5 erg s$^{-1}$) and were selected to have broad (FWHM $>$ 700 km s$^{-1}$), luminous emission components that must contribute $>$ 30\% of the flux. They too found extended broad emission with W$_{80}$ between 600-1500km s$^{-1}$. 

To further investigate the nature of outflows driven by AGN, we selected an unbiased sample of highly luminous local type II AGN to observe across a large spectral range at moderately high spectral resolution. Observations were performed with the SPIRAL integral field unit (IFU) on the Anglo-Australian Telescope (AAT). A luminous sample was chosen to investigate winds at the extreme of the distribution. Additionally, no requirement for broad emission lines was applied to the sample, as was done in \cite{Harrison2014}.
Nearby galaxies were selected to allow for the best possible spatial resolution. In future work, this sample will be expanded and compared to galaxies observed by the SAMI Galaxy Survey \citep[see][]{Croom2012, Allen2014, Bryant2014} to better understand how feedback varies with AGN power and how these galaxies differ from their quiescent counterparts. 

In this paper we use a much larger spectral range than previous studies, allowing for fitting of the H$\beta$, [O\textsc{iii}]$\lambda$5007, H$\alpha$ and [N\textsc{ii}]$\lambda$6583 lines with a higher spectral resolution of R$_{blue} =$ 2141 and R$_{red} =$ 5600. This large spectral range also allows for multiple line ratios to be considered, meaning ionization states of the winds can be derived using theoretical and observational cut-offs used in conjunction with diagnostic diagrams \citep{Baldwin1981, Kewley2001, Kauffmann2003, Veilleux1987}. Our spectral range also captures many stellar absorption lines including the Calcium H and K lines at 3968.5\AA\ and 3933.7\AA\ respectively. This enables measurements of the stellar populations and kinematics.

Primarily, the sample was selected and observed in order to learn more about AGN feedback. Using IFS data means that we are able to look for kinematic signatures of AGN feedback across the spatial extent of galaxies, to see how these hosts are effected by the presence of the AGN. A key observation linked to AGN feedback is high velocity gas, referred to as winds or outflows. To determine whether these are present we must look for several signatures; broad emission lines, asymmetric line profiles, and high excitation. We expect winds to cause broad emission lines due to the large range in velocities of the swept up gas, and the profile will likely be asymmetric as a result of the offset of the wind velocity. Additionally, we expect high ionization due to the winds being driven by the AGN or due to shocks caused as the outflow passes through the surrounding gas. In order to find these signatures, we fit multiple components to characterise the complex emission line profiles of our sample. We then use two approaches; the first is a non-parametric emission line analysis, and the second considers the multiple components separately. 
 
The outline of this paper is as follows; in Section 2 we describe the observations and data reduction process. Section 3 provides details of the kinematic analysis performed on the data, including the stellar and emission line kinematics, the statistics used to interpret our results, and the method used to obtain ionization diagnostics. In Section 4 we discuss the results of the analysis of our data. Section 5 considers the energetics of the outflows we observe, the implications of our results, and discusses a peculiar galaxy - J111100. In this paper we adopt an H$_{0}$ = 71 km s$^{-1}$ Mpc$^{-1}$, $\Omega_{M}$=0.27, $\Omega_{\Lambda}$=0.73 cosmology.

\renewcommand{\tabcolsep}{3pt}

\begin{table*}
  \centering
    \begin{tabular}{cccccccccccccc}
    \toprule
    Object Name & RA    & DEC   & z & S/N &  A$_{\rm{\oiii}}$ & L$_{\rm{\oiii}}$ & M$_{star}$ & D$_{em}$  & Seeing & FIRST & L$_{\rm{FIRST}}$ & M$_{BH}$ & L$_{Bol}$\\
    (1) & (2)  & (3) & (4) & (5) & (6) & (7) & (8) & (9) & (10) & (11) & (12) & (13) & (14)\\
    \midrule
    J095155.34+032900.3 & 147.980606 & \multicolumn{1}{c}{3.483436} & \multicolumn{1}{c}{0.0601} & \multicolumn{1}{c}{30} & \multicolumn{1}{c}{5.63} & \multicolumn{1}{c}{42.3} & \multicolumn{1}{c}{11.0} & \multicolumn{1}{c}{3.3} & \multicolumn{1}{c}{3.0} & 1.78  & 1.55$\times$10$^{22}$& 7.70  & 45.85 \\
    J101927.55+013422.4 & 154.864822 & \multicolumn{1}{c}{1.572914} & \multicolumn{1}{c}{0.0730} & \multicolumn{1}{c}{291} & \multicolumn{1}{c}{3.87} & \multicolumn{1}{c}{42.5} & \multicolumn{1}{c}{11.0} & \multicolumn{1}{c}{6.9} & \multicolumn{1}{c}{1.3} & 2.95  & 3.75$\times$10$^{22}$ & 8.52  & 46.01 \\
    J102143.30+011428.4 & 155.430420 & \multicolumn{1}{c}{1.241226} & \multicolumn{1}{c}{0.0787} & \multicolumn{1}{c}{269} & \multicolumn{1}{c}{3.21} & \multicolumn{1}{c}{42.5} & \multicolumn{1}{c}{10.8} & \multicolumn{1}{c}{6.3} & \multicolumn{1}{c}{2.7} & 1.97  & 2.93$\times$10$^{22}$ & 8.62  & 46.06 \\
    J103600.37+013653.5 & 159.001556 & \multicolumn{1}{c}{1.614885} & \multicolumn{1}{c}{0.1068} & \multicolumn{1}{c}{320} & \multicolumn{1}{c}{1.98} & \multicolumn{1}{c}{42.8} & \multicolumn{1}{c}{11.2} & \multicolumn{1}{c}{9.6} & \multicolumn{1}{c}{2.3} &  ---     &   ---    & 7.94  & 46.38 \\
    J103915.69--003916.9 & 159.815414 & \multicolumn{1}{c}{-0.654716} & \multicolumn{1}{c}{0.0770} & \multicolumn{1}{c}{208} & \multicolumn{1}{c}{2.98} & \multicolumn{1}{c}{42.4} & \multicolumn{1}{c}{10.5} & \multicolumn{1}{c}{7.2} & \multicolumn{1}{c}{2.1} & 30.34 & 4.31$\times$10$^{23}$ & 7.40  & 45.98 \\
    J111100.60--005334.8 & 167.752533 & \multicolumn{1}{c}{-0.893009} & \multicolumn{1}{c}{0.0904} & \multicolumn{1}{c}{121} & \multicolumn{1}{c}{3.12} & \multicolumn{1}{c}{42.8} & \multicolumn{1}{c}{11.4} & \multicolumn{1}{c}{16.6} & \multicolumn{1}{c}{1.7} & 7.30  & 1.46$\times$10$^{23}$ & 8.96  & 46.37 \\
    J124321.32+005923.7 & 190.838837 & \multicolumn{1}{c}{0.989920} & \multicolumn{1}{c}{0.0834} & \multicolumn{1}{c}{114} & \multicolumn{1}{c}{3.15} & \multicolumn{1}{c}{42.3} & \multicolumn{1}{c}{10.8} & \multicolumn{1}{c}{8.8} & \multicolumn{1}{c}{2.0} & 1.01  & 1.70$\times$10$^{22}$ & 7.53  & 45.87 \\
    J124859.92--010935.4 & 192.249680 & \multicolumn{1}{c}{-1.159850} & \multicolumn{1}{c}{0.0888} & \multicolumn{1}{c}{240,365} & \multicolumn{1}{c}{2.60} & \multicolumn{1}{c}{42.7} & \multicolumn{1}{c}{11.2} & \multicolumn{1}{c}{15.2} & \multicolumn{1}{c}{1.6,1.4} & 8.16  & 1.57$\times$10$^{23}$ & 8.89  & 46.27 \\
    J130116.09--032829.1 & 195.317062 & \multicolumn{1}{c}{-3.474768} & \multicolumn{1}{c}{0.0864} & \multicolumn{1}{c}{396} & \multicolumn{1}{c}{2.89} & \multicolumn{1}{c}{42.4} & \multicolumn{1}{c}{10.8} & \multicolumn{1}{c}{11.4} & \multicolumn{1}{c}{1.4} & 1.96  & 3.55$\times$10$^{22}$ & 7.67  & 45.94 \\
    J133152.88+020059.2 & 202.970367 & \multicolumn{1}{c}{2.016448} & \multicolumn{1}{c}{0.0861} & \multicolumn{1}{c}{234} & \multicolumn{1}{c}{5.83} & \multicolumn{1}{c}{43.4} & \multicolumn{1}{c}{10.9} & \multicolumn{1}{c}{8.0} & \multicolumn{1}{c}{3.6} & 9.81  & 1.77$\times$10$^{23}$ & 7.85  & 46.96 \\
    J141926.33+013935.8 & 214.859711 & \multicolumn{1}{c}{1.659962} & \multicolumn{1}{c}{0.0764} & \multicolumn{1}{c}{190} & \multicolumn{1}{c}{3.34} & \multicolumn{1}{c}{42.8} & \multicolumn{1}{c}{10.9} & \multicolumn{1}{c}{7.1} & \multicolumn{1}{c}{1.5} & 4.69  & 6.56$\times$10$^{22}$ & 7.59  & 46.35 \\
    J142237.91+044848.5 & 215.657990 & \multicolumn{1}{c}{4.813485} & \multicolumn{1}{c}{0.0871} & \multicolumn{1}{c}{182} & \multicolumn{1}{c}{3.36} & \multicolumn{1}{c}{42.6} & \multicolumn{1}{c}{11.0} & \multicolumn{1}{c}{9.2} & \multicolumn{1}{c}{1.3} & 1.79  & 3.30$\times$10$^{22}$ & 8.71  & 46.16 \\
    J143046.02+001451.4 & 217.691772 & \multicolumn{1}{c}{0.247628} & \multicolumn{1}{c}{0.0546} & \multicolumn{1}{c}{270} & \multicolumn{1}{c}{3.81} & \multicolumn{1}{c}{42.5} & \multicolumn{1}{c}{11.2} & \multicolumn{1}{c}{6.7} & \multicolumn{1}{c}{2.5} & 9.47  & 6.55$\times$10$^{22}$ & 8.38  & 46.05 \\
    J150754.38+010816.7 & 226.976593 & \multicolumn{1}{c}{1.137990} & \multicolumn{1}{c}{0.0610} & \multicolumn{1}{c}{182} & \multicolumn{1}{c}{1.95} & \multicolumn{1}{c}{42.3} & \multicolumn{1}{c}{10.8} & \multicolumn{1}{c}{8.3} & \multicolumn{1}{c}{1.8} & 3.90  & 3.40$\times$10$^{22}$ & 7.51  & 45.83 \\
    J151147.47+033827.2 & 227.947815 & \multicolumn{1}{c}{3.640911} & \multicolumn{1}{c}{0.0780} & \multicolumn{1}{c}{187} & \multicolumn{1}{c}{3.76} & \multicolumn{1}{c}{43.1} & \multicolumn{1}{c}{10.9} & \multicolumn{1}{c}{8.3} & \multicolumn{1}{c}{1.6} & 6.94  & 1.01$\times$10$^{23}$ & 7.65  & 46.61 \\
    J152133.35--003628.5 & 230.388977 & \multicolumn{1}{c}{-0.607926} & \multicolumn{1}{c}{0.0937} & \multicolumn{1}{c}{25} & \multicolumn{1}{c}{5.32} & \multicolumn{1}{c}{42.5} & \multicolumn{1}{c}{11.5} & \multicolumn{1}{c}{4.9} & \multicolumn{1}{c}{3.6} & 7.89  & 1.70$\times$10$^{23}$ & 8.00  & 46.07 \\
    J152637.67+003533.5 & 231.656967 & \multicolumn{1}{c}{0.592640} & \multicolumn{1}{c}{0.0507} & \multicolumn{1}{c}{187} & \multicolumn{1}{c}{3.09} & \multicolumn{1}{c}{42.5} & \multicolumn{1}{c}{10.9} & \multicolumn{1}{c}{6.3} & \multicolumn{1}{c}{2.3} & 3.80  & 2.25$\times$10$^{22}$ & 7.49  & 46.03 \\
            \bottomrule
    \end{tabular}%
  \label{obs_table}%
    \caption{A summary table of our observations. (1) IAU format name. (2)-(3) RA and DEC positions as taken from SDSS DR7. (4) SDSS redshift. (5) S/N of the H$\alpha$ line in the central spaxel as taken from our emission line fitting (described in Section \ref{fitting_sec}), Galaxy J124859 has two values here as it was observed twice. (6) \oiii\ extinction derived from the Balmer decrement in the SDSS spectrum. (7) Extinction corrected log(L$_{\rm{\oiii}}$) in erg s$^{-1}$. (8) Galaxy stellar mass in units of log solar mass. (9) The extent of the observed two component region, used to define the spatial extent of the outflow (see Section \ref{energetics}). (10) Average seeing during the observation measured in arc seconds. (11) Peak radio flux from FIRST measured in mJy. (12) Radio luminosity calculated from the FIRST flux, measured in W/Hz. (13) Mass of the central black hole in solar masses, as determined from the galaxy velocity dispersion. (14) Bolometric luminosity in erg s$^{-1}$, obtained from the \protect\oiii\ luminosity.}
\end{table*}%

\section[]{Observations and Data Reduction}
\subsection[]{Target Selection}

The sample presented in this paper was selected to investigate the
effects of feedback in luminous type II AGN by determining
the fraction of these objects that show evidence of winds or
outflows. We selected 17 highly luminous, local (z \textless\ 0.11) type II AGN
from the SDSS DR7 \citep{Abazajian2009} using
spectral measurements from the MPA/JHU database \citep{2003MNRAS.341...33K}.  According to
the unified AGN model, the nuclei of type II AGN are obscured by a
putative circum-nuclear torus of dust, meaning their spectra are dominated by
narrow emission lines, typically with $\sigma$ \textless\ 1200 km s$^{-1}$
\citep{Antonucci1993}. Evidence now points to the dust
obscuration being more clumpy \citep[e.g.][]{2008ApJ...685..160N, Riffel2014, Honig2013}.  However, this does not influence our
selection which was chosen to get a clear view of the host galaxies of
luminous AGN, relatively unhindered by the continuum emission from the
nucleus.  Using the luminosity of the \oiii\,\lam 5007 line within the
$3''$ diameter SDSS fibre as a surrogate for the luminosity of the nucleus, targets were
required to have $\log(\rm{L}_{\oiii}/\rm{L}_{\odot}) > 8.7 $ to achieve a sample
near the typical quasar-Seyfert luminosity 
boundary.   To correct to a bolometric luminosity we use $\rm{L}_{\rm
  bol}/\rm{L}_{\rm \oiii}=3500$ (Heckman et al., 2004), although we note
that the scatter  on this correction is large, $\sim0.38$ dex.  Using
this correction our \oiii\ luminosity limit corresponds to $\rm{L}_{\rm
  bol}=10^{12.2}$ \rm{L}$_{\odot} = 10^{45.8}$ erg\,s$^{-1}$.

\begin{figure}
\centering
\includegraphics[width=0.5\textwidth]{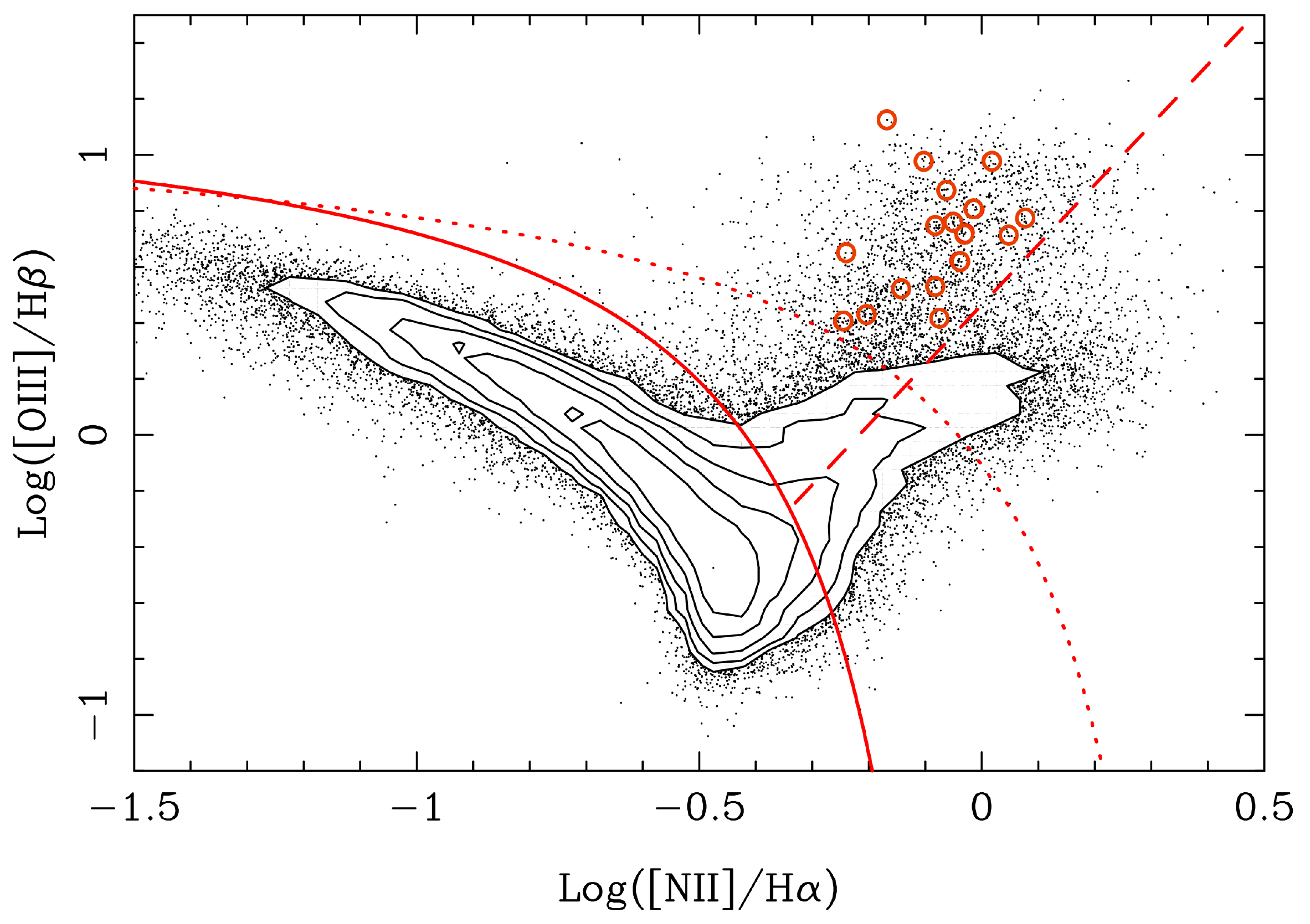}
\caption{A BPT
  plot showing the selection criteria imposed on the sample. Red circles
  are the observed targets. The contours (logarithmically spaced) and
  black points show location of the full SDSS parent sample, the red dotted line indicates the theoretical
  star forming limit, from \protect\cite{Kewley2001}. The solid red line is
  the empirical AGN limit above which galaxies are considered  to have
  an AGN contribution to their emission lines,
  and the dashed red line shows the division between true AGN
  and LINER-like emission, both from \protect\cite{Kauffmann2003}.}
\label{selection_plot}
\end{figure}

Baldwin-Phillips-Terlevich diagrams (henceforth BPT, \cite{Baldwin1981}) of $\rm{log}(\oiii/H\beta)$ against $\rm{log}(\nii/H\alpha)$, were used to distinguish AGN, star forming and LINER-like (Low Ionization Nuclear Emission-line Regions) systems (shown in Figure \ref{selection_plot}). 
Targets were required to be above the theoretical star-forming limit
described by \cite{Kewley2001} (red dotted line in Figure
\ref{selection_plot}), in order that the \oiii\ luminosity 
did not have a significant contribution from star formation. Any objects with
LINER-like line ratios were also rejected, meaning that they must lie
above the diagonal dashed red line shown in Figure
\ref{selection_plot} taken from \cite{Kauffmann2003}.  

The galaxies were required to have a redshift within the range $0.01 <
z < 0.11$ (all fall within the range, $0.05 < z < 0.11$).  This was chosen to obtain the best possible spatial
resolution, while still including sufficient high luminosity AGN.  A
limit to declination $\delta < 10^{\circ}$ was also required to allow
the objects to be observed from the Anglo-Australian Telescope.  The
luminosity, BPT, redshift and declination constraints resulted in
sample of 149 AGN.  We also prioritised objects with extinction
corrections to the \oiii\ luminosity $A_{\rm \oiii}<4$ (with the
median of the sample being $A_{\rm \oiii}=3.8$), to restrict
the impact of uncertainty in the extinction correction on our \oiii\ luminosity (3
objects are observed above this limit).  From this final sample we
observed 17 AGN selected to have suitable right ascension at the time
of observation and to be brighter than petrosian $\rm{r} < 17.0 $ in order to
achieve a signal--to--noise ratio of $\sim10$\,\AA$^{-1}$\,pixel$^{-1}$ at
$\lambda \approx $5500\AA\ at R$_{e}$.  The objects observed are shown
in Figure \ref{selection_plot} as the red circles, and details of each
object are given in Table \ref{obs_table}. 

\renewcommand{\tabcolsep}{4pt}
 
    \begin{table*}
  \centering
    \begin{tabular}{cccccrcccccccccc}
    \toprule
    Galaxy & Class & Comp & L$_{H\alpha}$ & W$_{80}$ & W$_{80, max}$ & $ |\Delta v |$ & $ |\Delta v | _{max}$ & v$_{1}$  & v$_{2}$  & v$_{3}$  & $\sigma_{1}$ & $\sigma_{2}$ & $\sigma_{3}$ & SRC   & Prob. \\
    (1) & (2)  & (3) & (4) & (5) & (6) & (7) & (8)  & (9)  & (10)  & (11) & (12) & (13) & (14) & (15) & (16) \\
    \midrule
    
    J095155 & P     & 2     & 43.18 & 317   & 415   & 37    & 195   & 133   & 68    &   ---    & 37    & 114   &  ---     & 0.43  & 0.24 \\
    J101927 & I     & 3     & 43.13 & 555   & 579   & 48    & 229   & -176  & 184   & 11    & 48    & 71    & 238   & 0.43  & 0.01 \\
    J102143 & I     & 3     & 42.72 & 576   & 1152  & 60    & 228   & -152  & 41    & 19    & 68    & 159   & 1072  & 0.61  & 3.18$\times$10$^{-5}$ \\
    J103600 & P     & 3     & 43.21 & 491   & 585   & 47    & 281   & 189   & -9    & -27   & 72    & 110   & 330   & 0.44  & 2.05$\times$10$^{-4}$ \\
    J103915 & I     & 3     & 43.21 & 673   & 1057  & 96    & 240   & 34    & 35    & -81   & 42    & 126   & 419   & 0.63  & 1.29$\times$10$^{-6}$ \\
    J111100 & T     & 3     & 43.46 & 736   & 1116  & 36    & 202   & -185  & 61    & -93   & 70    & 148   & 363   & 0.05  & 0.55 \\
    J124321 & I     & 2     & 43.33 & 406   & 645   & 72    & 299   & -6    & 8     &    ---   & 81    & 269   &   ---    & 0.52  & 1.76$\times$10$^{-5}$ \\
    J124859 & T     & 3     & 43.25 & 927   & 1569  & 48    & 178   & 22    & 33    & -67   & 68    & 202   & 515   & 0.23  & 0.01 \\
    J124859\_a & T     & 3     & 43.39 & 903   & 1617  & 48    & 250   & 34    & 45    & -44   & 60    & 198   & 539   & 0.32  & 9.08$\times$10$^{-5}$ \\
    J130116 & P     & 3     & 43.84 & 334   & 405   & 60    & 167   & 123   & -5    & -48   & 49    & 88    & 361   & 0.71  & 3.76$\times$10$^{-27}$ \\
    J133152 & T     & 3     & 43.50 & 834   & 1001  & 250   & 334   & -27   & 21    & -336  & 64    & 140   & 487   & 0.70  & 3.82$\times$10$^{-16}$ \\
    J141926 & T     & 3     & 43.40 & 385   & 529   & 60    & 204   & 79    & -11   & -151  & 55    & 99    & 342   & 0.64  & 3.63$\times$10$^{-16}$ \\
    J142237 & I     & 3     & 43.56 & 476   & 714   & 71    & 214   & -14   & -57   & -28   & 42    & 74    & 253   & 0.52  & 7.99$\times$10$^{-10}$ \\
    J143046 & P, T  & 3     & 43.39 & 565   & 687   & 25    & 245   & -247  & 53    & -48   & 71    & 142   & 415   & 0.24  & 0.04 \\
    J150754 & I     & 3     & 42.67 & 488   & 659   & 37    & 220   & 50    & -11   & 19    & 55    & 102   & 307   & 0.36  & 8.75$\times$10$^{-4}$ \\
    J151147 & I     & 3     & 43.43 & 408   & 528   & 36    & 240   & 102   & -2    & -18   & 65    & 94    & 283   & 0.61  & 5.48$\times$10$^{-12}$ \\
    J152133 & T     & 1     & 42.87 & 331   & 520   & 59    & 154   & -110  & 62    & -140  & 21    &   ---    &   ---    & 0.20  & 0.25 \\
    J152637 & I     & 3     & 42.88 & 468   & 517   & 37    & 148   & 106   & -107  & -36   & 44    & 98    & 226   & 0.08  & 0.34 \\

    \bottomrule
    \end{tabular}%
  \label{data_table}%
  \caption{A summary table of the results obtained from the sample. (1) Shortened galaxy name. (2) Class, a galaxy classification, I - Isolated, the galaxy has no near neighbours. P - An interacting pair, a galaxy undergoing a merger with a companion. T - Tidal tails, a galaxy with prominent tidal features. (3) Optimal number of components required in the central region of the galaxy (which is henceforth defined as the three or two component region of the galaxy depending on the value listed in column. (4) Log(\ha) luminosity in erg s$^{-1}$ derived from our multiple component fits to emission lines within D$_{em}$ as listed in Table \ref{obs_table}. (5) W$_{80}$ is the line width containing 80\% of the flux derived from fits to the H$\alpha$ line. The value listed is the median within the central region of the galaxy. (6) The maximum value of W$_{80}$ within the spatially coherent maximal component region. (7) $|\Delta v|$ is absolute value of the asymmetry parameter defined in equation 2, the value listed is the median within the central region of the galaxy. (8) The maximum of the absolute value of $\Delta$v within the central region of the galaxy. (9), (10), (11) Median velocity of component 1, 2, and 3 from the emission line fitting. If only 2 components are required, then only components 1 and 2 are listed. Errors on these values are typically 2\% for component 1, 6\% for component 2, and 30\% for component 3.  (12), (13), (14) Median dispersion for each component. Errors on these values are typically 4\% for component 1, 4\% for component 2, and 6\% for component 3. (15) The Spearman Rank Coefficient for the correlation between velocity dispersion and \nii/H$\alpha$ line ratio. (16) The probability of the observed correlation given the number of data points.}
\end{table*}%

\subsection{Observations}

The targets were observed in March and April of 2011 using the AAT's
SPIRAL IFU. The SPIRAL IFU is a 512 element 16 $\times$ 32 rectangular
array of fibres attached to the AAOmega spectrograph
\citep{Saunders2004, Sharp2006}. Each fibre subtends 0.7" on the sky,
approximately half the typical seeing. 

As a result of the large FoV of SPIRAL a main difference between our
data and the \cite{Harrison2014} data is the size of the IFU
pointing. Our field of view covers an average of 915 kpc$^{2}$
compared to their 100 kpc$^{2}$. This means that while our spatial
resolution is significantly poorer than theirs, we do capture the
entire galaxy within the frame rather than just the central
region. This difference will be important in later analysis as we
consider different regions of the galaxies.  

The AAOmega spectrograph splits
the data into two arms (blue and red), resulting in two datacubes for
each observed galaxy. The blue cubes encompass a wavelength range of
3700-5700\AA\ and the red from 6500-7600\AA. The red cubes have a
higher resolution than the blue, with $\rm{R}_{red}=5600$, while
$\rm{R}_{blue}=2141$, where $\rm{R}=\lambda/\Delta\lambda$ and
$\Delta\lambda$ is the full width at half maximum (FWHM) of the
spectroscopic point spread function. These resolutions correspond to a
velocity dispersion of $\sigma_{red}=23$\kmsec$~$ and $\sigma_{blue}=60$\kmsec.

\subsection{Data reduction}

The raw data from SPIRAL were reduced using {\small 2DFDR}, an
automatic data reduction program designed for use with the AAOmega
multi-object spectrograph \citep{Sharp2010}. This package turns the
raw CCD data received from the spectrograph into wavelength
calibrated, sky subtracted row-stacked spectra, with instrumental
signatures removed. 

The {\small 2DFDR} processing includes bias subtraction and
dark subtraction, which removes a number of cosmetic features of the
CCD in the blue arm of AAOmega.  Flat-fielding is carried out at two
levels, both using quartz lamp exposures.  First,  a set of flat-field
exposures with a defocussed spectrograph is used to provide relatively
uniform illumination across the whole CCD.  This is filtered to remove
large-scale variations (e.g. colour response) but leave in
pixel-to-pixel variations in CCD gain.  The flat-field is then applied to
the data prior to extraction of the spectra.  Once the
data is extracted to provide a spectrum per fibre (using a Gaussian
fit to the spatial PSF of each fibre), a second stage of flat-fielding
is applied, which uses an extracted fibre flat-field, which
corrects each fibre to the mean wavelength response of the system.
The relative calibration of the throughput of each fibre is done using
twilight sky exposures.  Wavelength calibration is carried out using
fits to emission lines in arc lamp exposures (containing copper, argon
and iron), and results in residuals typically at the level of
$\sim0.1$ pixels.  Sky subtraction is carried out by constructing a
mean sky spectrum from regions of the IFU around the edge of the array
which  do not contain flux from the target galaxy.  Typically these
were rows 1,2,31 and 32 in the $16\times32$ format of the SPIRAL
IFU (although in some cases this was tuned to the geometry of the
galaxy).  The field-of-view of SPIRAL meant that we did not require
extra offset sky exposures.  After this processing, which results in
row-stacked spectra, the data are aligned and combined into a
data cube which places spectra at their observed $x$ and $y$
coordinates on the sky using a bespoke IDL routine. At all stages of
the reduction the variance information is propagated, resulting in a
final variance cube to match the data cube. 

Flux calibration and telluric correction were performed using standard
stars which were observed together with the targets each night,
allowing the shape of the spectra to be corrected for any instrumental
or atmospheric effects.  The range of airmasses for the observed
targets was small, with a median of 1.25 and a maximum of 1.4.  As a
result differential atmospheric refraction effects between the \ha\
and \hb\ regions are significantly smaller (typically 0.25 arcsec,
with a max of 0.44 arcsec) than both the IFU spaxel size and the
atmospheric seeing (median 2.0 arcsec).  Given this, and the extra
spatial covariance in the noise properties of the data introduced by
resampling \citep[e.g.][]{Sharp2014}, we did not apply any
correction for differential atmospheric correction.  To account for
transmission variations through the night we then 
compare our datacubes to the SDSS spectra for each object.  We
create a summed spectrum from an aperture of diameter $3"$ to match
the SDSS fibre aperture and then apply a simple scaling of our cubes to
match the normalization of the SDSS spectra.

\section{Spectral analysis}

In this section we will discuss the analysis of the kinematic properties of the galaxies within our sample. As previously discussed, we expect the kinematics to contain information about the environment within these galaxies and how the AGN is affecting it. In particular, we want to find evidence for disturbances within the galaxies to show that outflows may be present. To do this we examine the stellar and gas kinematics firstly to observe how the galaxies are rotating, and further to see if the gas kinematics show signs of disruption by the AGN (as the stars will not be effected by winds). In Section \ref{stellar_sec}, we discuss the fitting of the stellar populations of the galaxies which were used to constrain the stellar motion allowing for the continuum of the galaxies to be subtracted. In Section \ref{non_param_sec} we discuss the first method used to look for signatures of winds, focussing on line broadness and asymmetry. Finally in Section \ref{fitting_sec} we will explain the method used by our fitting routine {\sc lzifu} (Ho et al., in preparation), and how we use it to separate out broad components in the spectra that are likely to represent winds. 

\subsection{Stellar kinematics}
\label{stellar_sec}

We fit the stellar kinematics of the sample using the Penalised Pixel Fitting ({\sc ppxf}) algorithm developed by \cite{Cappellari2004}. This routine uses simple stellar population (SSP) synthesis to fit both the velocity and velocity dispersion of the stars within the galaxy. 

In this paper we made use of the high resolution Gonzalez Delgado SSP template library from \cite{Martetal04} and \cite{GDetal04}.\footnote{The library was based on atmosphere models from PHOENIX \citep{HB99,Aetal01} and ATLAS9 computed with SPECTRUM \citep{GC94}. The ATLAS9 library is computed with SYNSPEC, and TLUSTY \citep{LH03}.} We selected nine solar metallicity templates aged between 0 and 17 Gyr to fit to our spectra. We run {\sc ppxf} on a spatially smoothed (3 pixel boxcar kernel) version of our data as the low S/N in the outskirts of the galaxies means that the velocities are poorly constrained. To retrieve the best possible fit to the velocities (disregarding all other information) we use a multiplicative polynomial of order eight and an additive polynomial of order three in addition to the templates which results in the most spatially continuous velocity maps possible. From this fit only stellar velocities are extracted, we do not consider the stellar dispersions. {\sc ppxf} is then run in conjunction with our emission line fitting (within the  {\sc lzifu} software) on the unsmoothed data to produce a best-fitting stellar template made up of a linear combination of the input templates with an appropriate velocity and velocity dispersion. During this run only the reddening keyword is used to account for dust, as in this case we are concerned with the exact continuum absorption. This information was then used to subtract the stellar component of the spectra, a necessary step to correct emission line fluxes for stellar absorption (particularly the Balmer lines). We note that because our emission lines are typically very strong, our results are not sensitive to the details of the continuum correction.

Derivation of the stellar kinematics also allowed us to observe the differences between the stellar and gaseous kinematics. This is useful as gas and stars react differently to disturbances within the host galaxy, such as feedback from the AGN or interactions with nearby galaxies. Due to their non-collisional nature, the stars are more immune to these effects and tend to reflect the overall gravitational potential of the galaxy, while gas is easily disrupted from ordered rotation. Since a wind will have no impact on the kinematics of the stars, these differences may be used to determine whether the wind is significantly disturbing the gas rotation.  

\subsection{Emission line analysis}

It is well established that the structure of emission lines in the central regions of AGN is very complex, often possessing apparent multiple velocity components along the line of sight \citep[e.g.][]{Villar-Martin2011,Veilleux2013}. An example of this complex structure may be seen in the bottom panel of Figure \ref{paraplot2}, where three distinct kinematic components are evident (two narrow components making up the horned profile, and underlying broad wings). While it has been suggested that emission lines with this double-peaked signature were due to the presence of two distinct AGN \citep{Comerford2013}, it has been demonstrated that an outflow with an uneven velocity distribution can lead to this profile \citep[e.g.][]{Fischer2011}. In order to gain better understanding of outflows we must disentangle these distinct components.  

\begin{figure}
\centering
 \includegraphics[width=0.48\textwidth]{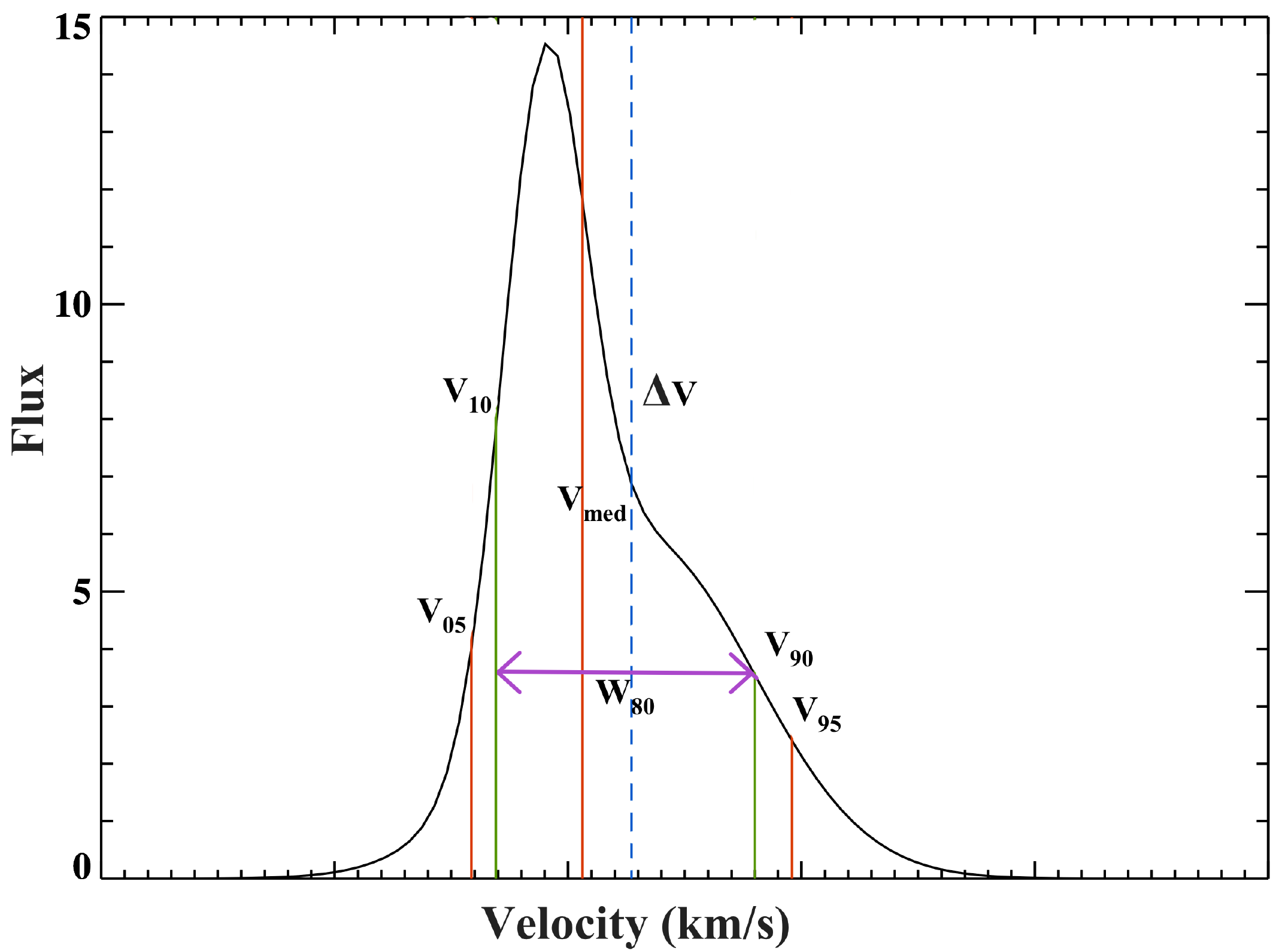}
\caption{A visualisation of the non-parametric measurements used in this paper. Plotted in black is an example fit to an emission line. The over-plotted lines represent the calculated velocities for the fit. The first is v$_{05}$ (red) - the velocity at 5\% flux, v$_{10}$ (green) - the velocity at 10\% flux, v$_{med}$ (red) - the median velocity, $\Delta$v (dashed blue line) - as defined in Equation 2 is a measure of the asymmetry of the line profile, v$_{90}$ (green) - the velocity at 90\% flux, and v$_{95}$ (red) - the velocity at 95\% flux. W$_{80}$ is defined as v$_{90}-\rm{v}_{10}$ which is the width that contains 80\% of the flux, shown by the purple arrow.}
\label{params}
\end{figure}

Two commonly used methods will be used to analyse fits to the AGN emission lines, the first is a non-parametric approach \citep[see Figure \protect\ref{params};][]{Harrison2014, Liu2013b} and the second considers these multiple components as signatures of the physical gas dynamics \citep[e.g.][]{Ho2014}. To model the emission profiles present in the data we used up to three Gaussian components for each emission line, requiring statistical significance above a chosen threshold to justify the higher order model (see Section \ref{stats_sec}). 

\subsubsection{Non-parametric emission line measures}
\label{non_param_sec}

\begin{figure*}
\centering
 \includegraphics[width=0.76\textwidth]{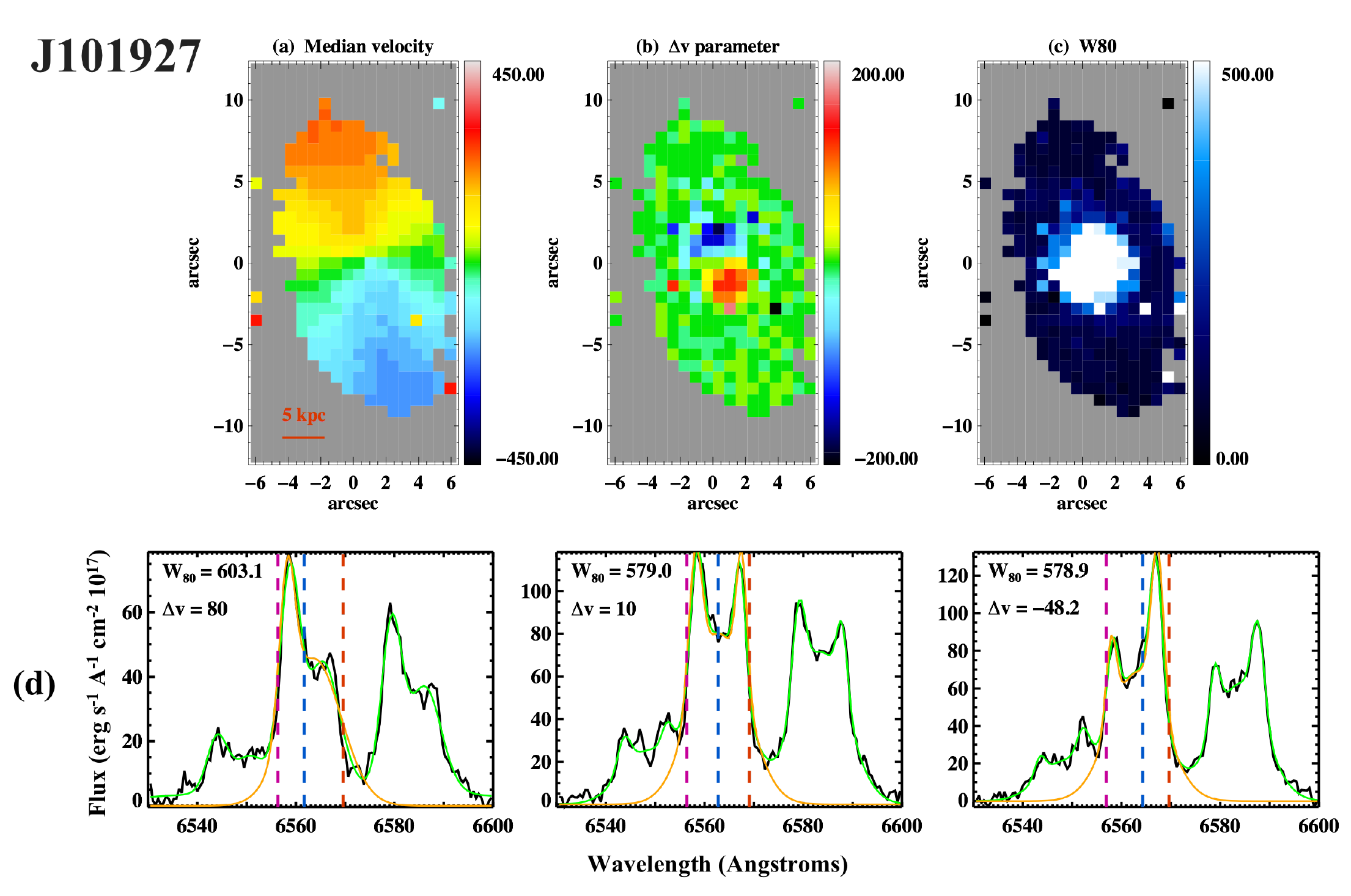}
\caption{All figure placements are the same as the above. This example serves to highlight how non-parametric measures do not fully capture the double peaked features of the spectra. In {\bf (d)} the double peaked emission lines can clearly be seen, here by only using the summed fit we are discarding extra information contained in the three component fit to the horned profile.}
\label{paraplot2}
\end{figure*}

Emission line properties may be derived using non-parametric line measures derived from the summed model fit produced by our multiple component fitting routine. We use the model spectra to counter low S/N in the outer parts of the galaxies, and to separate out the H$\alpha$ and \nii\ emission which is often blended. When using non-parametric measures we do not separate out components within the spectra (i.e. we do not separate the multiple Gaussian components), but instead use them to characterise the profile by making the measurements based on the summed model spectra. To do this several definitions from the literature are used, allowing us to compare our results to those found previously. A visualisation of these measures is shown in Figure \ref{params}. The definitions are as follows:

\begin{enumerate}
\item The median velocity of the primary emission line (H$\alpha$) is defined as the velocity that bisects the area under the fit, and is denoted v$_{\rm{med}}$. Asymmetry of the line profile is measured from this value. 
\item v$_{05}$, and v$_{10}$,  v$_{90}$, and v$_{95}$ are defined as the velocities at 5\%, 10\%, 90\%, and 95\% of the line flux, respectively.
\item The non-parametric line width, W$_{80}$, is defined as the width that contains the central 80\% of the flux: 

\begin{equation}
\rm{W}_{80} = \rm{v}_{90} - \rm{v}_{10}
\end{equation}
 
For a Gaussian it is approximately the FWHM (W$_{80} = 1.09$ FWHM for a Gaussian), and is used as a more sensitive measure of the broadness of the emission lines. 

\item Finally, an asymmetry parameter \citep[modified from][]{Harrison2014} is used to aid in comparison to their results. We define $\Delta$v as: 

\begin{equation}
\centering
\Delta \rm{v} =  \frac{v_{05} + v_{95}}{2} - v_{med}
\end{equation}
This asymmetry parameter allows for the velocity of the underlying broad component to be traced across the galaxy to examine the direction of the outflow.\footnote{A similar asymmetry parameter is used by \cite{Liu2013b}, originally defined by \cite{Whittle1985}. It produces qualitatively the same results, but is a ratio of velocities rather than a velocity. We chose to use $\Delta$v as it is conceptually easier to understand as a velocity.}

The  \cite{Harrison2014} definition of $\Delta$v uses the peak velocity of their primary emission line, \oiii. In place of this we use the median velocity as the zero velocity point, as done in \cite{Liu2013b}. We chose to do this as the peak velocity is not smoothly varying with spatial location in the majority of our sample due to the prominent double peaked emission lines in the centre of the galaxies (e.g. the spectra shown in Figure \ref{paraplot2}). As a result of this, the peak velocity shows non-physical gradients as the double peaked profiles vary rapidly in flux. We do not, however, claim that the median velocity of the line traces the velocity of the outflow as stated in \cite{Liu2013b}. Comparison with the stellar kinematics (see appendix) demonstrates that this value is in fact qualitatively closer to the overall rotation of the galaxies in most of our sample. In 9/17 galaxies we see evidence for large scale consistent rotation in both the gas and stars (e.g. J101927 and J124321, shown in Figure \ref{J101927} and Figure \ref{J124321} in the appendix). 8/17 galaxies show different gas and stellar rotation, of these six were found to be in visible mergers (e.g. J143046, Figure \ref{J143046} or J111100, Figure \ref{J111100}), while one is a barred spiral (J142237, Figure \ref{J142237}), and one is a regular looking spheroidal.
 
\end{enumerate}

\begin{figure*}
\centering
 \includegraphics[width=0.76\textwidth]{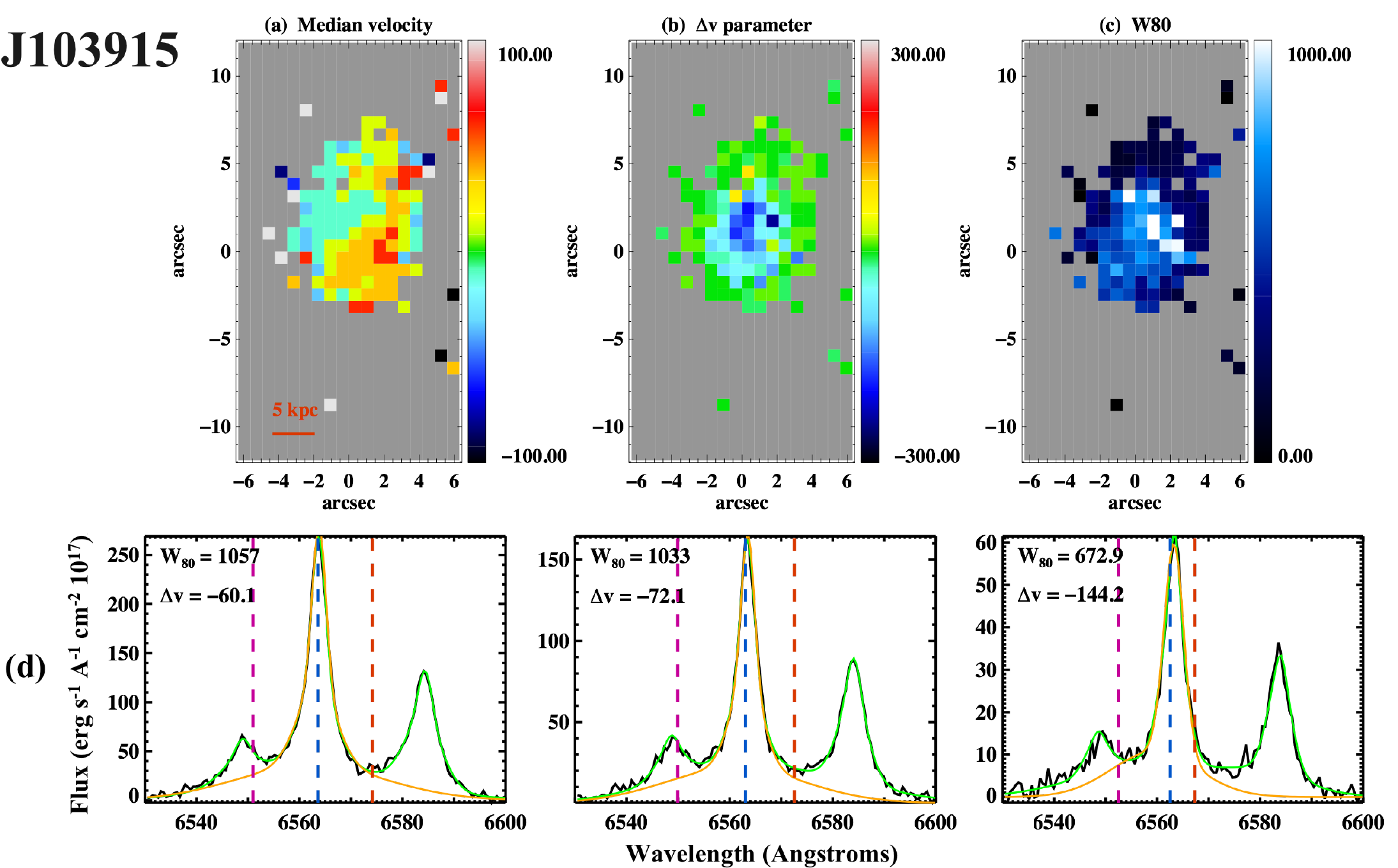}
\caption{An example of non-parametric emission line measures for J103915. {\bf(a)} v$_{med}$ is the median velocity (km/s) of the emission line, which is shown graphically on the spectra below as the blue dotted line. It is evident here that this galaxy shows only slight rotation (with velocity varying to approximately $\pm 45$ \kmsec\ on either side of the galaxy), meaning it may be relatively face-on. {\bf(b)} $\Delta$v parameter, blue represents a blue shifted asymmetry in the line profile, while green indicates no asymmetry. {\bf(c)} W$_{80}$ is mapped in blue with increasing line width as the colour lightens. It is clear that this object is broader in the centre, with W$_{80}$ exceeding 1000 \kmsec. {\bf(d)} The data and the H$\alpha$ fit for three adjacent central spaxels are plotted. The data is in black, the total fit is over plotted in green, the H$\alpha$ component of the fit only is plotted in orange. Several velocity measures are also plotted, purple is v$_{10}$ (velocity that bounds 10\% of the flux), blue is the median velocity (velocity at 50\% flux), and in red is v$_{90}$. Values for W$_{80}$ and $\Delta$v are printed in the top left corner of the plots.}
\label{paraplot1}
\end{figure*}

\begin{figure*}
\centering
\includegraphics[width=0.9\textwidth]{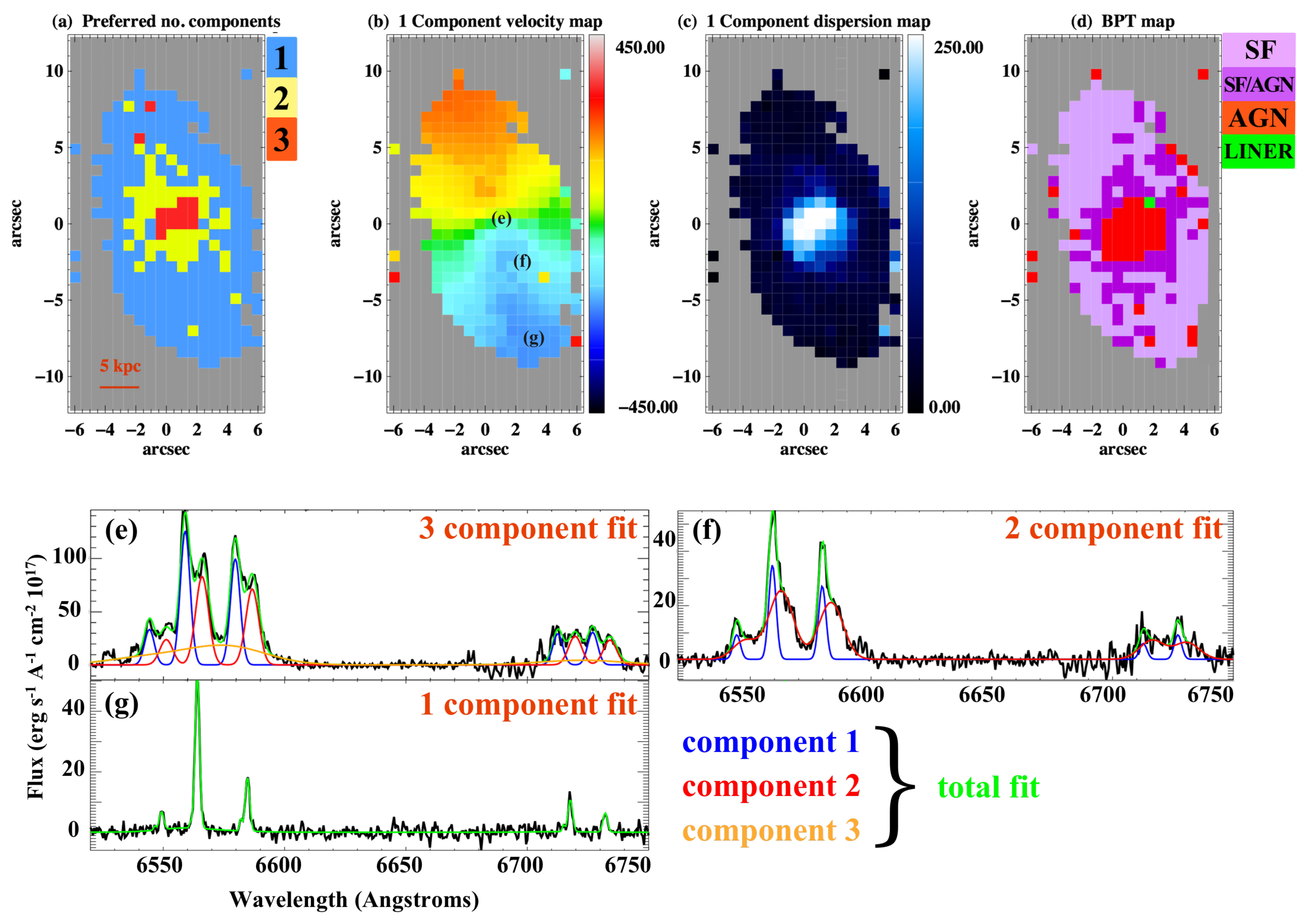}
\caption{An example of the results of a multiple component fit for galaxy J101927. {\bf(a)} A component map showing the number of components required. Grey represents data that does not meet the S/N cut (S/N $>$ 3 in \nii\ and H$\alpha$), blue is regions that require only 1 component, yellow is regions that require 2 components, and red is regions that require 3 components. {\bf(b)} - {\bf(c)} 1 component velocity and velocity dispersion maps for the whole galaxy. These are derived from the single Gaussian fit that is performed on all the data. While this is qualitatively similar to $v_{med}$, it is not the same and is derived in a different way. {\bf(d)} A BPT map, made using a BPT diagnostic plot from the line ratios log(\nii/H$\alpha$) and log(\oiii/H$\beta$), by classifying each spectrum then projecting the classification back onto the spatial dimensions of the galaxy. Here light purple represents star formation dominated ionization, purple is the composite region of the BPT diagram, and red is AGN. Shown are 3 example spectra for each component region of the galaxy. Approximate locations of the spaxel containing each spaxel are marked on the velocity map. {\bf(e)} 3 component fit where components 1, 2, and 3 are shown in blue, red, and orange respectively with the total fit over plotted in green. {\bf(f)} 2 component fit with components 1 and 2 plotted in blue and red, with the total fit in green. {\bf(g)} 1 component fit with only the green line representing the total fit. Distinct differences in the spectra are seen as the complexity of the emission lines increases greatly in the 3 component region.}
\label{multiplot}
\end{figure*}

\subsubsection{Emission line fitting procedure}
\label{fitting_sec}

When dealing with complex emission or absorption line spectra the standard approach is to use a function made up of multiple Gaussians. However, it is still an open question as to what should be done with these multiple component fits and how they should be interpreted. One approach states that nothing physically meaningful should be extracted from these fits, and that they are only needed to account for our inability to model non-Gaussian emission lines. If this is the stance taken, the appropriate solution is to use non-parametric emission line measures as done in much of the literature \citep[e.g.][]{Harrison2014, Liu2013b, Veilleux2013}. However, examination of our sample shows that distinct spatially varying velocity components are present in the majority (e.g. double peaked lines). As a result, we decided to model these components separately and to investigate whether they possessed any meaningful physical information (centroids, ionization, velocity, dispersion) as was done in \cite{Ho2014}.  

We used {\sc lzifu}, an emission and absorption line fitting program developed by Ho et al., (in preparation.) to model the emission line kinematics using single and multiple Gaussian profiles. {\sc lzifu} first subtracts the stellar continuum of the galaxies using {\sc ppxf} (in this case using only the reddening keyword to correct for the presence of dust), then fits the emission lines using {\sc mpfit}. This software has the capability to simultaneously fit the red and blue spectra taking into account the differences in resolution. Additionally, this feature allows for the typical BPT line diagnostics to be performed on the multiple component fits.

Each set of emission lines was modelled with a set of Gaussian functions constructed using the known wavelength values of each line. The central wavelength of the most prominent line (\oiii5007 in the blue, or $\rm{H}\alpha$ in the red) was allowed to vary up to a velocity of $600$\kmsec$~$ and the velocity dispersion up to $2000$\kmsec. We then add up to two more sets of Gaussians to the fit, meaning that each emission line is modelled by up to three Gaussian components. To constrain the fit, each set of Gaussians was required to have the same velocity and velocity dispersion (meaning their relative wavelengths were fixed), while their fluxes were allowed to vary. The reasoning for this is that each component represents a kinematically distinct part of the gas, meaning they may have different ionization states determined by their relative fluxes. This meant that, for example, the single component set of Gaussians fit to the red data (encompassing the \nii, H$\alpha$, and \sii\ lines) instead of having fifteen free parameters, only had six (a single velocity and velocity dispersion, flux of the $\rm{H}\alpha$ line, one flux defining the stronger \nii\ line, with the weaker line being locked to 1/3 of the stronger line's flux, and two fluxes for the \sii\ doublet).

It should be noted that {\sc lzifu} sorts components by dispersion. So for every spaxel component 1 is always the narrowest, component 2 is always the medium dispersion component, and component 3 is the broadest. Because of this, whenever we are discussing a wind or outflowing component we are primarily referring to component 3 (or component two if only two components are required as in J095155 and J124321) as this will have the greatest wind contribution.

\subsubsection{Model Selection}
\label{stats_sec}

To determine if the multiple components detected are statistically
significant, and not a better fit purely by virtue of the extra model
parameters, we performed a series of f-tests. The f-test is a standard
statistical test  to gauge whether a higher order model is preferable
to a simpler model when fitting a particular dataset. We set the false
rejection probability for the lower-order model to $10^{-5}$
(corresponding to a $\sim 3.9\sigma$ requirement that the more complex
model is preferred). This choice of probability results in spatially
contiguous regions of preferred number of components. Adjusting it
higher results in nosier component maps, while adjusting it lower only
slowly reduces the regions where the more complex models are
preferred. This means that the higher this threshold is set, the harder it is 
to justify the more complex model. 

Once the preferred number of components for each spaxel of the galaxy
has been determined, maps of this preference in each spaxel can be
made as shown in the top left of Figure \ref{multiplot}. By producing this we
are able to see where spatially these multiple components lie, and to
only examine the appropriate multiple component fits in these regions.

\begin{figure*} 
\centering
 \includegraphics[width=0.99\textwidth]{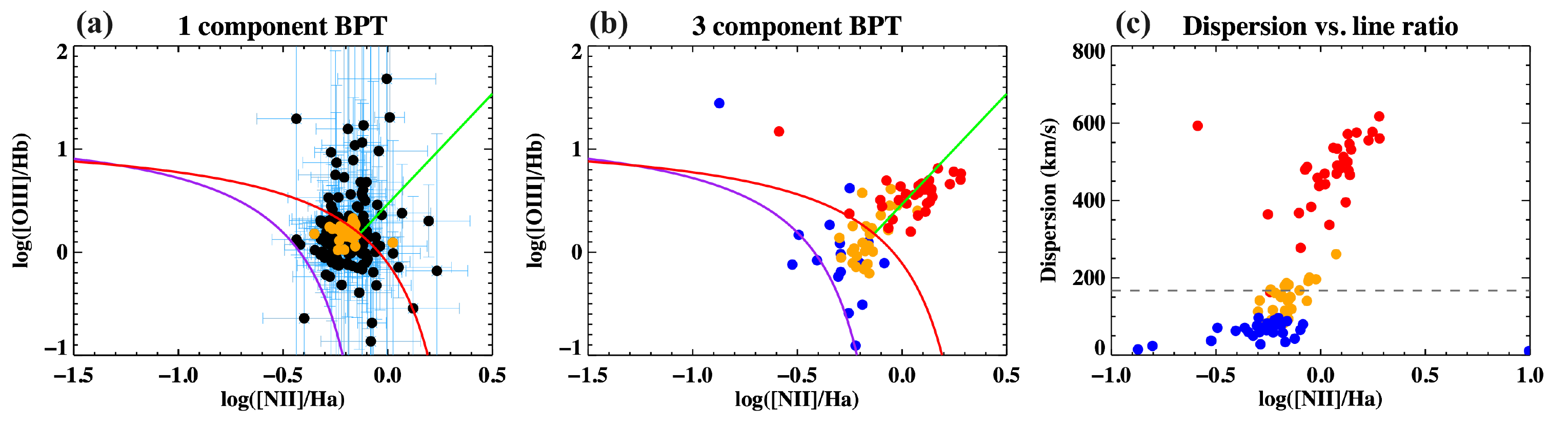}
\caption{{\bf(a)} A standard 1 component BPT diagram for J133152 with diagnostic lines over-plotted. Points from the three component region are colour-coded orange. The observed star formation cut-off from \protect\cite{Kauffmann2003} is plotted in purple, while their LINER classification line is in green. The red line shows the maximum theoretical starburst ionization calculated by \protect\cite{Kewley2001}. The orange points are the spaxels from the 3 component region, but fit with a single Gaussian, whose three component fits are then used in the following plots. We see that their 1 component fits place them clustered on the border of the AGN, LINER, and composite regions, but when the three component fits are examined we see clear trends between dispersion and ionization. {\bf(b)} A BPT diagram showing the location of each of the three components fitted in the centre of J133152. The blue points are component 1, orange are component 2, and red are component 3. {\bf(c)} Velocity dispersion plotted against $\rm{log}(\nii/H\alpha)$ line ratio, demonstrating the trend of increasing line ratio with dispersion indicative of a contribution from shocks. The colour coding of the points is the same as in (b) . The dashed grey line shows the median stellar dispersion within the 3 component region to demonstrate where gas dispersions are significantly higher than this value.}
\label{bptfig}
\end{figure*}

\subsubsection{Ionization diagnostics}

The ratios of the intensity of prominent emission lines may be used to determine the dominant ionising source, whether it be ionization from young stars, from a power-law accretion disk, or from shock heating \citep{Baldwin1981}. Due to their differing ionising spectra softer ionising sources such as star formation tend to excite permitted emission such as the hydrogen Balmer lines, and harder ionising sources excite greater emission in forbidden lines such as \nii\ and \oiii\ leading to greater flux in these lines. 

From the single component emission line fits discussed in the previous section we can construct BPT diagrams for each of the galaxies, as shown in Figure \ref{bptfig}. In addition to this, BPT maps of the ionization states across the spatial projection of the galaxies can be made, seen in the top right of Figure \ref{multiplot}. 

We are also able to determine the ionization states of the multiple components present in the central regions of the galaxies. We produce BPT diagrams and maps for each of the three components. We also plot the velocity dispersion of the three components versus their $\rm{log}(\nii/H\alpha)$ line ratio following  \cite{Rich2011} and \cite{Ho2014} to see how the components separate out in this parameter space, as shown on the right of Figure \ref{bptfig}. By comparing dispersion to ionization state we are testing for evidence of shocks as they are known to cause a strong correlation between the two \citep{Dopita1995}.

\section{Results}

\subsection{Non-parametric emission line measures}

We measured the value of W$_{80}$ for the emission lines in the central regions of our galaxies (defined as either the 2 or 3 component region dependent on the maximum number of components required for each galaxy, as listed in Table \ref{data_table}). The results of this are shown in columns (6) and (7) of Table 1. Column (6) contains the mean W$_{80}$ in the central region of the galaxy, and column (7) shows the maximum width within the spatially contiguous maximum component region. From this we find that our sample shows W$_{80, max}$ $\approx$ 400-1600 \kmsec, with an average of 794 $\pm\ 90$ \kmsec. This is significantly higher than is expected for quiescent galaxies whose central regions would generally have W$_{80} < \, 400 \: \rm{km\,s}^{-1}$ \citep[e.g.][]{Vega2001}. Since this broadness may not be attributed to normal rotation, we must conclude that is plausibly due either outflowing gas driven by the AGN or dynamical disturbances caused in merging systems (though we do not expect this to be so centrally peaked if it is due to merging, as we observe in our sample).

We may directly compare these results to previous studies performed by \cite{Harrison2014} and \cite{Liu2013b} who measure W$_{80}$ of the \oiii\ emission lines. \cite{Harrison2014} find very similar results, with W$_{80, max}$ ranging from 720 - 1600 \kmsec. Since our galaxies have similar luminosities we expect our results to be in agreement, and this is mostly true though Harrison's results are slightly higher. This may be attributed to their selection methods, which required a significant broad component with FWHM \textgreater\ 700 \kmsec\ to be present leading to their lowest W$_{80, max}$ being around that value (as W$_{80}$ = 1.09 FWHM). Since we did not make such a selection, our sample also contains some galaxies with narrower emission lines. 

\cite{Liu2013b} also find similar, but slightly larger, values; W$_{80, max} \approx $ 525-2142 \kmsec. Here it should be noted that this sample is at a considerably higher redshift than ours, z $\approx$ 0.55, and they were therefore able to select more luminous targets.  Our widths were measured using the H$\alpha$ rather than the \oiii\ emission line, which may introduce a slight difference as we are comparing a permitted emission line width to a forbidden width. However, since in the fitting code the velocity dispersion of the Balmer and forbidden lines are tied we do not expect this to introduce a large difference except in objects with significant contributions from the BLR (e.g. J102142). 

\begin{figure*}
\centering
 \includegraphics[width=0.75\textwidth]{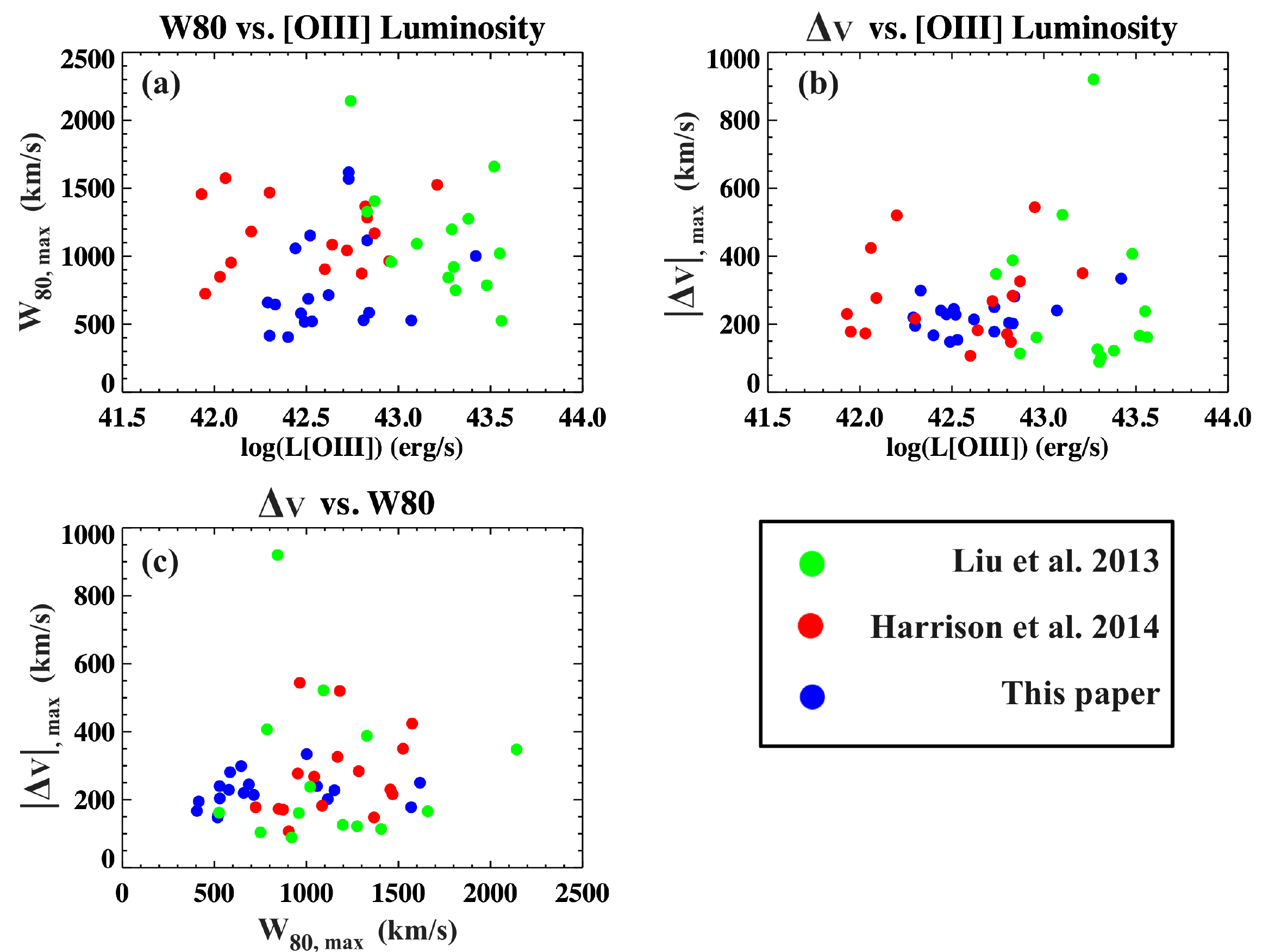}
\caption{Plots comparing emission line measures from this sample, \protect\cite{Harrison2014}, and \protect\cite{Liu2013b}. {\bf (a)} Plots log(L\oiii) luminosity against W$_{80, max}$, it is apparent that there is no correlation between the two. {\bf (b)} $|\Delta v|_{max}$ is plotted against log(L\oiii), again no correlation is seen.  {\bf (c)} Shows $|\Delta v|_{max}$ vs. W$_{80, max}$, there is still no clear correlation. }
\label{comparison}
\end{figure*}

In Figure \ref{comparison} we plot the results from our sample (blue), \cite{Harrison2014} (red), and \cite{Liu2013b} (green). The three values we consider are W$_{80, max}$ (the maximum value of W$_{80}$ within the spatially contiguous 3 component region), $|\Delta v|_{max}$ (the maximum value of $\Delta$v within the spatially contiguous 3 component region), and the \oiii\ luminosity from SDSS. It is apparent from these plots that there is no correlation between the parameters in the available data, despite the large range of luminosities across the three samples. This is not wholly unexpected as \cite{Mullaney2013} show that while there is a strong trend between line width (FWHM) and \oiii\ luminosity across a large range of luminosities, at the highest luminosities ($L_{\rm{\oiii}} \, > 10^{42}$) objects show a large variety of line widths. Since we made no selection to preferentially observe objects known to have moderately broad lines, we likely selected a larger number of these high luminosity, low line width objects than \cite{Harrison2014}.

Three galaxies show strong blue-shifted wings in their centre indicative of an outflow directed towards the observer. An example of this is shown in the centre of the top row of Figure \ref{paraplot1} ($\Delta v$ map) where velocities close to zero are represented by green, negative offset in velocity is bluer, and positive offset in velocity would be redder. It is clear in this example that there is a central blue asymmetry in the emission lines, which is coincident with very broad emission lines (evident from the centrally high W$_{80}$ shown to the right of the top row of Figure \ref{paraplot1}), which indicates an outflow. This observation is supported by these galaxies showing low amplitude (\textless 100 \kmsec) rotation potentially due to a close to face-on inclination, meaning that an outflow perpendicular to the disk would be directed along the line of sight to the observer.

\renewcommand{\tabcolsep}{3pt}
\begin{table}
  \centering
    \begin{tabular}{lcccc}
    \toprule
    Sample & Large-scale & $\Delta$v  & Blue  & Double\\
      & rotation & gradient  & $\Delta$v & peaked \\
    \midrule
    Harrison et al. 2014 & 50$\pm$13\%     & 31$\pm$12\%  &  31$\pm$12\% & 19$\pm$10\% \\
    Liu et al. 2013 & 71$\pm$14\%    &   43$\pm$15\% &  28$\pm$14\%  & 0\% \\
    This paper & 76$\pm$10\%    &  65$\pm$12\% &  17$\pm$9\% & 41$\pm$12\% \\
    \bottomrule
    \end{tabular}%
  \label{comparison_table}%
  \caption{A table showing the occurrence of rotation, gradients in $\Delta$v, blue-shifted $\Delta$v, and double-peaked emission lines within the three samples. The sample discussed in this paper has the highest occurrence large-scale rotation, $\Delta$v gradients, and double-peaked emission lines. While it also shows the lowest occurrence of blue-shifted $\Delta$v. Binomial errors are shown in each case. It should be noted that the much smaller IFU pointing that only captures the central region of many targets in \protect\cite{Harrison2014} may be the reason for the lower number of galaxies that show evidence for rotation. This is because without seeing the extended disk or outskirts of the galaxies it can be difficult to tell if there is large-scale rotation. }
\end{table}%

Both \cite{Harrison2014} and \cite{Liu2013b} find a larger fraction of galaxies with blue-shifted asymmetries (around 30\% of their samples), and a slightly lower fraction showing rotation-like gradients (between 31 - 43\%). Due to the very similar selection methods we expect to see the same trends, however asymmetry is more closely tied to the geometry and orientation of the target galaxies on the sky. In the case of the \cite{Harrison2014} sample this may be the result of their selection criteria, which required a broad (FWHM $>$ 700 \kmsec) component. Since the greatest apparent dispersion will be seen when an outflow is viewed face-on (if it is not spherically symmetric), selecting for broadness may bias a sample towards face-on outflows. When the errors on the percentages shown in Table \ref{comparison_table} are considered the difference between our conclusions becomes less stark, as the errors are large due to the small sample sizes. Our values then mostly agree, but we do still see a greater fraction of $\Delta$v gradients. It is noted that the lower number of galaxies showing large scale rotation in the \cite{Harrison2014} sample is potentially a result of their much smaller FoV.

In Table \ref{comparison_table} we list the percentages of galaxies
from this sample, \cite{Harrison2014} and \cite{Liu2013b} that have
apparent large-scale rotation, $\Delta$v gradients in the opposite
direction to the galactic rotation, and double-peaked emission
lines. We see $\Delta$v gradients opposite to the rotation across the
major axis of a large fraction of our sample (65$\pm$12\%). This is
not what we expect from an outflow perpendicular to the disk due to
the varied inclinations of the sample all showing the same
signature. The gradients in $\Delta$v are only seen in galaxies that
have what appears to be large scale rotation (the same is true of the
\cite{Harrison2014} and \cite{Liu2013b} samples).  This signature
would be a natural consequence of beam-smearing, due to atmospheric
seeing combined with a velocity gradient across the field and a
strong centrally-peaked flux distribution.  However, beam-smearing
cannot be used to explain the complex multi-component structure in the
central, most luminous spaxels, as these would naturally be less
impacted by beam-smearing.  The galaxies that show $\Delta$v gradients
also tend to have resolved multiple components in their central (and
surrounding) spaxels.  It is the changing contributions of these
different components that appears to drive the $\Delta$v gradients.  

Due to the uncertainty in the meaning of $\Delta$v, we do not use it
in any calculations or reasoning. Since W$_{80}$ is centrally peaking
it is much less effected by this issue, and the observed broadness is
much greater than could be achieved by beam-smearing of a regularly
rotating disk. So when discussing outflow velocities we rely on our
W$_{80}$ measurements.  Full kinematic modelling of these galaxies including the
effects of disk rotation, winds, seeing, and beam-smearing will be
undertaken in a later paper. The typical seeing is $\approx$ 2" and
exact values for each object are tabulated in Table \ref{obs_table}.

We find that all of the galaxies show abnormalities in their emission in the form of extremely broad and asymmetric lines. These properties are not expected of normal rotating discs, or even of luminous high redshift mergers whose line widths are generally $ < 600$ \kmsec \citep[e.g.][]{Harrison2012}. These properties are generally taken to be signatures of high velocity outflows in the gas \citep[see][]{Heckman1981, Liu2013b, Harrison2014}.

\subsection{Multiple components}

\begin{figure*}
\centering
 \includegraphics[width=0.99\textwidth]{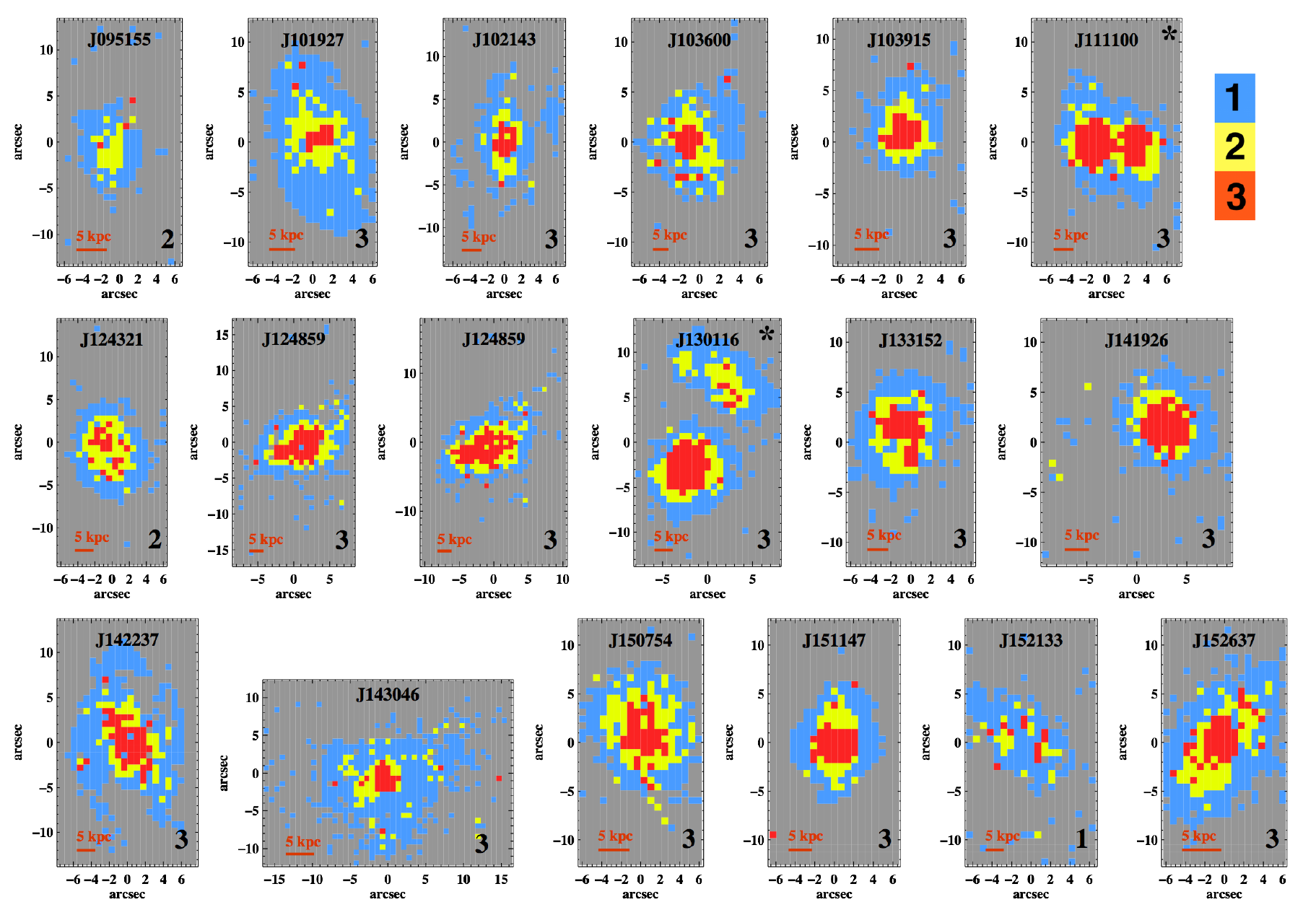}
\caption{Maps showing the preferred number of components in each region of every galaxy in the sample. The number in the bottom right corner indicates the maximum number of components required (and considered in the analysis) for each galaxy. As the legend on the side states, blue is 1 component regions, yellow is 2 components, and red is 3 components. The grey regions represent data that do not meet the required S/N cut (S/N $>$ 3 in H$\alpha$ and \nii) and are ignored throughout the analysis. Here it is clear that the vast majority of galaxies require 3 components in their central regions near the AGN, and then fewer as we move out into the disk. While this is likely partially S/N dependant, it appears to be primarily driven by line width, and complex structure. J111100 and J130116 are marked with asterisks as they will be discussed individually in Section 5.}
\label{compmaps}
\end{figure*}

Evidence for multiple components that meet both the requirements for statistical significance (outlined in Section \ref{stats_sec}), and are extended on scales greater than the spatial resolution (almost 3 spaxels, or $\approx$ 2") were found to be present in all but one of the sample, with the majority preferring a 3 component fit in their central region.  A visualisation of this can be seen in Figure \ref{compmaps} where the spatial maps of where each number of components are required for the whole sample are shown. It should be noted that the one galaxy that does not show evidence for multiple components has the poorest data of the entire sample, with a central S/N ten times lower than the average (J152133 in Figure \ref{compmaps}). Additionally the galaxies that show mainly two components, seen in Figure \ref{compmaps} (J095155 and J124321), also have low S/N (tabulated central S/N values may be found in Table \ref{obs_table}). The poor S/N of the data for these galaxies is likely partially due to the observing conditions, with seeing as high as 3.6". 

Several special cases may be picked out by examining the component maps in Figure \ref{compmaps}. We highlight J111100 and J130116 (marked with asterisks in Figure \ref{compmaps}). J111100 has two apparent nuclear regions with similar flux requiring 3 components. This galaxy is discussed in detail in the discussion section, and a full set of maps and spectra are presented in Figure \ref{J111100}. Throughout the analysis of this galaxy the two nuclei are not distinguished, therefore plots containing information from the three component region include both sources. J130116 is a merging pair of galaxies, the AGN is the galaxy in the bottom left of the frame. Complex emission regions are found in both the AGN and its quiescent companion (as evidenced by the two component regions in the companion). Examination of the spectra shows that this is valid and that there are complex emission line regions in the companion, likely as a result of the merger. When the three component region of J130116 is discussed we are referring to the central three component region of the AGN host, disregarding the scattered three component spaxels in the companion. 

While the maps in Figure \ref{compmaps} are convincing evidence of complex emission line profiles near the AGN, we cannot conclude that these galaxies possess winds from these alone. We carried out several tests to ascertain whether these multiple components were present due to winds and to conclude whether or not these different components may be rightly considered as separate physical entities. 

For multiple components to be considered they must contribute a significant amount of flux to the fit, and not be only low level corrections to wings on the emission lines. The narrowest component contributes the least flux to the fit, while components two and three contribute similar levels of flux. Typically the average contribution from the broad component is 36\% across the three component region, with all falling in the range 17 - 59\%.

\subsubsection{Coherence of the broad component}

\begin{figure*}
    \centering
    \includegraphics[width=1.0\textwidth]{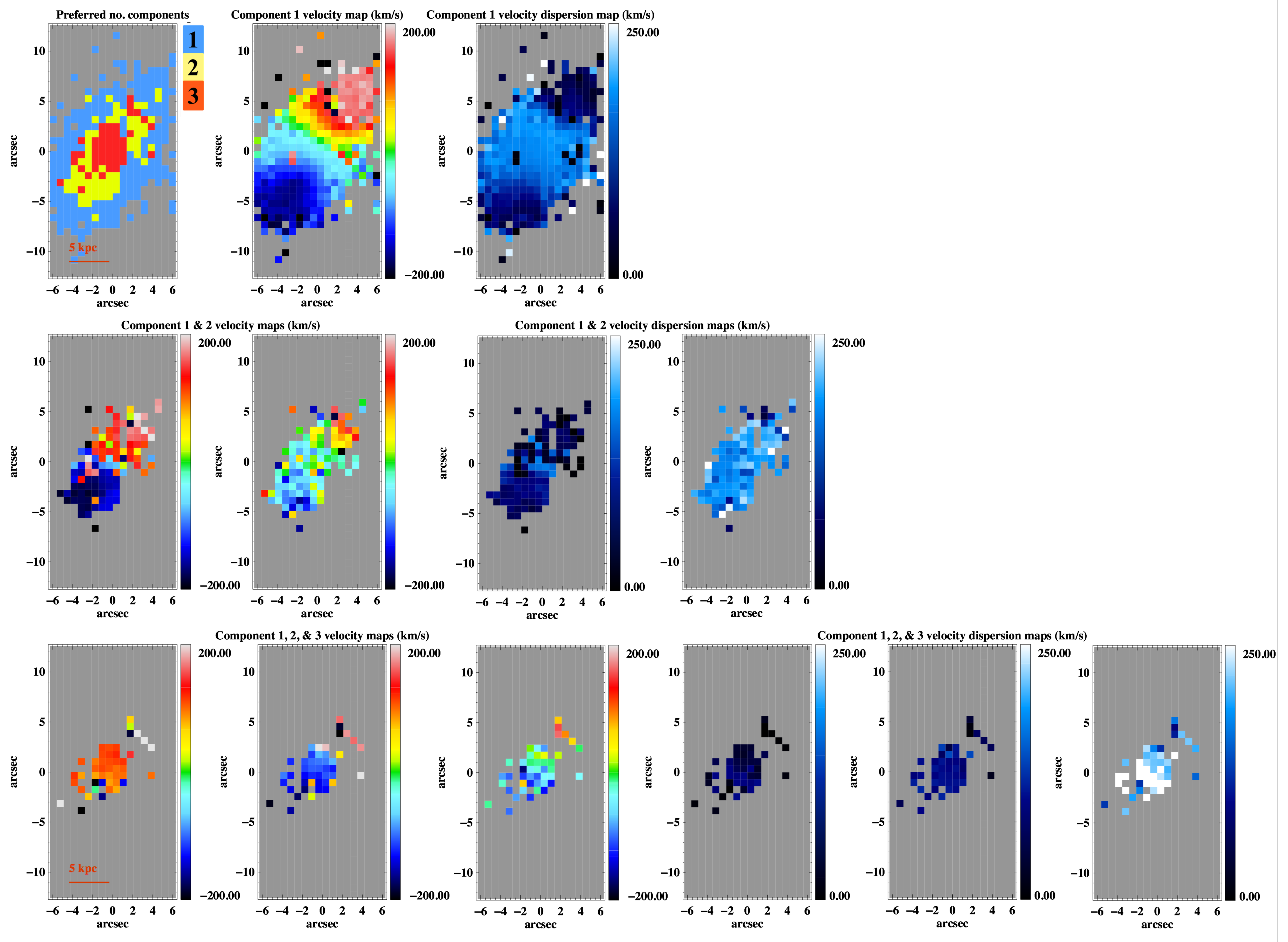}
    \caption{Example velocity and velocity dispersion maps for all fitted components of J152637. A map of where each number of components is the preferred model is shown in the top left, with the 1 component velocity and velocity dispersion maps next to it. The 1 component fits are plotted for data from the entire galaxy to provide a continuous view of the bulk velocity of the galaxy. The second row shows the velocity and velocity dispersion maps of the two component fits, which are plotted for the two and three component regions shown in the component map. Since the components are sorted by dispersion the velocity map on the left is the narrow component, and the one to the right is the broader component. The narrow component shows a velocity discontinuity in the centre due to the rapidly changing flux of the double peaked emission lines. The broader component appears to follow the overall rotation of the galaxy. The third row shows the velocity and velocity dispersion maps from the three component fits for only the three component region. Due to the double peaked nature of the spectra in the three component region we see a narrow redshifted component, a slightly broader blue shifted component and a broad component with a velocity gradient across the major axis of the galaxy. These components show spatial coherence. We see that each of the components are spatially smooth and show very little noise in the central region.}
    \label{components_ex}
\end{figure*}

We expect the wind to be most clearly represented by the high velocity dispersion component that is found. The narrow components may be partly affected by the wind, but the high velocity dispersion component must be dominated by wind (or Broad Line Region - BLR, if it is not completely obscured) emission. This means that for galaxies requiring 3 components the wind is most dominant in component 3, and for galaxies with 2 component it is component 2. If the broad component is merely accounting for the non-Gaussianity in the emission lines, then we expect the velocity of the broad component to largely trace the velocity of the narrow component, essentially following the bulk rotation of the galaxy. Alternatively, if the velocity of the broad component appears random and noisy showing no structure or coherence then it likely is not physically meaningful. However, if the velocity is coherent and different from the rotation of the galaxy then it may represent a real and separate kinematic component of the gas.

To determine whether a component is coherent we look for spatial smoothness in the velocity and velocity dispersion maps produced by the fitting code. We found all of the broad components appear coherent in both velocity and velocity dispersion, except J102143 which has an extremely broad underlying contribution from the BLR whose extreme broadness ($\sigma$ of up to 1800 \kmsec) leads to a poorly constrained velocity. The bottom row of Figure \ref{components_ex} shows example maps of velocity and velocity dispersion for the three component fit to J152637. From these we see that the velocities and velocity dispersions are spatially smooth.\footnote{Since {\sc lzifu} sorts component by dispersion rather than tracking them through velocity separation there are a few spaxels where components one and two switch due to changing dispersion and lead to a discontinuous velocity. This does not generally effect the broad component as it is much broader than components one and two.}

 \subsubsection{Ionization states of the multiple components}
 \label{ionization_results}
 
If the wind component (taken to be the third component, which will be the most dominated by the wind) is merely random or just the non-Gaussianity of the emission lines then the ionization state should either be incoherent and noisy or the same as the rotational component. However, if the ionization is both coherent and different from that of the narrow components then this is compelling evidence that the broad component traces a kinematically distinct gas component. 

\begin{figure*}
\centering
 \includegraphics[width=0.99\textwidth]{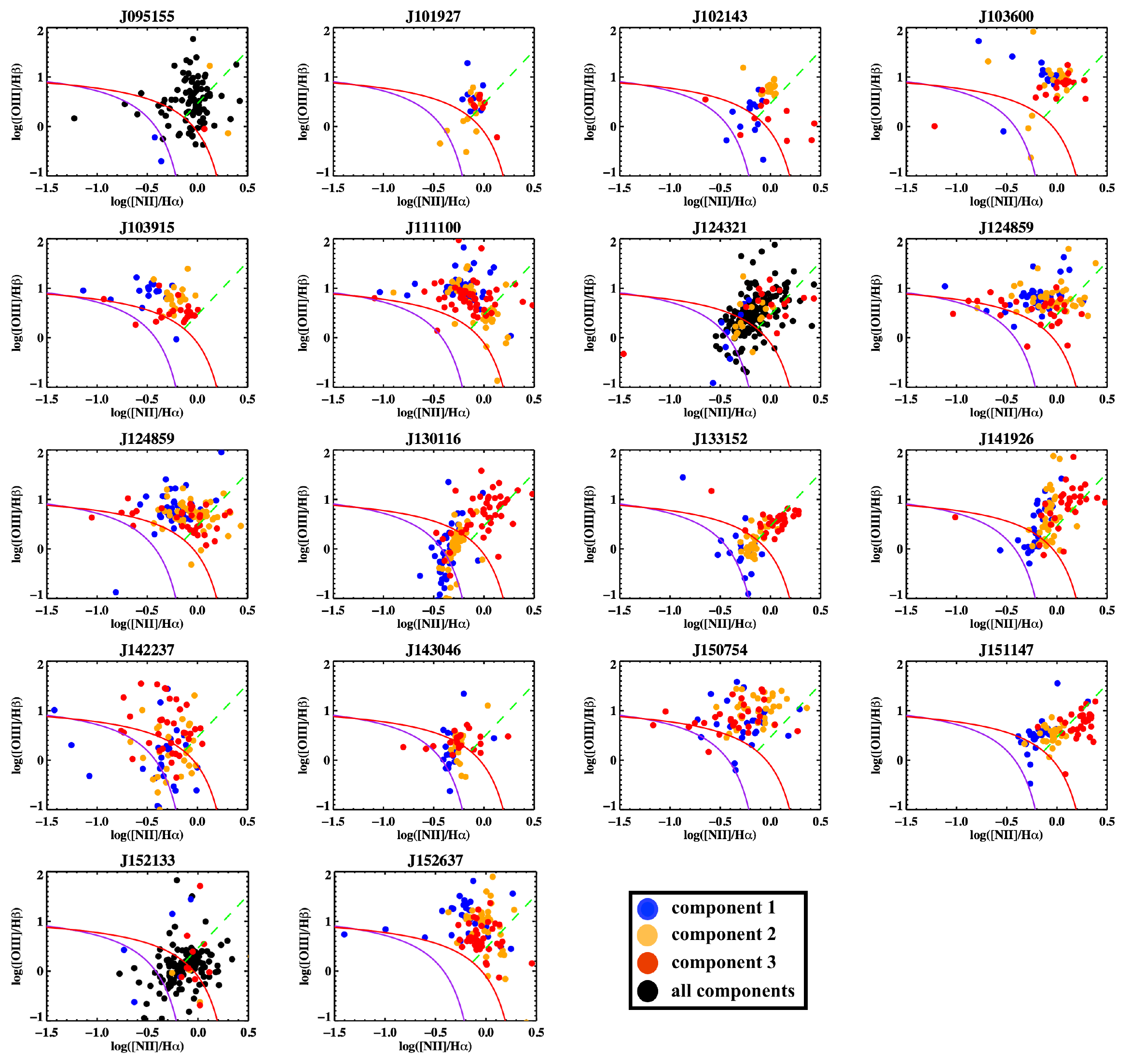}
\caption{BPT diagrams for all galaxies within the sample. As shown in the legend, blue points are the narrowest component, orange are the moderate component, and red is the broad third component. For galaxies without significant resolved 3 component regions points from the one and two component fits, where each is preferred, are plotted in black in addition to the 3 component points where they apply. The purple line is the observed maximum star formation line \protect\citep{Kewley2001}, red is the theoretical star forming limit \protect\citep{Kauffmann2003}, and the green line represents the division between Seyferts and LINERs.}
\label{bptall}
\end{figure*}

\begin{figure*}
\centering
 \includegraphics[width=0.99\textwidth]{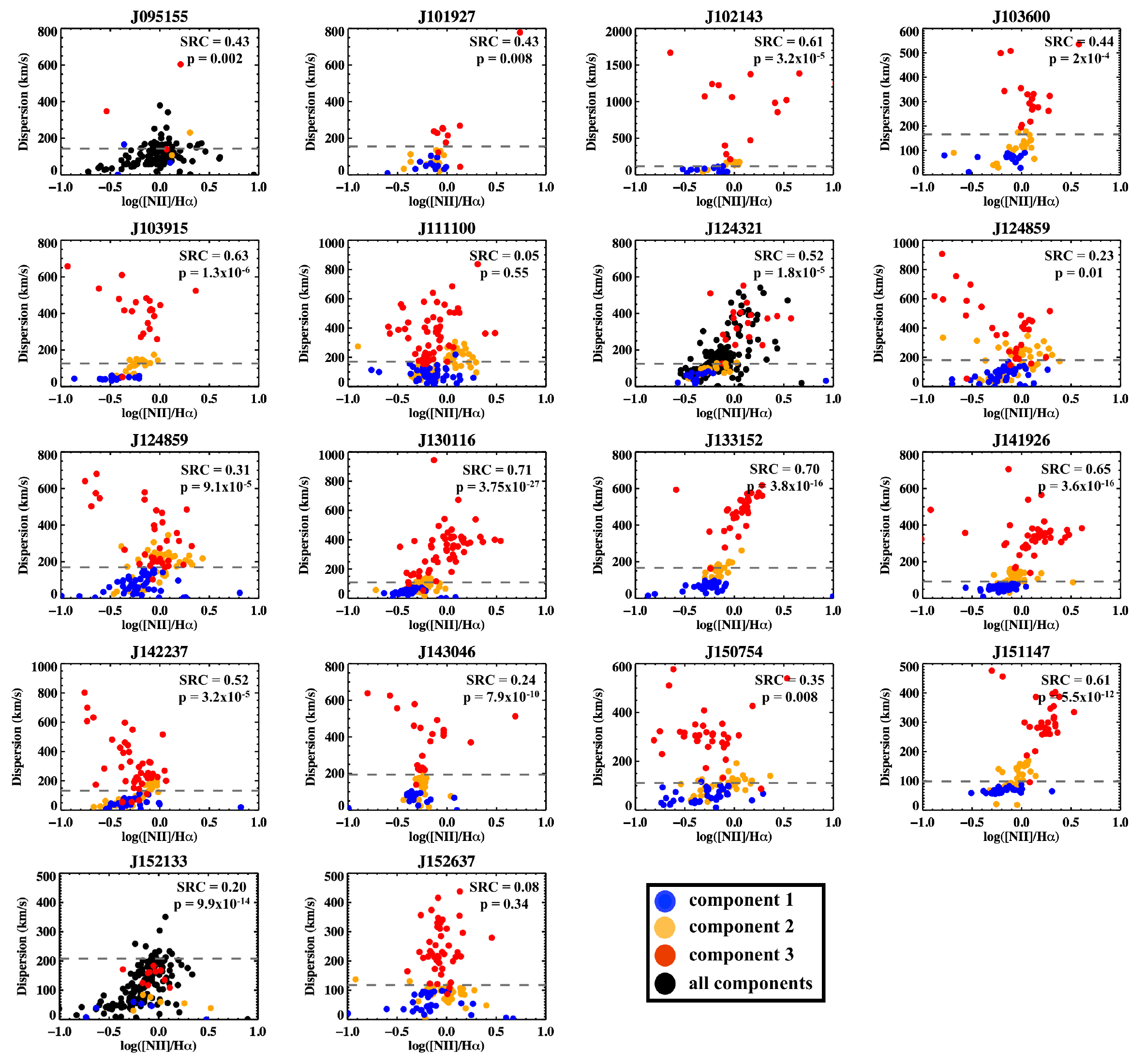}
\caption{Dispersion vs. $\rm{log}(\nii/H\alpha)$ for the entire sample. The legend is as in the previous plot, with the narrowest component plotted in blue, the moderate component in orange, the broad component in red, and 1 and 2 component fits where they are preferred in the 3 galaxies without significant 3 component regions. The horizontal dark grey dashed line shows the median central stellar dispersion. The Spearman rank correlation (SRC) coefficient and corresponding probabilities (p) are shown on each plot, the probability corresponding to the probability of the observed correlation in the data occurring by chance. In galaxies where the \nii\ emission drops at the highest dispersions we make a dispersion cut based on where the turn begins prior to calculating the correlation as including this would unnecessarily skew the results. This value was generally around 400 \kmsec.}
\label{dispratio}
\end{figure*}

To examine the ionization states of the multiple components found in our spectra we look at two diagnostic plots. The first of these is the typical \nii\ BPT diagram (shown in Figure \ref{bptall}), and the second is $\rm{log}(\nii/H\alpha)$ line ratio versus velocity dispersion (shown in Figure \ref{dispratio}). For galaxies with 3 components only points from the three component region are plotted. The blue points represent the narrowest component, the orange is the moderate dispersion component, and the red is the broad component. The three galaxies that did not show significant 3 component regions have all their high S/N spaxels plotted in black in Figure \ref{dispratio} (S/N $>$ 3 in \nii\ and H$\alpha$, spaxels where the fits have failed are also removed). The black points are from the 1 component fit in the the 1 component region and the two component fit in the two component region. Points from the scattered three component regions are over plotted in the same way as the other galaxies (as defined by the component maps in Figure \ref{compmaps}). This was done in order to retrieve a reasonable number of points.

Two general trends may be seen in these plots; the first is an increasing line ratio with dispersion which is seen in the vast majority. We used a Spearman rank correlation coefficient to determine the significance of this trend and found that 15/17 had a probability of $<$ 0.05 of the correlation being observed by chance. The exact values are shown in the top right corner of each panel in Figure \ref{dispratio} and are listed in Table \ref{data_table}. To calculate the significance of this trend we only consider points up to a limiting dispersion (typically 400 \kmsec, but tuned to each galaxy between 200 - 600 \kmsec). If the fits tend to lower \nii\ emission as dispersion increases then the limiting dispersion is set to the effective turning point (e.g. for J124859 in Figure \ref{dispratio} this limit is set to 400\kmsec). This is necessary as the behaviour of decreasing \nii\ with dispersion is likely due to BLR contribution, and is a separate correlation from the one considered here (though, it is discussed in detail below). The most prominent example of a strong increasing trend of line ratio vs. dispersion may be seen in J133152 (third from the left of the third row of Figure \ref{dispratio}). This trend implies that the kinematics and ionization state are coupled, suggesting that they both have the same physical cause. A correlation between these two quantities is often interpreted as strong evidence for shock excitation as photoionization is not known to cause such a trend \citep[see][]{Dopita1995,Monreal2006,Monreal2010, Rich2011, Arribas2014}. As the velocity of a shock front increases so does the excitation and observed line ratio \citep[e.g.][]{Allen2008}. For idealised planar shock fronts the observed shock velocity and velocity dispersion will not necessarily have similar values. This is because we would only be observing the shock at a single velocity. However, at a sufficiently wide opening angle the shock velocity and the measured velocity dispersion of the emission can be similar. We would be viewing shock fronts travelling in a range of directions relative to our line of sight, and at different velocities as a result of the inhomogeneous density of the medium.  This gives rise to the trend observed between line ratio and dispersion. \cite{Ho2014} show that given this assumption they are able to exactly reproduce their observed correlation between line ratios and velocity dispersion using shock modelling of their star forming galaxy. From this, we conclude that we are seeing shock excitation due to an AGN driven wind. We note that even though the galaxies emission is AGN-like in the central regions, we still see this trend likely resulting from to AGN/shock emission mixing. To fully understand what is happening in the future we will pursue full modelling of these mixing sequences.

\begin{figure}
\centering
\includegraphics[width=0.45\textwidth]{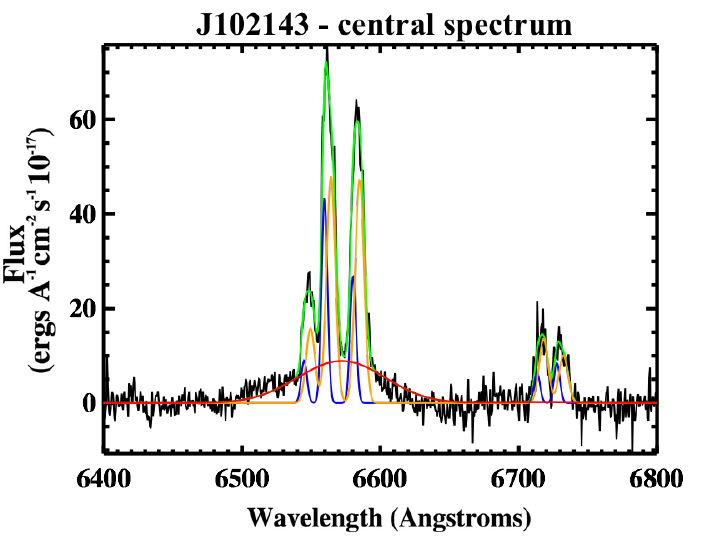}
\caption{A central spectrum for J102143, a galaxy showing a clear negative trend between \nii/H$\alpha$ and dispersion at the highest dispersions (see Figure \protect\ref{dispratio}). When the spectrum is examined, an underlying broad component in only H$\alpha$ is easily seen. This component has a velocity dispersion $>$ 1000 \kmsec and shows no emission in either \nii\ or \sii, which supports the theory that it is in fact coming from the BLR.}
\label{J102143_cent}
\end{figure}

Several galaxies show a correlation between line ratio and dispersion coupled with a turn at the higher dispersions back to lower \nii\ emission. Extremely broad emission lines with little to no forbidden line flux are suggestive of emission from the BLR, much closer to the SMBH. In order to determine whether this trend is genuine rather than due to low S/N we inspect the central spectra of the 5/17 galaxies that exhibit this behaviour (J102143 - an example spectrum is shown in Figure \ref{J102143_cent}, J103915, J124859, J142237, J143046). Of these five, three (J102143, J103915, J124859) show some BLR-like emission in their spectra (very broad emission only present in the Balmer lines). For two galaxies, J142237 and J143046, the fits to the spectra were found to be spatially unstable with some spaxels being fit with an extremely broad component and some without. This is due to the degeneracies involved in fitting multiple components to these galaxies. It should be noted that these inspections were performed on the entire sample, and others were found to show consistent and stable fits. While we do not claim to be viewing the BLR directly or alone in the remaining 3 galaxies, as we still see some forbidden line emission from the \nii\ lines, it is possible that we are seeing a portion of broad line emission alongside the dominant emission from the NLR in these type II AGN. 

\subsubsection{What do the components mean?}

If component three is taken to represent a kinematic component primarily due to the wind, then what are components 1 and 2 representative of? There are several possibilities, they may be due to the disk rotation and therefore be a quiescent star forming component, or an inflow towards the AGN, or another component of the AGN driven wind. In this section we will discuss these possibilities. 

While we would expect that, were they present, inflows would also be observed as separate offset kinematic components, we do not find evidence for them in our data. There are several reasons for this, most obviously we expect an inflow to be low ionization as it has yet to reach the AGN. We generally do not observe a low ionization component. Additionally we would expect an inflow to have low energy, and therefore to have relatively low velocity and velocity dispersion. Even the narrowest components of our fits have velocity dispersion $\approx$ 60 \kmsec. Finally, the kinematic signatures of inflows are likely only visible very close to the AGN and we are therefore unlikely to be able to resolve them due to our spatial resolution. In order to see signatures of inflows analysis of the central regions of much lower redshift galaxies would be required \citep[e.g.][]{Riffel2008}. 

Rather than an inflow it is likely that even the narrow components are at least in part due to the outflow. We may gather this from their generally high ionization. Through examination of the BPT diagrams and ionization versus dispersion plots (Figure \ref{dispratio}, \ref{bptall}) we can see that the three components tend to form a continuous trend in both dispersion and ionization. This leads to the conclusion that we have increasing contribution from the outflow. Component 1 has the lowest ionization and dispersion, however in most galaxies it still does not fall into the star forming region of the BPT diagram. Component two seems to be a transition component of both moderate dispersion and ionization likely due to increasing contribution to this component from the outflow. Component three then represents our wind dominated component with the highest dispersion and ionization.  

In Figure \ref{components_ex} we show all of the fitted Gaussian components for J152637. In this particular example we see that the broad components seen in component 2 of the two component fit (middle row, second from the left) and component 3 of the three component fit (bottom row, third from the left) are more like the overall rotation of the galaxy than the narrow components. When we examine the BPT diagram for this source (J152637, Figure \ref{bptall}) we see that the narrow components have higher \oiii\ ionization that the broadest component, which tends towards the shock region. From this we can conclude that none of these components are quiescent and all are due to the AGN and the outflow. This conclusion is supported by the velocities of two narrow components in the three component region (bottom left of Figure \ref{components_ex}). One of these components is blue-shifted and one is redshifted, which suggests that what we are in fact seeing is two sides of a biconical outflow alongside a broad underlying outflowing or shock related component represented by component 3.

\section{Discussion}

In this paper we have presented a sample of optically selected, local,
luminous type II AGN observed using the AAT's SPIRAL IFU. We have
demonstrated that winds are present throughout the sample by analysing
the data using non-parametric emission line measures and by considering
multiple kinematic components within the spectra. Now we will discuss
the implications of these results and the energies and masses involved.

There are several mechanisms thought to drive winds within
galaxies. The primary candidates are star formation
and black hole activity. We rule out radio jets as a likely cause due
to the radio-quiet nature of the galaxies.  Of our 17 sources 16 are
detected in FIRST \citep{1995ApJ...450..559B} at 1.4\,GHz (see Table
\ref{obs_table}).  In all cases
these are unresolved and the range in radio powers for the detected
objects is $1.5\times10^{22} - 4.3\times10^{23}$\,W\,Hz$^{-1}$.  Most
are fainter than the break in the local 1.4\,GHz luminosity function
at $P_{1.4}\sim10^{23}$W\,Hz$^{-1}$ \citep{2007MNRAS.375..931M} which
separates the regions where star forming galaxies and 
AGN dominate.  This leads us to infer that the kinetic energy from a jet
is not driving the outflows we observe.  In our multi-component fits
to the central regions of our galaxies we find that in most cases all
components sit in either the AGN or composite part of the BPT diagram
(see Figure \ref{bptall}).  Only one galaxy (J130116) shows evidence
of a component (the narrowest, blue points in Figure \ref{bptall})
which is dominated by star formation, that is, it lies below the
empirical Kauffmann et al.\ (2003) line.  While it is likely that at
least some of our targets also contain ongoing star formation in their
central regions, star formation does not dominate the ionizing
radiation field in any of them.

Using the central velocity dispersions of our galaxies as measured
from the SDSS spectra we estimate the BH masses in our objects.  We
use the $M_{\rm BH}-\sigma$ relation from Tremaine et al.\ (2002)  of
the form
\begin{equation}
\log(M_{\rm BH}/M_\odot)=8.13 + 4.03\log(\sigma/200{\rm km\,s}^{-1}).
\end{equation}  
Combining this with the bolometric correction factor for \oiii\ of 3500
\citep{2004ApJ...613..109H} we can estimate the accretion efficiency
in our galaxies.  The median $L_{\rm   Bol}/L_{\rm Edd}=1.3$ with a range of $0.2-7$.  There is large
uncertainty in this number due to scatter or uncertainties in bolometric corrections,
the relationship between galaxy dispersion and $M_{\rm BH}$ and the
extinction correction of \oiii, however, within an order of magnitude, we see
that our targets must be accreting near the Eddington limit.  As a
result it is unsurprising that they are able to drive strong
outflows.  The winds could be driven directly by radiation
pressure from the AGN accretion disk or indirectly through heating or
inflating bubbles.

\subsection{How much energy is in the winds?}
\label{energetics}

In this section we consider how much energy may be involved in the
winds present within the sample. Due to the nature of our data, we are
only able to approximate the energy present in the optical emission
lines. Since winds and outflows are generally thought to be made up of
not only ionised but also molecular and neutral gas
\citep[e.g.][]{Rupke2013}, we are calculating what is likely only a
small part of the energy involved.  In calculating the ionized gas
mass we follow \cite{Osterbrock2006}.  However, thanks to our broader
wavelength coverage, encompassing both \ha\ and \hb, we are able to
use the luminosity in the \ha\ line and explicitly correct this for
extinction using the Balmer decrement.  This is somewhat different to
other recent authors \cite[e.g.]{Liu2013b,Harrison2014} who have used
\hb\ and have only been able to use population average extinction
corrections.  The relation we use is
\begin{equation}
\frac{M_{\rm gas}}{0.98\times10^9 \,M_{\odot}} = \left(\frac{L_{\rm{H}\alpha}}{10^{43}\,\rm{erg\,s}^{-1}}\right) \left(\frac{\rm{n}_e}{100\,\rm{cm}^{-3}}\right)^{-1}
\end{equation}
where $L_{\rm H\alpha}$ is the extinction corrected \ha\ luminosity.
We assume that the entrained gas is confined to the complex emission
line regions near the centres of the galaxies, and therefore take the
extent of the ionised gas involved in the outflow to be defined by the
outer edge of the two component region (values for this are shown in
Table \ref{obs_table}). In this calculation we use the flux measured in all 
fitted Gaussian components. We do this because all of the components 
generally have high ionization and as a result are likely, at least partially, associated with 
the outflow.

We calculate the electron density, n$_e$, based on the measured ratio
of the \sii\ emission doublet. We find an average ratio of 1.3
(typically varying between 1 - 1.7) that gives an n$_e \, \approx 100
\rm{cm}^{-3}$ according to the relation provided by
\cite{Osterbrock2006}, which is consistent with other results
\citep{Greene2011}. Using this we find an ionised gas mass,
M$_{\rm{gas}} = (1-20) \,\times\ $10$^9$ \mdot, with a median of 6
$\times\ $10$^9$ \mdot. This value for gas mass may then be converted to an energy
simply using; 

\begin{equation}
\rm{E}_{kin} = \frac{1}{2} \, \rm{M}_{gas} \, \rm{v}^2_{gas}
\end{equation}

We take the gas velocity is taken to be W$_{80}$/1.3 as done by
\cite{Harrison2014} and \cite{Liu2013b} to approximate outflow
velocities of wide opening angle bi-conical outflow models.\footnote{Note that use of the traditional $E = \frac{1}{2}M(v^2 + \sigma ^2)$ yields results of the same order of magnitude, but we do not use this due to the dependance of velocity on orientation which leads to a wide variety of observed velocities.}  From this
we find E$_{kin} = (1 - 36) \times$ 10$^{57}$ erg, with a median of 7
$\times$ 10$^{57}$ erg. This value is slightly higher than what \cite{Harrison2014} find, likely since they do not correct for extinction. This is supported by the fact that \cite{Liu2013b} find a very similar number using an average extinction correction.

Using the extent of the ionised gas regions within
the sample and the gas velocity used above (W$_{80}$/1.3 ) we may then
use; $v_{\rm out} = D_{\rm em} / 2t_{\rm out}$ to determine a time-scale for
the outflow (where $D_{\rm em}$ is the diameter of the region
containing gas ionized by the AGN). This leads to a value of t$_{out} \, \approx 6$ Myrs,
allowing for a time averaged wind power (in ionized gas) to be
calculated, resulting in $\dot{\rm{E}}_{out} \approx (5 - 190) \times
10^{42}$ erg s$^{-1}$.  If we then calculate the median ratio of wind
power to bolometric luminosity we find it to be $\sim0.02$ percent.  This
demonstrates that the AGN is easily able to drive the outflows, even
accounting for the extra mass in neutral atomic or molecular gas not
seen in our study.

To calculate an order of magnitude mass outflow rate we assume that
the outflows are spherically symmetric. Following \cite{RZ2013} and
\cite{Harrison2014}, and assuming our outflow velocity is
approximately equal to W$_{80}$/1.3 (we do not
use $\Delta$v here due to the uncertainty in what this quantity truly
represents in our sample, and the unknown degree to which it is
effected by beam smearing) we find outflow rates, $\rm{M}_{out} \approx 370 - 2700 $ \mdot
yr$^{-1}$. These values are well below the upper limit of 1.9 $\times\ 10^4$\mdot yr$^{-1}$ derived by \cite{Liu2013b}.
Our derived outflow rates are dramatically larger than outflow rates typical of the nearest low luminosity AGN, 
which are generally on order 1.0 \mdot yr$^{-1}$ \citep[e.g.][]{Riffel2013, SB2010}, this difference is likely due to the much greater luminosity of our sample. 

\subsection{Outflows and shocks}

We showed in Section \ref{ionization_results} that most of the
sample have a significant correlation between dispersion and
excitation. This trend is generally taken to indicate the
presence of shocks, as it is evident that in shocked gas, excitation
must be linked to the velocity of the shock front
\citep{Dopita1995}. Further, in wide-angle outflows a range of
velocities will be observed due to a range of outflow directions
compared to our line-of-sight.  As a result we expect the measured
line-widths to correlate well with the typical shock velocities in the
outflows (i.e.\ we always see some gas with its velocity parallel to
our line-of-sight).  Therefore, the natural explanation for the observed
correlation between excitation and dispersion is the presence of
shocks in the gas.  It is worth noting that we see
correlations between line dispersion and ionization even when most or
all of the spaxels in a galaxy are above both the LINER and
theoretical maximal star formation line in the BPT diagram.  We
therefore expect the ionization in these objects to be due to a mix of
photo-ionization from the AGN and shock ionization.

Having seen high velocities indicative of outflows it should not be
surprising that we find evidence of shocks in the surrounding gas, as
a high velocity outflow propagating through a medium is expected to
cause shocks. Evidence supporting this has been found in several
different kinds of galaxy such as luminous infrared galaxies
\citep{Monreal2006, Monreal2010}, mergers \citep{Rich2011}, and star
forming galaxies \citep{Ho2014}.  

We finally note that our selection method potentially biases against
the most shock dominated AGN, as we exclude objects with central (SDSS)
spectra than indicate they lie below the LINER/Seyfert divide in the
BPT diagram.  That said, we note that above $\log[L_{\rm
  \oiii}/L_{\odot}]=8.7$ only $\simeq20$ percent of objects lie below this
line.

\begin{figure}
\centering
\includegraphics[width=0.48\textwidth]{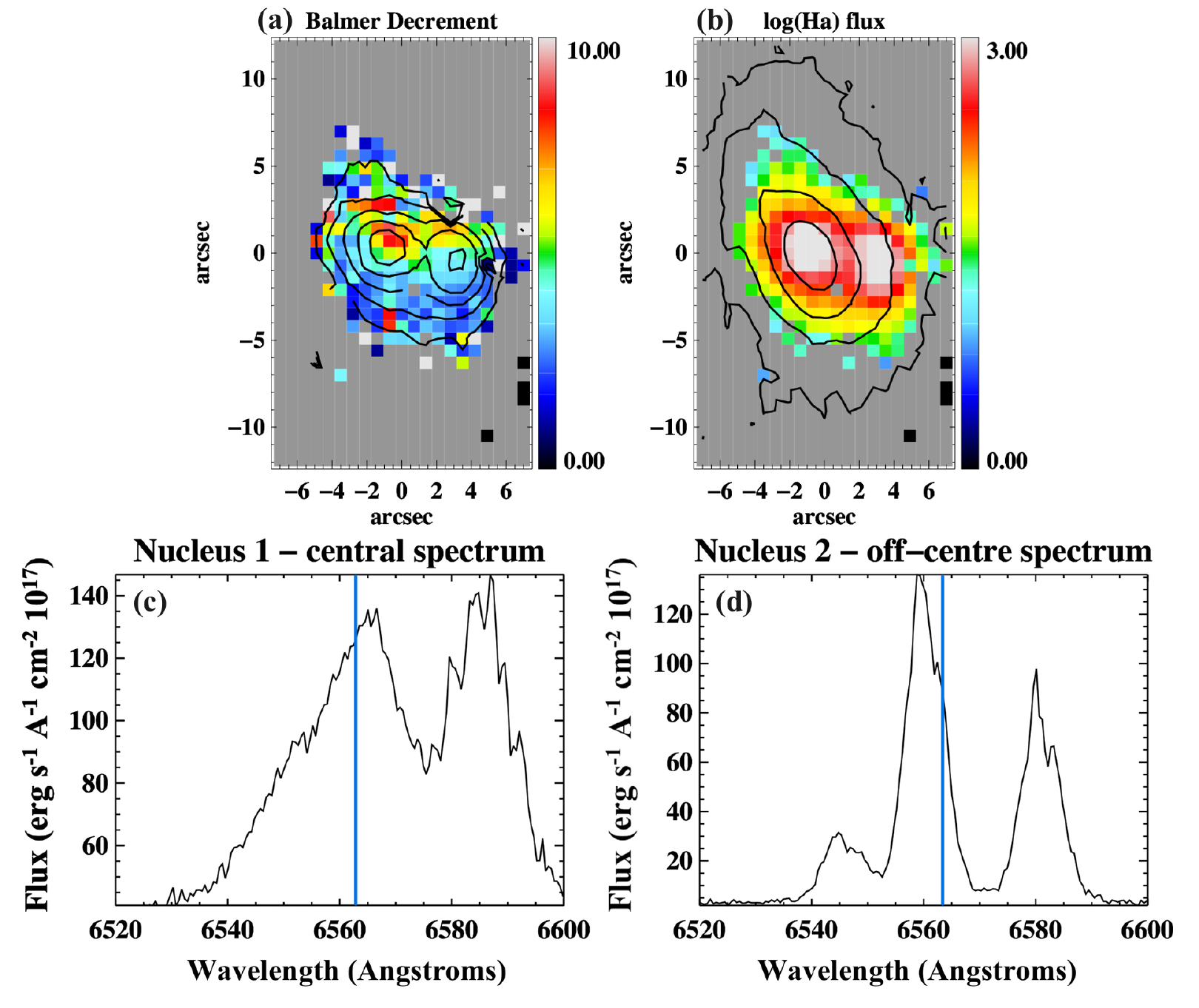}
\caption{{\bf (a)} Balmer decrement map for J1111000 with contours
  spaced by one magnitude intervals of \ha\ flux overlaid.  {\bf (b)} \ha\ flux
  map in 10$^{17}$erg s$^{-1}$ \AA$^{-1}$cm$^{-2}$ with stellar continuum contours from the blue data overlaid.  {\bf (c) - (d)} spectrum of central emission line
  peak.  Bottom right: spectrum of offset emission line peak.  For
  both spectra the vertical line shows the location of \ha\ at rest.}
\label{J111100_extra}
\end{figure}

\subsection{J111100}

The galaxy J111100 shows a prominent double peak in its emission line
flux.  The galaxy appears morphologically disturbed with tidal
features and is a possible merger remnant.  The distribution of the
\ha\ flux compared to the continuum light from the blue data is
shown in Figure \ref{J111100_extra} (top right).  The \ha\ flux peak on the
left hand side aligns well with the continuum (contours) suggestive of
a typical centrally located AGN.  The \ha\ flux peak on the right hand
side has no significant continuum peak associated with it.  When we
examine the galaxy's kinematics we find that the stars are rotating
regularly, but that the gas shows no large scale rotation and has
several discontinuities (see Figure \ref{J111100}).

An average spectrum from a 3 $\times$ 3 spaxel bin is plotted in
Figure \ref{J111100_extra} centred on the two \ha\ flux peaks. The central
spectrum is very broad with blended emission lines, this is indicative
of an outflow driven by the central AGN or potentially a contribution from
the BLR. The second source, however, shows much narrower emission
lines (though the dispersion is still $\approx$ 200 \kmsec).  When we look
at the relative fluxes of the two sources, we find that the off-centre
source is in fact brighter (observed \ha\ flux) than the central
source by $\approx$ 20 percent.  This is somewhat surprising, but can
be explained by the greater extinction see towards the line of sight
of the central source (see the Balmer decrement map in Figure
\ref{J111100_extra}).

There are several possible explanations for the structure in this source, which
we discuss in turn.  Given that both peaks show line ratios consistent
with AGN ionization, we could argue that we are seeing two separate
AGN (i.e. two accreting black holes).  However, the lack of any significant
continuum associated with the offset emission line source means that
this is unlikely unless one black hole is ejected from its host during
the merger process.  While we cannot rule out such an ejection
hypothesis, the narrow-emission-line gas we see will not
in general be gravitationally bound to the black  hole, so an ejected
black hole would likely have had to be ejected in the direction
of a sufficiently large gas cloud that it could then ionize. 
Alternatively we are seeing ionization from a single black hole.  In
this case the morphology we are seeing may be driven by other
characteristics of the galaxy.  For example, could a dust lane cause
the two peak nature of this source?  We find this to be unlikely as a
map of Balmer decrements (Figure \ref{J111100_extra}) does not show an
excess between the two peaks.  We are thus left with the hypothesis
that the offset emission line flux is due to gas ionized by the
central AGN.  The surprising feature here is the high luminosity of the
offset emission-line peak at a distance of at least 6 kpc (projected) from the nucleus. The central AGN has a log(H$\alpha$) luminosity of 42.76 ergs$^{-1}$ before extinction correction, and 43.53 ergs$^{-1}$ after correction for extinction. 
The off-centre source has a similar luminosity prior to extinction correction, log(H$\alpha$) of 42.6 ergs$^{-1}$, and due to the lower extinction at the location of this source this becomes 42.99 ergs$^{-1}$ after correction. This object is discussed in \cite{Keel2012} as a part of a search for AGN that show emission line clouds similar to the Hanny's Voorwerp discovered by the Galaxy Zoo project. This luminosity ratio between the central AGN and the cloud is very similar to the objects in the Keel et al. survey that have sufficient data for these calculations. For J111100, they are not able to resolve the cloud and suggest that follow-up spectroscopy will be required to decipher what is going on in this system. 
An ionised gas cloud is the most plausible explanation, but we will discuss this galaxy further in a
separate paper, including additional data to further  diagnose its
physical nature. 

\section{Conclusions}

In this paper we have presented a kinematic analysis of 17 local, luminous, type II AGN observed with the SPIRAL IFU on the AAT. By selecting this sample, our aim was to determine how widespread winds and outflows are in luminous AGN through analysis of the detailed information provided by spatially resolved spectroscopy. Through analysis of the kinematics we determined that winds were detected in some form in the entire sample. Here we present a summary of our conclusions:

\begin{enumerate}
\item We find complex emission features suggestive of winds in 100\%
  of the sample, while 16/17 galaxies (and all those with high S/N)
  show strong evidence for significant multiple velocity components in
  their spectra.  
\item Of the galaxies that possess multiple components all have
  properties that suggest they have a wind component; high
  dispersion, high excitation.  
\item Examination of the ionization states of our multiple components
  showed a correlation between the kinematics and the line ratio in
  15/17 of the galaxies, implying that both are being driven by the
  same source. The majority (12/17) showed a positive trend,
  indicating the presence of shocks. A further 3 galaxies show this
  trend alongside a turn back to lower \nii\ emission at the highest
  dispersions which is evidence for a contribution from BLR emission (which was
  confirmed through examination of their spectra).  

In future work we will aim to more fully characterize the outflows in
these galaxies by combining dynamical models of the stellar kinematics
with modelling of the outflows, including the contribution from
shocks.  Galaxies from the SAMI Galaxy Survey \citep[see][]{Croom2012}
will also be used as a control sample to see how AGN feedback varies
with AGN power by selecting lower luminosity AGN. 

\end{enumerate}

\section*{Acknowledgments}

We thank the anonymous referee for their helpful comments which improved the clarity of the paper. 
This research was based on data taken with the Anglo-Australian
Telescope.  Parts of this research were conducted by the Australian
Research Council Centre of Excellence for All-sky Astrophysics
(CAASTRO), through project number CE110001020.  SMC acknowledges the
support of an ARC future fellowship (FT100100457).

\bibliography{library_edit}

\begin{thebibliography}{64}
\expandafter\ifx\csname natexlab\endcsname\relax\def\natexlab#1{#1}\fi

\bibitem[{{Abazajian} {et~al}\mbox{.}(2009){Abazajian}, {Adelman-McCarthy},
  {Ag{\"u}eros}, {Allam}, {Allende Prieto}, {An}, {Anderson}, {Anderson},
  {Annis}, {Bahcall}, \& et~al.}]{Abazajian2009}
{Abazajian} K.~N. {et~al.}, 2009, \apjs, 182, 543

\bibitem[{{Allard} {et~al}\mbox{.}(2001){Allard}, {Hauschildt}, {Alexander},
  {Tamanai}, \& {Schweitzer}}]{Aetal01}
{Allard} F., {Hauschildt} P.~H., {Alexander} D.~R., {Tamanai} A., {Schweitzer}
  A., 2001, \apj, 556, 357

\bibitem[{{Allen} {et~al}\mbox{.}(2014){Allen}, {Croom}, {Konstantopoulos},
  {Bryant}, {Sharp}, {Cecil}, {Fogarty}, {Foster}, {Green}, {Ho}, {Owers},
  {Schaefer}, {Scott}, {Bauer}, {Baldry}, {Barnes}, {Bland-Hawthorn}, {Bloom},
  {Brough}, {Colless}, {Cortese}, {Couch}, {Drinkwater}, {Driver}, {Goodwin},
  {Gunawardhana}, {Hampton}, {Hopkins}, {Kewley}, {Lawrence}, {Leon-Saval},
  {Liske}, {L{\'o}pez-S{\'a}nchez}, {Lorente}, {Medling}, {Mould}, {Norberg},
  {Parker}, {Power}, {Pracy}, {Richards}, {Robotham}, {Sweet}, {Taylor},
  {Thomas}, {Tonini}, \& {Walcher}}]{Allen2014}
{Allen} J.~T. {et~al.}, 2014, ArXiv e-prints

\bibitem[{{Allen} {et~al}\mbox{.}(2008){Allen}, {Groves}, {Dopita},
  {Sutherland}, \& {Kewley}}]{Allen2008}
{Allen} M.~G., {Groves} B.~A., {Dopita} M.~A., {Sutherland} R.~S., {Kewley}
  L.~J., 2008, \apjs, 178, 20

\bibitem[{Antonucci(1993)}]{Antonucci1993}
Antonucci R., 1993, ARAA, 31, 473

\bibitem[{{Arribas} {et~al}\mbox{.}(2014){Arribas}, {Colina}, {Bellocchi},
  {Maiolino}, \& {Villar-Mart{\'{\i}}n}}]{Arribas2014}
{Arribas} S., {Colina} L., {Bellocchi} E., {Maiolino} R.,
  {Villar-Mart{\'{\i}}n} M., 2014, \aap, 568, A14

\bibitem[{Baldwin, Phillips \& Terlevich(1981)Baldwin, Phillips, \&
  Terlevich}]{Baldwin1981}
Baldwin J.~A., Phillips M.~M., Terlevich R., 1981, PASP, 93, 5

\bibitem[{{Barrows} {et~al}\mbox{.}(2013){Barrows}, {Sandberg Lacy},
  {Kennefick}, {Comerford}, {Kennefick}, \& {Berrier}}]{Barrows2013}
{Barrows} R.~S., {Sandberg Lacy} C.~H., {Kennefick} J., {Comerford} J.~M.,
  {Kennefick} D., {Berrier} J.~C., 2013, \apj, 769, 95

\bibitem[{{Becker}, {White} \& {Helfand}(1995){Becker}, {White}, \&
  {Helfand}}]{1995ApJ...450..559B}
{Becker} R.~H., {White} R.~L., {Helfand} D.~J., 1995, \apj, 450, 559

\bibitem[{{Bower} {et~al}\mbox{.}(2006){Bower}, {Benson}, {Malbon}, {Helly},
  {Frenk}, {Baugh}, {Cole}, \& {Lacey}}]{Bower2006}
{Bower} R.~G., {Benson} A.~J., {Malbon} R., {Helly} J.~C., {Frenk} C.~S.,
  {Baugh} C.~M., {Cole} S., {Lacey} C.~G., 2006, \mnras, 370, 645

\bibitem[{{Bryant} {et~al}\mbox{.}(2014){Bryant}, {Owers}, {Robotham}, {Croom},
  {Driver}, {Drinkwater}, {Lorente}, {Cortese}, {Scott}, {Colless}, {Schaefer},
  {Taylor}, {Konstantopoulos}, {Allen}, {Baldry}, {Barnes}, {Bauer},
  {Bland-Hawthorn}, {Bloom}, {Brooks}, {Brough}, {Cecil}, {Couch}, {Croton},
  {Davies}, {Ellis}, {Fogarty}, {Foster}, {Glazebrook}, {Goodwin}, {Green},
  {Gunawardhana}, {Hampton}, {Ho}, {Hopkins}, {Kewley}, {Lawrence},
  {Leon-Saval}, {Leslie}, {Lewis}, {Liske}, {Lopez-Sanchez}, {Mahajan},
  {Medling}, {Metcalfe}, {Meyer}, {Mould}, {Obreschkow}, {O'Toole}, {Pracy},
  {Richards}, {Shanks}, {Sharp}, {Sweet}, {Thomas}, {Tonini}, \&
  {Walcher}}]{Bryant2014}
{Bryant} J.~J. {et~al.}, 2014, ArXiv e-prints

\bibitem[{{Cappellari} \& {Emsellem}(2004)}]{Cappellari2004}
{Cappellari} M., {Emsellem} E., 2004, \pasp, 116, 138

\bibitem[{{Comerford} {et~al}\mbox{.}(2013){Comerford}, {Schluns}, {Greene}, \&
  {Cool}}]{Comerford2013}
{Comerford} J.~M., {Schluns} K., {Greene} J.~E., {Cool} R.~J., 2013, \apj, 777,
  64

\bibitem[{{Cowie} {et~al}\mbox{.}(1995){Cowie}, {Songaila}, {Kim}, \&
  {Hu}}]{Cowie1995}
{Cowie} L.~L., {Songaila} A., {Kim} T.-S., {Hu} E.~M., 1995, \aj, 109, 1522

\bibitem[{{Croom} {et~al}\mbox{.}(2012){Croom}, {Lawrence}, {Bland-Hawthorn},
  {Bryant}, {Fogarty}, {Richards}, {Goodwin}, {Farrell}, {Miziarski}, {Heald},
  {Jones}, {Lee}, {Colless}, {Brough}, {Hopkins}, {Bauer}, {Birchall}, {Ellis},
  {Horton}, {Leon-Saval}, {Lewis}, {L{\'o}pez-S{\'a}nchez}, {Min}, {Trinh}, \&
  {Trowland}}]{Croom2012}
{Croom} S.~M. {et~al.}, 2012, \mnras, 421, 872

\bibitem[{{Croton} {et~al}\mbox{.}(2006){Croton}, {Springel}, {White}, {De
  Lucia}, {Frenk}, {Gao}, {Jenkins}, {Kauffmann}, {Navarro}, \&
  {Yoshida}}]{Croton2006}
{Croton} D.~J. {et~al.}, 2006, \mnras, 365, 11

\bibitem[{{Dopita} \& {Sutherland}(1995)}]{Dopita1995}
{Dopita} M.~A., {Sutherland} R.~S., 1995, \apj, 455, 468

\bibitem[{Fischer {et~al}\mbox{.}(2011)Fischer, Crenshaw, Kraemer, Schmitt,
  Mushotsky, \& Dunn}]{Fischer2011}
Fischer T.~C., Crenshaw D.~M., Kraemer S.~B., Schmitt H.~R., Mushotsky R.~F.,
  Dunn J.~P., 2011, ApJ, 727, 71

\bibitem[{Ganguly \& Brotherton(2008)}]{Ganguly2008}
Ganguly R., Brotherton M.~S., 2008, ApJ, 672, 102

\bibitem[{{Gonz{\'a}lez Delgado} {et~al}\mbox{.}(2005){Gonz{\'a}lez Delgado},
  {Cervi{\~n}o}, {Martins}, {Leitherer}, \& {Hauschildt}}]{GDetal04}
{Gonz{\'a}lez Delgado} R.~M., {Cervi{\~n}o} M., {Martins} L.~P., {Leitherer}
  C., {Hauschildt} P.~H., 2005, \mnras, 357, 945

\bibitem[{{Gray} \& {Corbally}(1994)}]{GC94}
{Gray} R.~O., {Corbally} C.~J., 1994, AJ, 107, 742

\bibitem[{{Greene} {et~al}\mbox{.}(2011){Greene}, {Zakamska}, {Ho}, \&
  {Barth}}]{Greene2011}
{Greene} J.~E., {Zakamska} N.~L., {Ho} L.~C., {Barth} A.~J., 2011, \apj, 732, 9

\bibitem[{{Harrison} {et~al}\mbox{.}(2014){Harrison}, {Alexander}, {Mullaney},
  \& {Swinbank}}]{Harrison2014}
{Harrison} C.~M., {Alexander} D.~M., {Mullaney} J.~R., {Swinbank} A.~M., 2014,
  \mnras, 441, 3306

\bibitem[{{Harrison} {et~al}\mbox{.}(2012){Harrison}, {Alexander}, {Swinbank},
  {Smail}, {Alaghband-Zadeh}, {Bauer}, {Chapman}, {Del Moro}, {Hickox},
  {Ivison}, {Men{\'e}ndez-Delmestre}, {Mullaney}, \& {Nesvadba}}]{Harrison2012}
{Harrison} C.~M. {et~al.}, 2012, \mnras, 426, 1073

\bibitem[{{Hauschildt} \& {Baron}(1999)}]{HB99}
{Hauschildt} P.~H., {Baron} E., 1999, Journal of Computational and Applied
  Mathematics, 109, 41

\bibitem[{{Heckman} {et~al}\mbox{.}(2004){Heckman}, {Kauffmann}, {Brinchmann},
  {Charlot}, {Tremonti}, \& {White}}]{2004ApJ...613..109H}
{Heckman} T.~M., {Kauffmann} G., {Brinchmann} J., {Charlot} S., {Tremonti} C.,
  {White} S.~D.~M., 2004, \apj, 613, 109

\bibitem[{{Heckman} {et~al}\mbox{.}(1981){Heckman}, {Miley}, {van Breugel}, \&
  {Butcher}}]{Heckman1981}
{Heckman} T.~M., {Miley} G.~K., {van Breugel} W.~J.~M., {Butcher} H.~R., 1981,
  \apj, 247, 403

\bibitem[{{Ho} {et~al}\mbox{.}(2014){Ho}, {Kewley}, {Dopita}, {Medling},
  {Allen}, {Bland-Hawthorn}, {Bloom}, {Bryant}, {Croom}, {Fogarty}, {Goodwin},
  {Green}, {Konstantopoulos}, {Lawrence}, {L{\'o}pez-S{\'a}nchez}, {Owers},
  {Richards}, \& {Sharp}}]{Ho2014}
{Ho} I.-T. {et~al.}, 2014, \mnras, 444, 3894

\bibitem[{{H{\"o}nig} {et~al}\mbox{.}(2013){H{\"o}nig}, {Kishimoto},
  {Tristram}, {Prieto}, {Gandhi}, {Asmus}, {Antonucci}, {Burtscher}, {Duschl},
  \& {Weigelt}}]{Honig2013}
{H{\"o}nig} S.~F. {et~al.}, 2013, \apj, 771, 87

\bibitem[{{Hopkins} {et~al}\mbox{.}(2006){Hopkins}, {Hernquist}, {Cox}, {Di
  Matteo}, {Robertson}, \& {Springel}}]{Hopkins2006}
{Hopkins} P.~F., {Hernquist} L., {Cox} T.~J., {Di Matteo} T., {Robertson} B.,
  {Springel} V., 2006, \apjs, 163, 1

\bibitem[{{Kauffmann} {et~al}\mbox{.}(2003{\natexlab{a}}){Kauffmann},
  {Heckman}, {Tremonti}, {Brinchmann}, {Charlot}, {White}, {Ridgway},
  {Brinkmann}, {Fukugita}, {Hall}, {Ivezi{\'c}}, {Richards}, \&
  {Schneider}}]{Kauffmann2003}
{Kauffmann} G. {et~al.}, 2003{\natexlab{a}}, \mnras, 346, 1055

\bibitem[{{Kauffmann} {et~al}\mbox{.}(2003{\natexlab{b}}){Kauffmann},
  {Heckman}, {White}, {Charlot}, {Tremonti}, {Brinchmann}, {Bruzual}, {Peng},
  {Seibert}, {Bernardi}, {Blanton}, {Brinkmann}, {Castander}, {Cs{\'a}bai},
  {Fukugita}, {Ivezic}, {Munn}, {Nichol}, {Padmanabhan}, {Thakar}, {Weinberg},
  \& {York}}]{2003MNRAS.341...33K}
{Kauffmann} G. {et~al.}, 2003{\natexlab{b}}, \mnras, 341, 33

\bibitem[{{Keel} {et~al}\mbox{.}(2012){Keel}, {Chojnowski}, {Bennert},
  {Schawinski}, {Lintott}, {Lynn}, {Pancoast}, {Harris}, {Nierenberg},
  {Sonnenfeld}, \& {Proctor}}]{Keel2012}
{Keel} W.~C. {et~al.}, 2012, \mnras, 420, 878

\bibitem[{{Kewley} {et~al}\mbox{.}(2001){Kewley}, {Dopita}, {Sutherland},
  {Heisler}, \& {Trevena}}]{Kewley2001}
{Kewley} L.~J., {Dopita} M.~A., {Sutherland} R.~S., {Heisler} C.~A., {Trevena}
  J., 2001, \apj, 556, 121

\bibitem[{{Lanz} \& {Hubeny}(2003)}]{LH03}
{Lanz} T., {Hubeny} I., 2003, \apjs, 146, 417

\bibitem[{Liu {et~al}\mbox{.}(2013)Liu, Zakamska, Greene, Nesvadba, \&
  Liu}]{Liu2013b}
Liu G., Zakamska N.~L., Greene J.~E., Nesvadba N. P.~H., Liu X., 2013, MNRAS,
  436, 2576

\bibitem[{{Martins} {et~al}\mbox{.}(2005){Martins}, {Gonz{\'a}lez Delgado},
  {Leitherer}, {Cervi{\~n}o}, \& {Hauschildt}}]{Martetal04}
{Martins} L.~P., {Gonz{\'a}lez Delgado} R.~M., {Leitherer} C., {Cervi{\~n}o}
  M., {Hauschildt} P., 2005, \mnras, 358, 49

\bibitem[{{Mauch} \& {Sadler}(2007)}]{2007MNRAS.375..931M}
{Mauch} T., {Sadler} E.~M., 2007, \mnras, 375, 931

\bibitem[{{Monreal-Ibero}, {Arribas} \& {Colina}(2006){Monreal-Ibero},
  {Arribas}, \& {Colina}}]{Monreal2006}
{Monreal-Ibero} A., {Arribas} S., {Colina} L., 2006, \apj, 637, 138

\bibitem[{{Monreal-Ibero} {et~al}\mbox{.}(2010){Monreal-Ibero}, {Arribas},
  {Colina}, {Rodr{\'{\i}}guez-Zaur{\'{\i}}n}, {Alonso-Herrero}, \&
  {Garc{\'{\i}}a-Mar{\'{\i}}n}}]{Monreal2010}
{Monreal-Ibero} A., {Arribas} S., {Colina} L., {Rodr{\'{\i}}guez-Zaur{\'{\i}}n}
  J., {Alonso-Herrero} A., {Garc{\'{\i}}a-Mar{\'{\i}}n} M., 2010, \aap, 517,
  A28

\bibitem[{{Mullaney} {et~al}\mbox{.}(2013){Mullaney}, {Alexander}, {Fine},
  {Goulding}, {Harrison}, \& {Hickox}}]{Mullaney2013}
{Mullaney} J.~R., {Alexander} D.~M., {Fine} S., {Goulding} A.~D., {Harrison}
  C.~M., {Hickox} R.~C., 2013, \mnras, 433, 622

\bibitem[{Murray, Quataert \& Thompson(2005)Murray, Quataert, \&
  Thompson}]{Murray2005}
Murray N., Quataert E., Thompson T.~A., 2005, ApJ, 569

\bibitem[{{Nenkova} {et~al}\mbox{.}(2008){Nenkova}, {Sirocky}, {Nikutta},
  {Ivezi{\'c}}, \& {Elitzur}}]{2008ApJ...685..160N}
{Nenkova} M., {Sirocky} M.~M., {Nikutta} R., {Ivezi{\'c}} {\v Z}., {Elitzur}
  M., 2008, \apj, 685, 160

\bibitem[{{Osterbrock} \& {Ferland}(2006)}]{Osterbrock2006}
{Osterbrock} D.~E., {Ferland} G.~J., 2006, {Astrophysics of gaseous nebulae and
  active galactic nuclei}

\bibitem[{{Reyes} {et~al}\mbox{.}(2008){Reyes}, {Zakamska}, {Strauss}, {Green},
  {Krolik}, {Shen}, {Richards}, {Anderson}, \& {Schneider}}]{Reyes2008}
{Reyes} R. {et~al.}, 2008, AJ, 136, 2373

\bibitem[{{Rich}, {Kewley} \& {Dopita}(2011){Rich}, {Kewley}, \&
  {Dopita}}]{Rich2011}
{Rich} J.~A., {Kewley} L.~J., {Dopita} M.~A., 2011, \apj, 734, 87

\bibitem[{{Riffel}, {Storchi-Bergmann} \& {Riffel}(2014){Riffel},
  {Storchi-Bergmann}, \& {Riffel}}]{Riffel2014}
{Riffel} R.~A., {Storchi-Bergmann} T., {Riffel} R., 2014, \apjl, 780, L24

\bibitem[{{Riffel}, {Storchi-Bergmann} \& {Winge}(2013){Riffel},
  {Storchi-Bergmann}, \& {Winge}}]{Riffel2013}
{Riffel} R.~A., {Storchi-Bergmann} T., {Winge} C., 2013, \mnras, 430, 2249

\bibitem[{{Riffel} {et~al}\mbox{.}(2008){Riffel}, {Storchi-Bergmann}, {Winge},
  {McGregor}, {Beck}, \& {Schmitt}}]{Riffel2008}
{Riffel} R.~A., {Storchi-Bergmann} T., {Winge} C., {McGregor} P.~J., {Beck} T.,
  {Schmitt} H., 2008, \mnras, 385, 1129

\bibitem[{{Rodr{\'{\i}}guez Zaur{\'{\i}}n}
  {et~al}\mbox{.}(2013){Rodr{\'{\i}}guez Zaur{\'{\i}}n}, {Tadhunter}, {Rose},
  \& {Holt}}]{RZ2013}
{Rodr{\'{\i}}guez Zaur{\'{\i}}n} J., {Tadhunter} C.~N., {Rose} M., {Holt} J.,
  2013, \mnras, 432, 138

\bibitem[{Rupke \& Veilleux(2013)}]{Rupke2013}
Rupke D. S.~N., Veilleux S., 2013, ApJ, 768, 75

\bibitem[{{Saunders} {et~al}\mbox{.}(2004){Saunders}, {Bridges}, {Gillingham},
  {Haynes}, {Smith}, {Whittard}, {Churilov}, {Lankshear}, {Croom}, {Jones}, \&
  {Boshuizen}}]{Saunders2004}
{Saunders} W. {et~al.}, 2004, in Society of Photo-Optical Instrumentation
  Engineers (SPIE) Conference Series, Vol. 5492, Ground-based Instrumentation
  for Astronomy, {Moorwood} A.~F.~M., {Iye} M., eds., pp. 389--400

\bibitem[{Scannapieco \& Oh(2004)}]{Scannapieco2004}
Scannapieco E., Oh S.~P., 2004, ApJ, 608, 62

\bibitem[{{Sharp} {et~al}\mbox{.}(2014){Sharp}, {Allen}, {Fogarty}, {Croom},
  {Cortese}, {Green}, {Nielsen}, {Richards}, {Scott}, {Taylor}, {Barnes},
  {Bauer}, {Birchall}, {Bland-Hawthorn}, {Bloom}, {Brough}, {Bryant}, {Cecil},
  {Colless}, {Couch}, {Drinkwater}, {Driver}, {Foster}, {Goodwin},
  {Gunawardhana}, {Ho}, {Hampton}, {Hopkins}, {Jones}, {Konstantopoulos},
  {Lawrence}, {Leslie}, {Lewis}, {Liske}, {Lorente}, {Medling}, {Mahajan},
  {Mould}, {Parker}, {Pracy}, {Obreschkow}, {Owers}, {Schaefer}, {Sweet},
  {Thomas}, {Tonini}, \& {Walcher}}]{Sharp2014}
{Sharp} R. {et~al.}, 2014, ArXiv e-prints

\bibitem[{{Sharp} \& {Birchall}(2010)}]{Sharp2010}
{Sharp} R., {Birchall} M.~N., 2010, \pasa, 27, 91

\bibitem[{{Sharp} {et~al}\mbox{.}(2006){Sharp}, {Saunders}, {Smith},
  {Churilov}, {Correll}, {Dawson}, {Farrel}, {Frost}, {Haynes}, {Heald},
  {Lankshear}, {Mayfield}, {Waller}, \& {Whittard}}]{Sharp2006}
{Sharp} R. {et~al.}, 2006, in Society of Photo-Optical Instrumentation
  Engineers (SPIE) Conference Series, Vol. 6269, Society of Photo-Optical
  Instrumentation Engineers (SPIE) Conference Series

\bibitem[{Sharp \& Bland-Hawthorn(2010)}]{SharpJBH2010}
Sharp R.~G., Bland-Hawthorn J., 2010, ApJ, 711, 818

\bibitem[{{Storchi-Bergmann} {et~al}\mbox{.}(2010){Storchi-Bergmann}, {Lopes},
  {McGregor}, {Riffel}, {Beck}, \& {Martini}}]{SB2010}
{Storchi-Bergmann} T., {Lopes} R.~D.~S., {McGregor} P.~J., {Riffel} R.~A.,
  {Beck} T., {Martini} P., 2010, \mnras, 402, 819

\bibitem[{{Tremaine} {et~al}\mbox{.}(2002){Tremaine}, {Gebhardt}, {Bender},
  {Bower}, {Dressler}, {Faber}, {Filippenko}, {Green}, {Grillmair}, {Ho},
  {Kormendy}, {Lauer}, {Magorrian}, {Pinkney}, \& {Richstone}}]{Tremaine2002}
{Tremaine} S. {et~al.}, 2002, \apj, 574, 740

\bibitem[{{Vega Beltr{\'a}n} {et~al}\mbox{.}(2001){Vega Beltr{\'a}n},
  {Pizzella}, {Corsini}, {Funes}, {Zeilinger}, {Beckman}, \&
  {Bertola}}]{Vega2001}
{Vega Beltr{\'a}n} J.~C., {Pizzella} A., {Corsini} E.~M., {Funes} J.~G.,
  {Zeilinger} W.~W., {Beckman} J.~E., {Bertola} F., 2001, \aap, 374, 394

\bibitem[{Veilleux {et~al}\mbox{.}(2013)Veilleux, Mel\'{e}ndez, Sturm,
  Gracia-Carpio, Fischer, Gonz\'{a}lez-Alfonso, Contursi, Lutz, Poglitsch,
  Davies, Genzel, Tacconi, de~Jong, Sternberg, Netzer, Hailey-Dunsheath, Verma,
  Rupke, Maiolino, Teng, \& Polisensky}]{Veilleux2013}
Veilleux S. {et~al.}, 2013, ApJ, 776, 27

\bibitem[{{Veilleux} \& {Osterbrock}(1987)}]{Veilleux1987}
{Veilleux} S., {Osterbrock} D.~E., 1987, \apjs, 63, 295

\bibitem[{Villar-Mart\'{\i}n {et~al}\mbox{.}(2011)Villar-Mart\'{\i}n, Humphrey,
  Delgado, Colina, \& Arribas}]{Villar-Martin2011}
Villar-Mart\'{\i}n M., Humphrey a., Delgado R.~G., Colina L., Arribas S., 2011,
  MNRAS, 418, 2032

\bibitem[{Whittle(1985)}]{Whittle1985}
Whittle M., 1985, MNRAS, 213, 1

\end{thebibliography}
\bibliographystyle{mn2e}

\appendix
\section{Galaxy by galaxy figures}

In this appendix a standard set of figures is presented for each galaxy, and all have the same format as is described in Figure \ref{J095155}.  A sample of this appendix is presented in the printed journal, the rest of the appendix is available in the online version.

\begin{figure*}
\centering
 \includegraphics[width=0.8\textwidth]{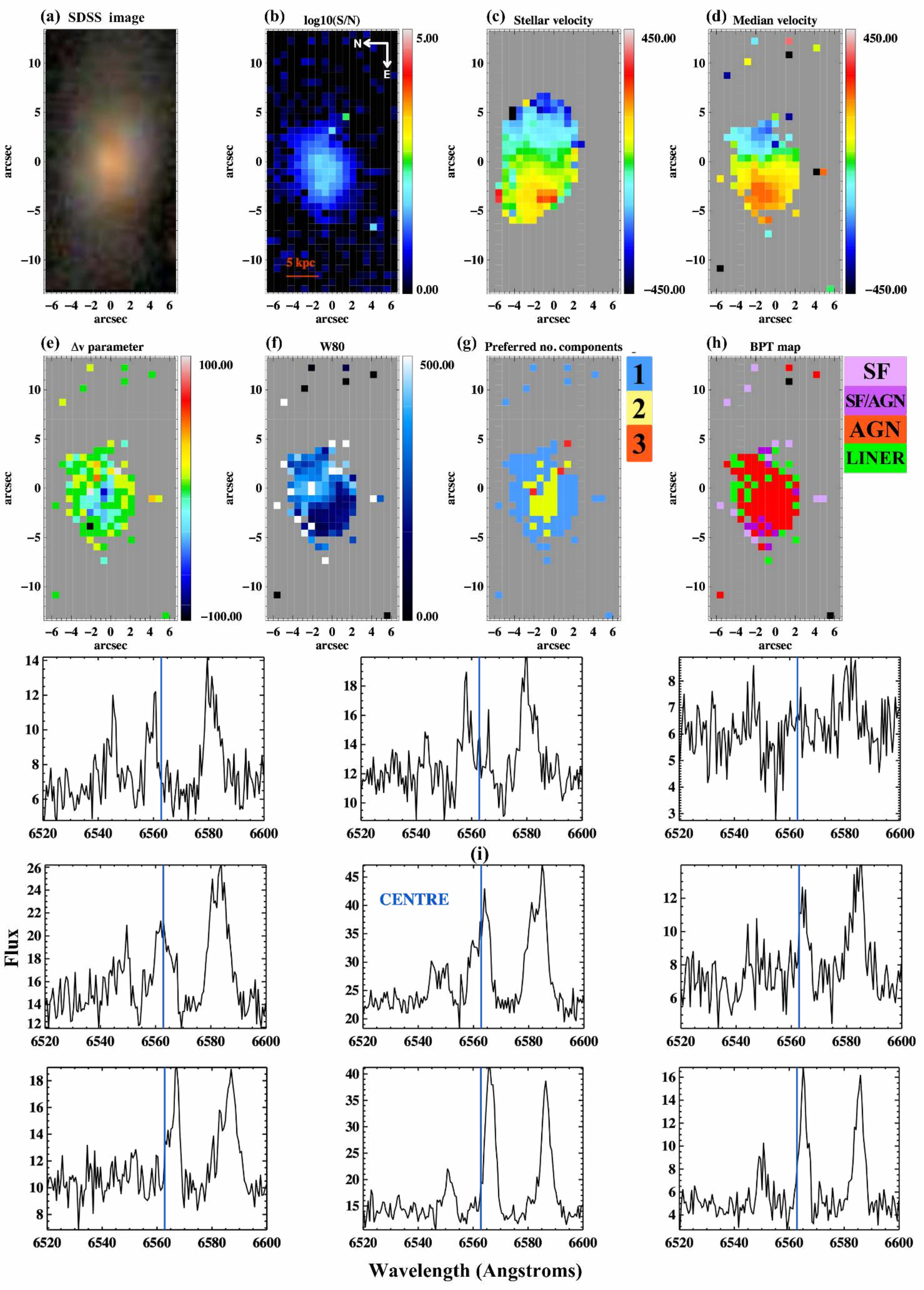}
\caption{J095155: {\bf (a)} The SDSS image at the same scale as our IFU pointing. {\bf (b)} A map of log$_{10}$(S/N) in H$\alpha$ to show the extended structure of the galaxy. {\bf (c)} The stellar velocity map derived from {\sc ppxf} fits to the data, as described in Section \ref{stellar_sec}. {\bf (d)} The median velocity map obtained by measuring the median velocity of fits to the emission lines as described in Section \ref{fitting_sec}. {\bf (e)} $\Delta$v, the asymmetry parameter. {\bf (f)} W$_{80}$, the width of the emission line that contains the central 80\% of the flux. {\bf (g)} The preferred number of components across the galaxy, where grey represents low S/N data, blue represents 1 component, yellow represents 2 components, and red represents 3 components. {\bf (h)} A BPT map, showing the dominant ionising source across the galaxy where light purple is star forming (SF), purple represents the combination region of the BPT diagram (SF/AGN), red represents AGN, and green the LINER region. The few scattered black points are bad spaxels that survived the S/N cut, but do not possess meaningful line ratios. {\bf (i)} An array of spectra from different regions of the galaxy, these were acquired by taking the median across a 3 $\times$ 3 box of spaxels to obtain a higher S/N spectrum from a larger region. The central spectrum is from a 3 $\times$ 3 bin centred on the galactic centre, and the adjacent spectra are from adjacent 3 $\times$ 3 bins. The vertical blue line on each spectrum represents the zero velocity position of H$\alpha$ emission. All figures in the appendix are presented in this manner.}
\label{J095155}
\end{figure*}

\begin{figure*}
\centering
 \includegraphics[width=0.8\textwidth]{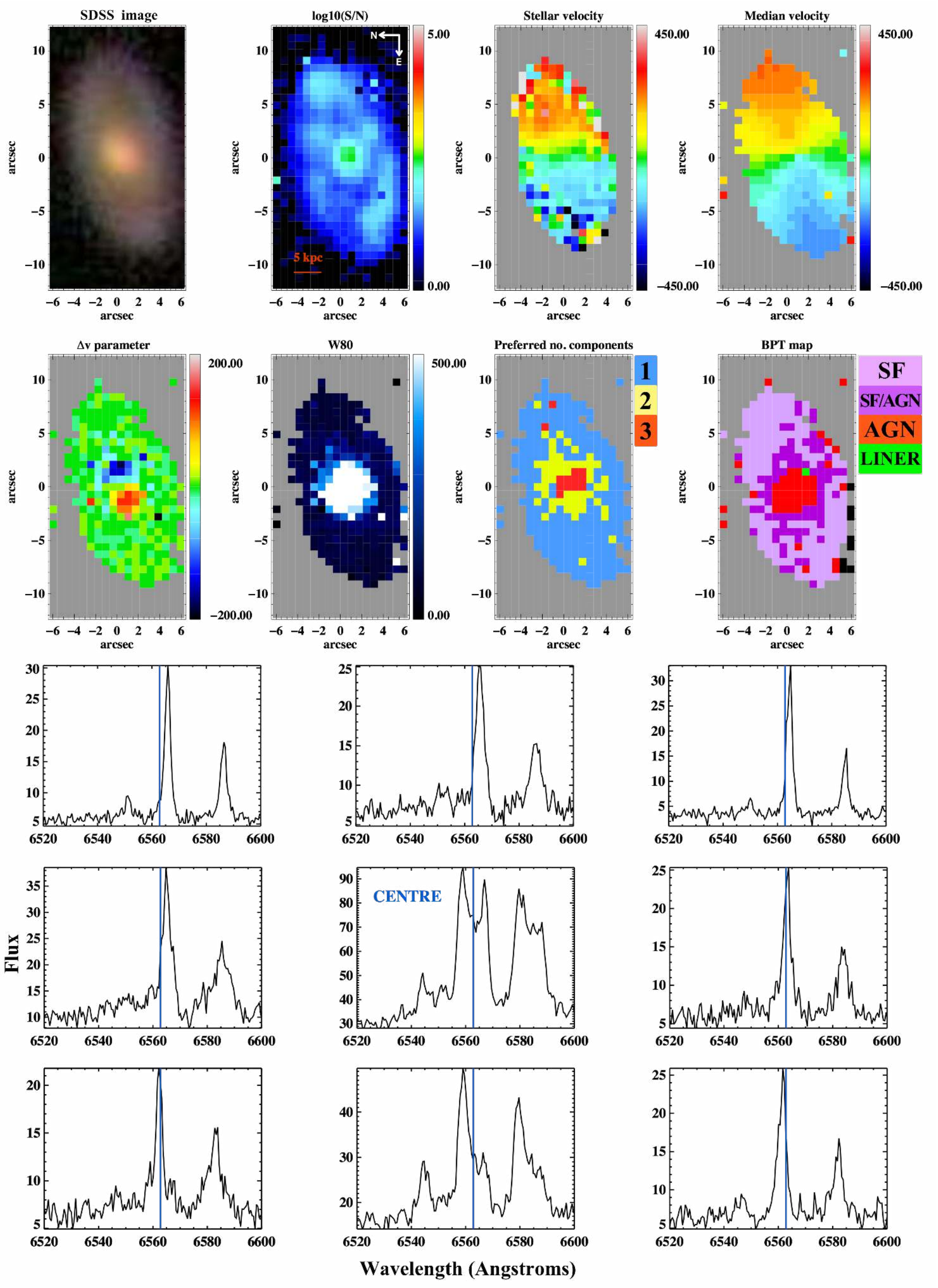}
\caption{J101927}
\label{J101927}
\end{figure*}

\begin{figure*}
\centering
 \includegraphics[width=0.8\textwidth]{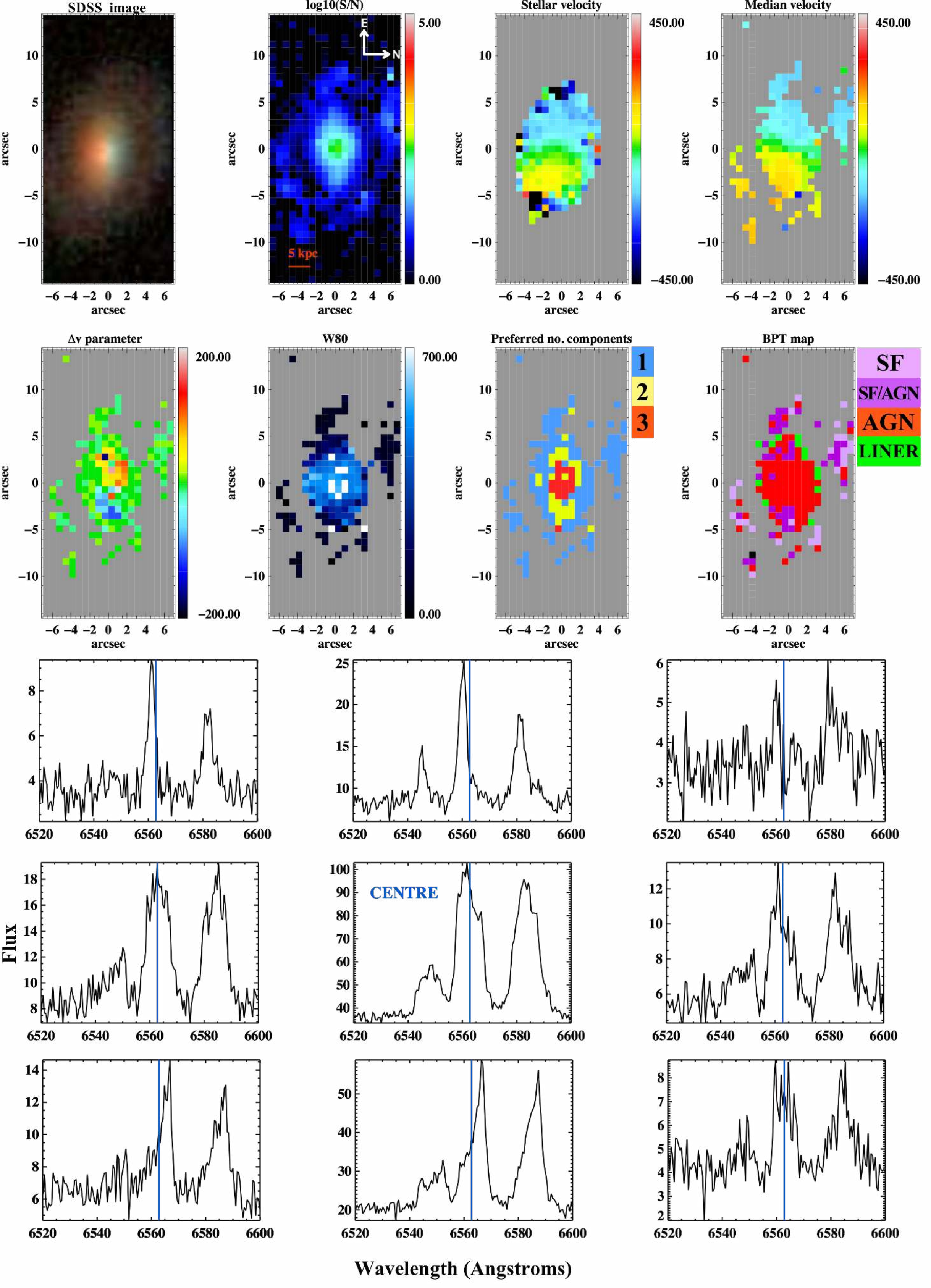}
\caption{J102143}
\label{J102143}
\end{figure*}

\begin{figure*}
\centering
 \includegraphics[width=0.8\textwidth]{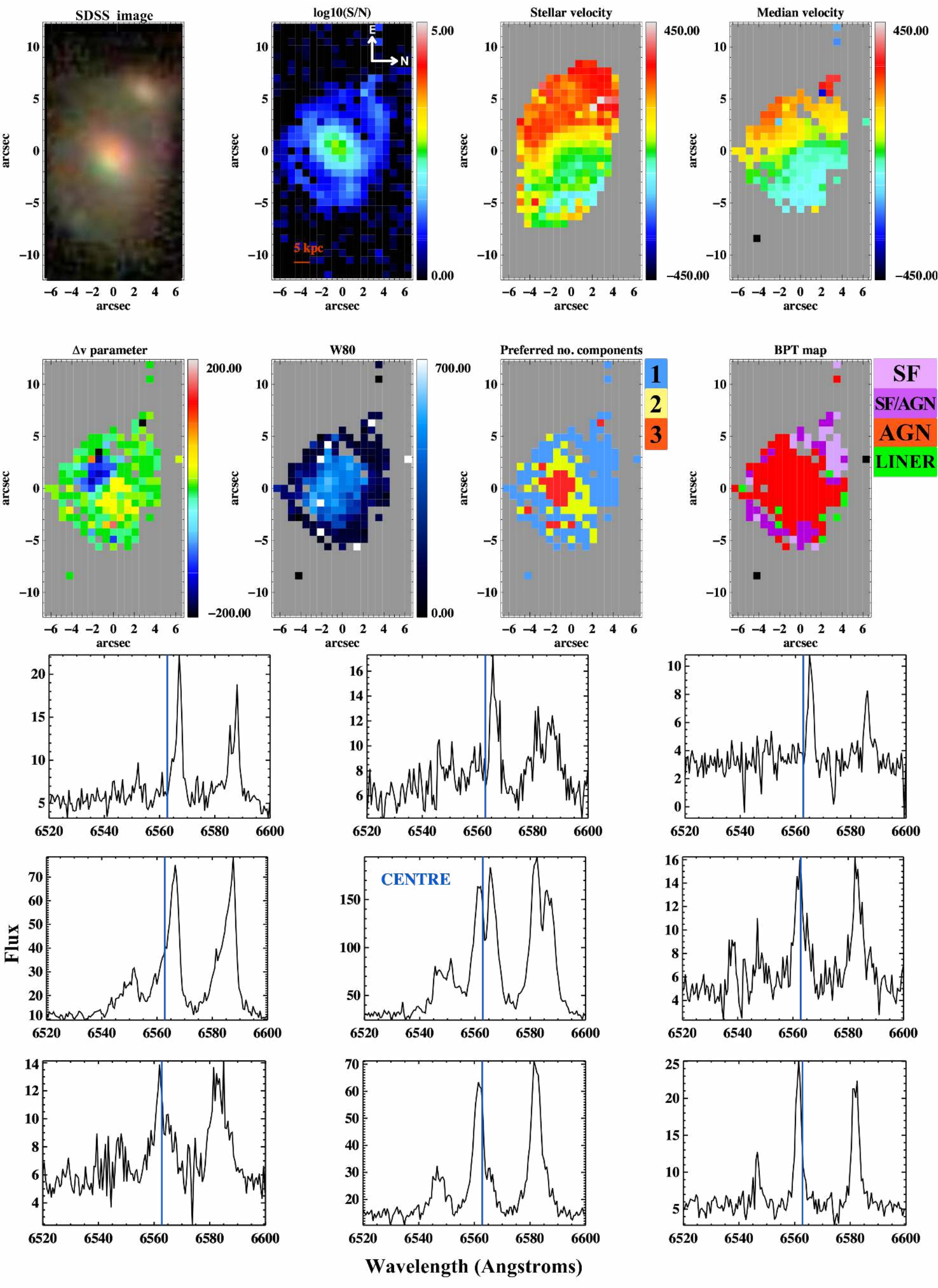}
\caption{J103600}
\label{J103600}
\end{figure*}

\begin{figure*}
\centering
 \includegraphics[width=0.8\textwidth]{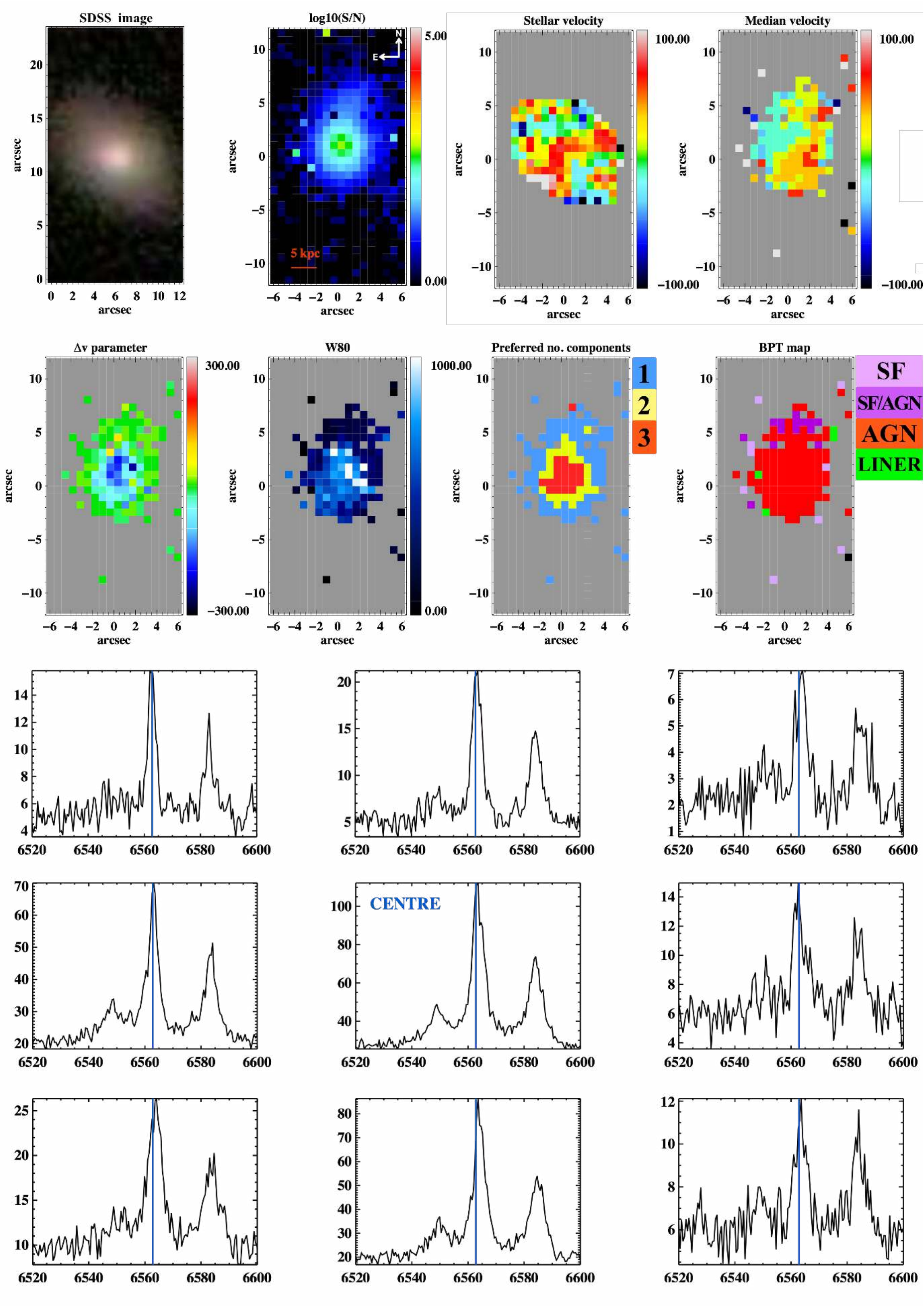}
\caption{J103915}
\label{J103915}
\end{figure*}

\begin{figure*}
\centering
 \includegraphics[width=0.8\textwidth]{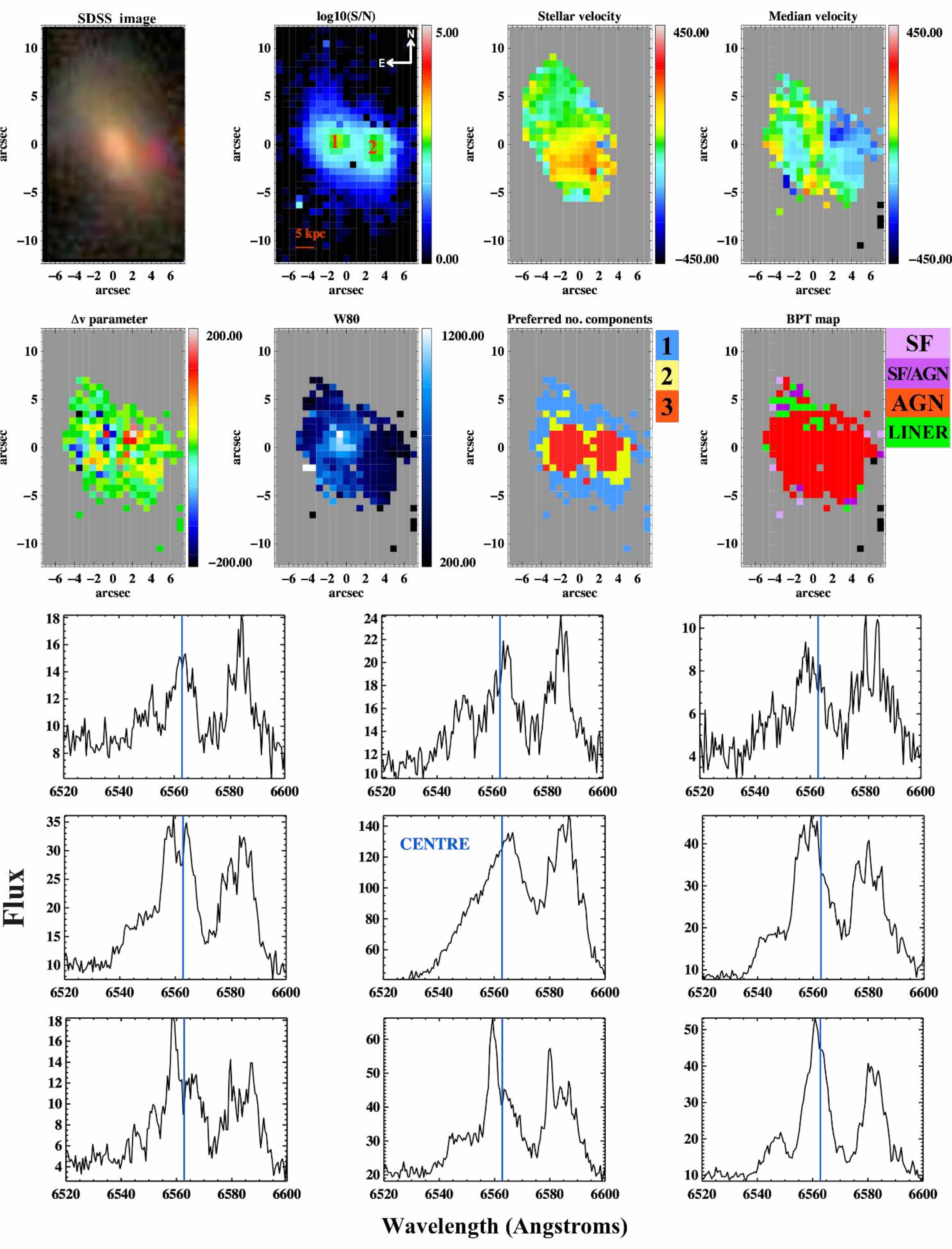}
\caption{J111100 }
\label{J111100}
\end{figure*}

\begin{figure*}
\centering
 \includegraphics[width=0.8\textwidth]{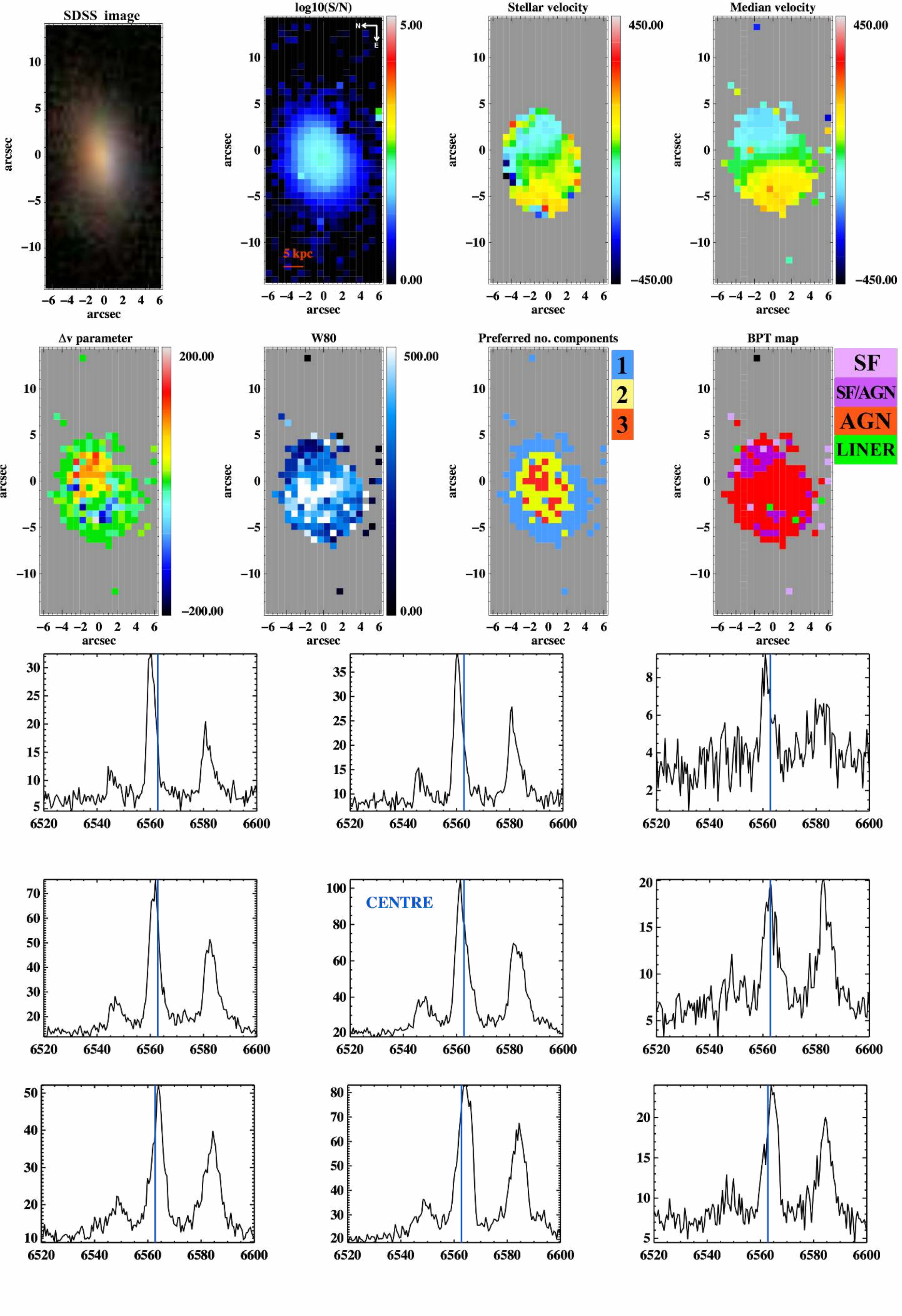}
\caption{J124321}
\label{J124321}
\end{figure*}

\begin{figure*}
\centering
 \includegraphics[width=0.8\textwidth]{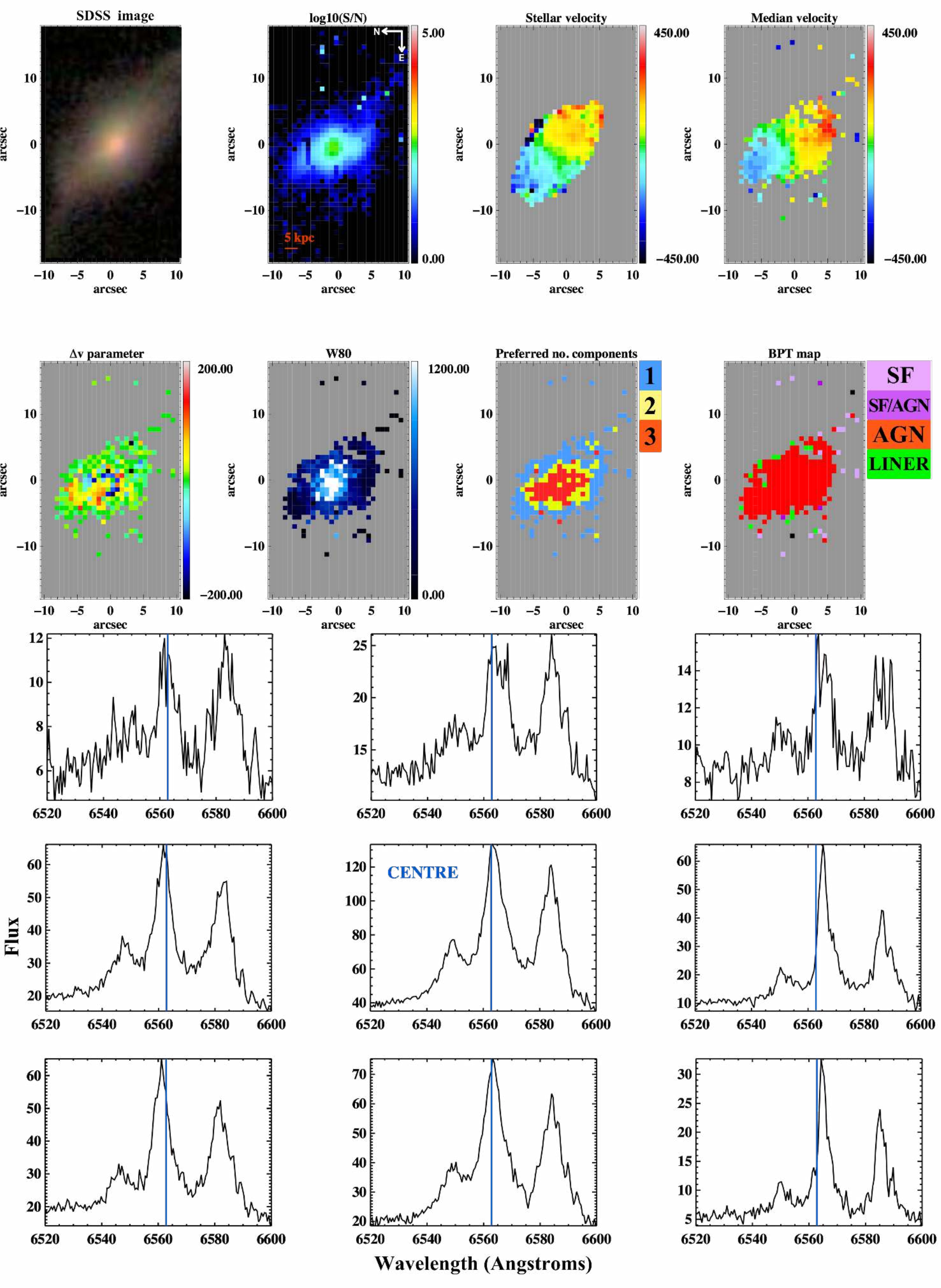}
\caption{J124859}
\label{J124859}
\end{figure*}

\begin{figure*}
\centering
 \includegraphics[width=0.8\textwidth]{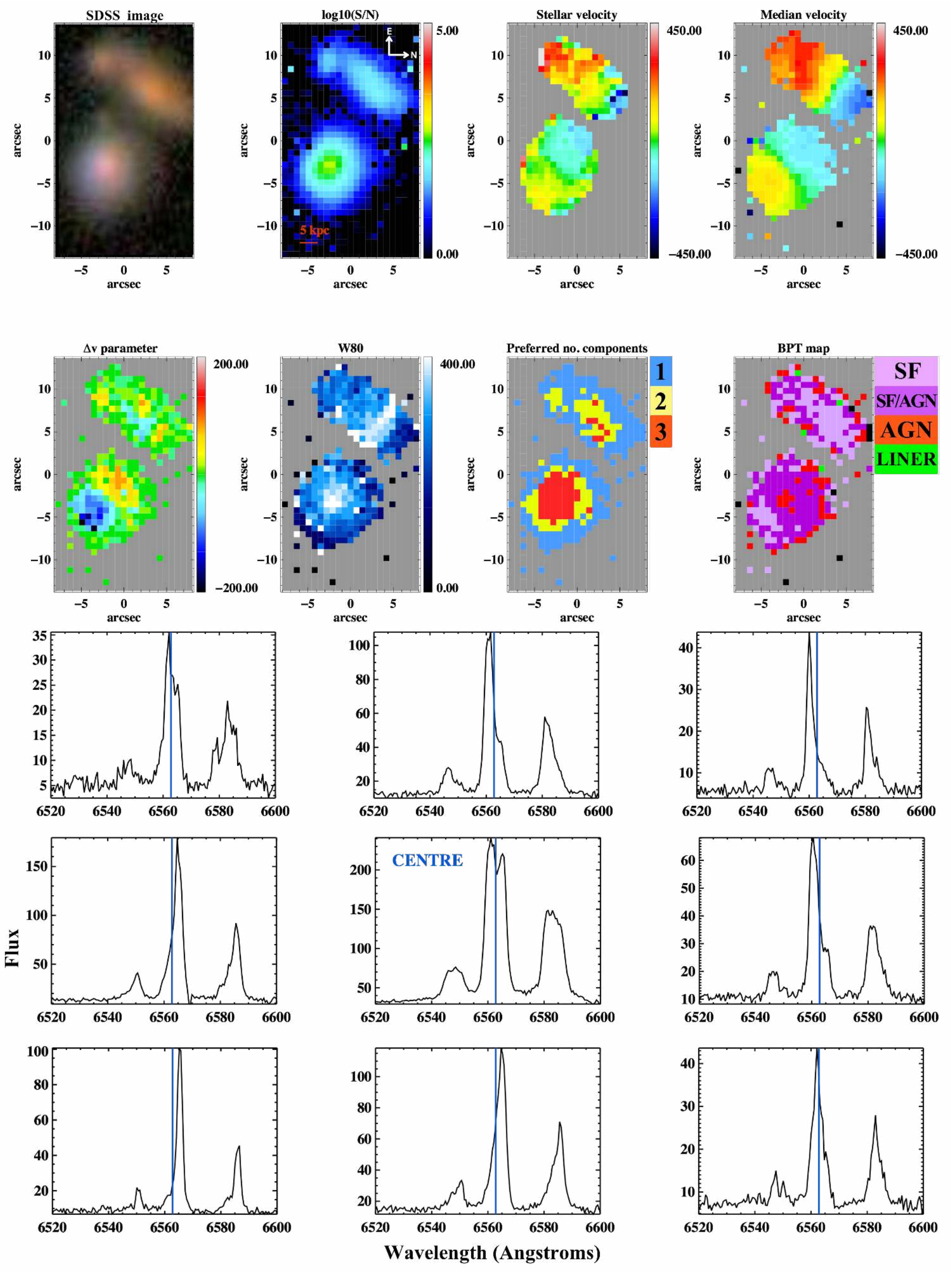}
\caption{J130116}
\label{J130116}
\end{figure*}

\begin{figure*}
\centering
 \includegraphics[width=0.8\textwidth]{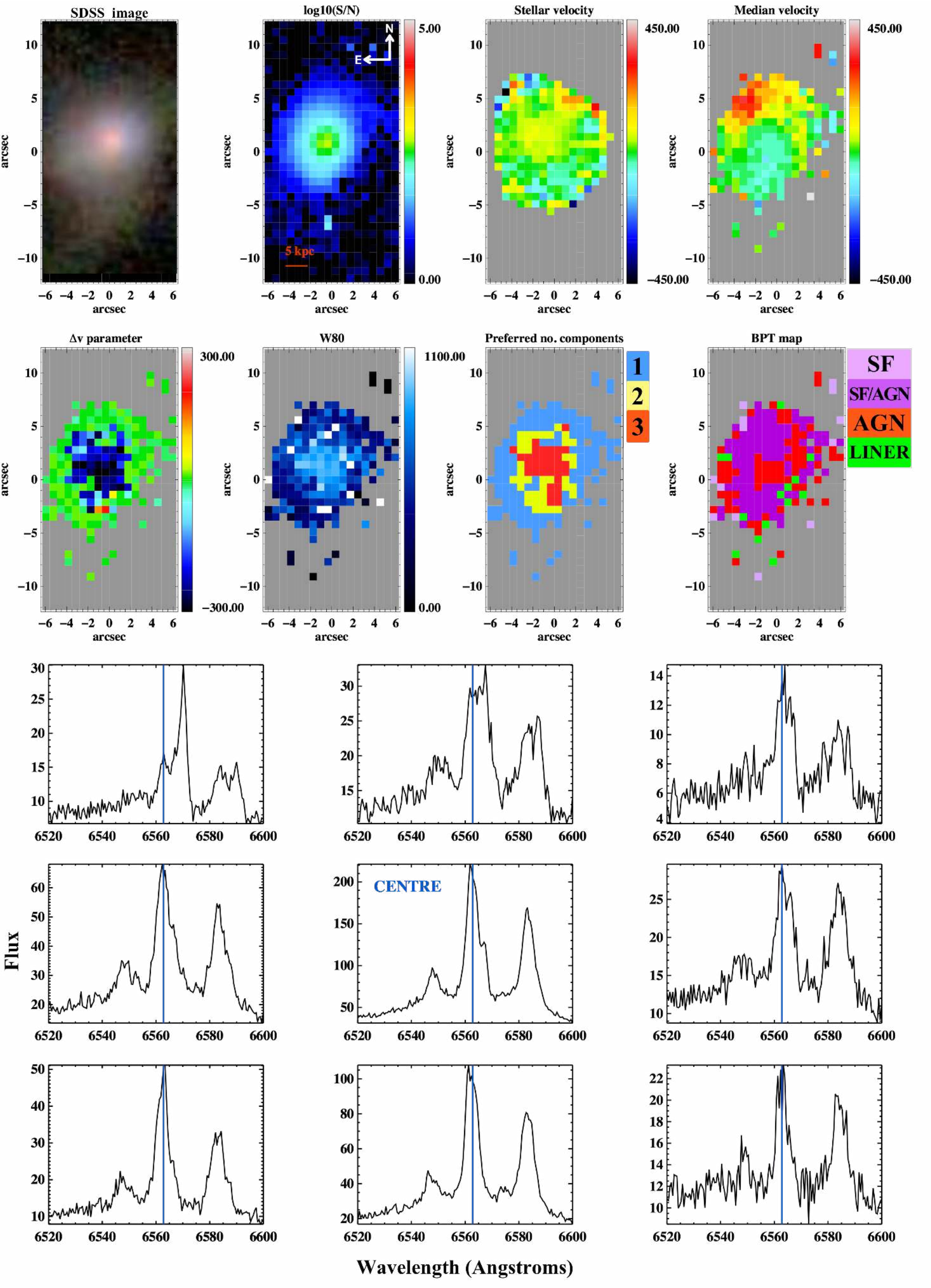}
\caption{J133152}
\label{J133152}
\end{figure*}

\begin{figure*}
\centering
 \includegraphics[width=0.8\textwidth]{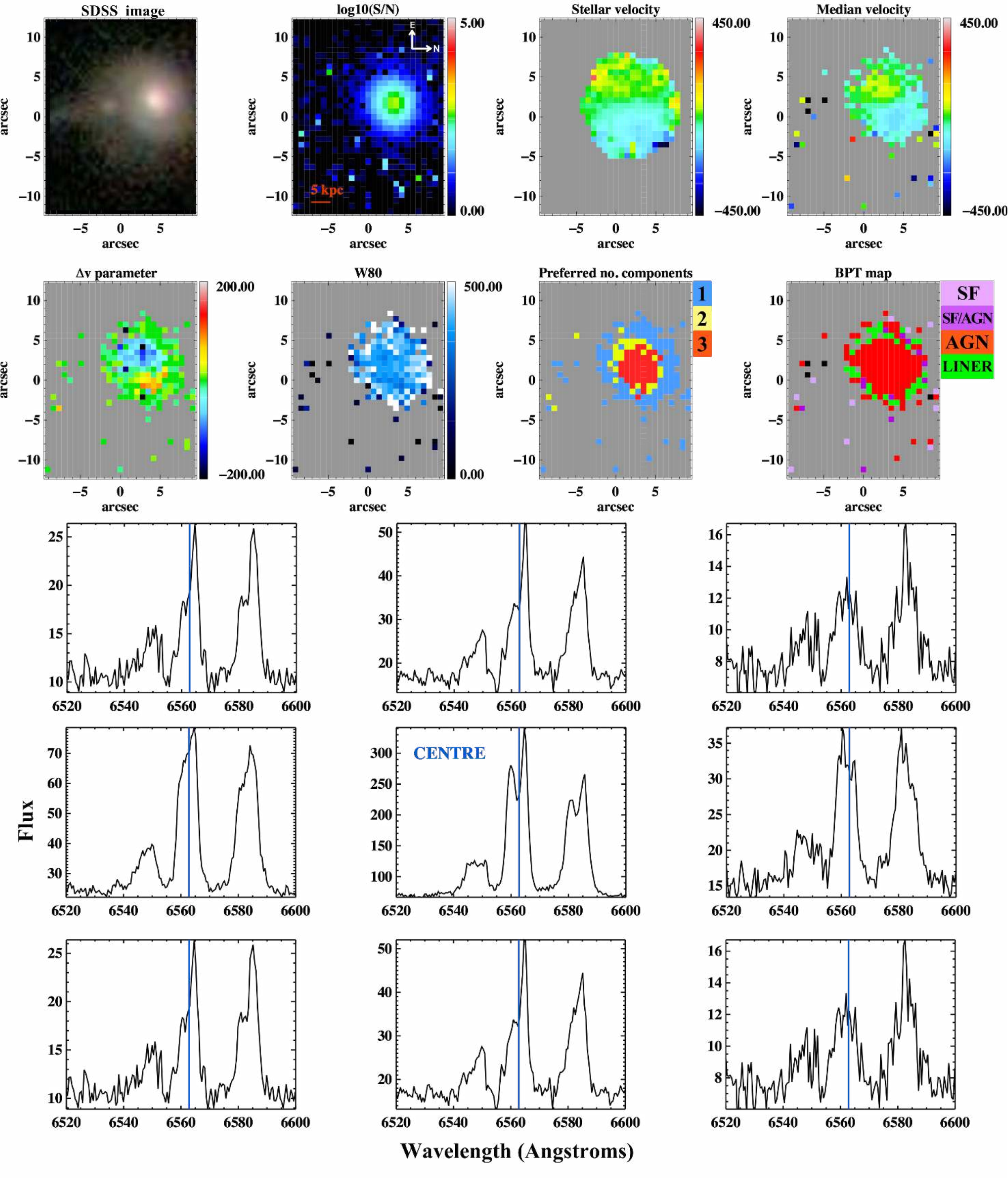}
\caption{J141926}
\label{J141926}
\end{figure*}

\begin{figure*}
\centering
 \includegraphics[width=0.8\textwidth]{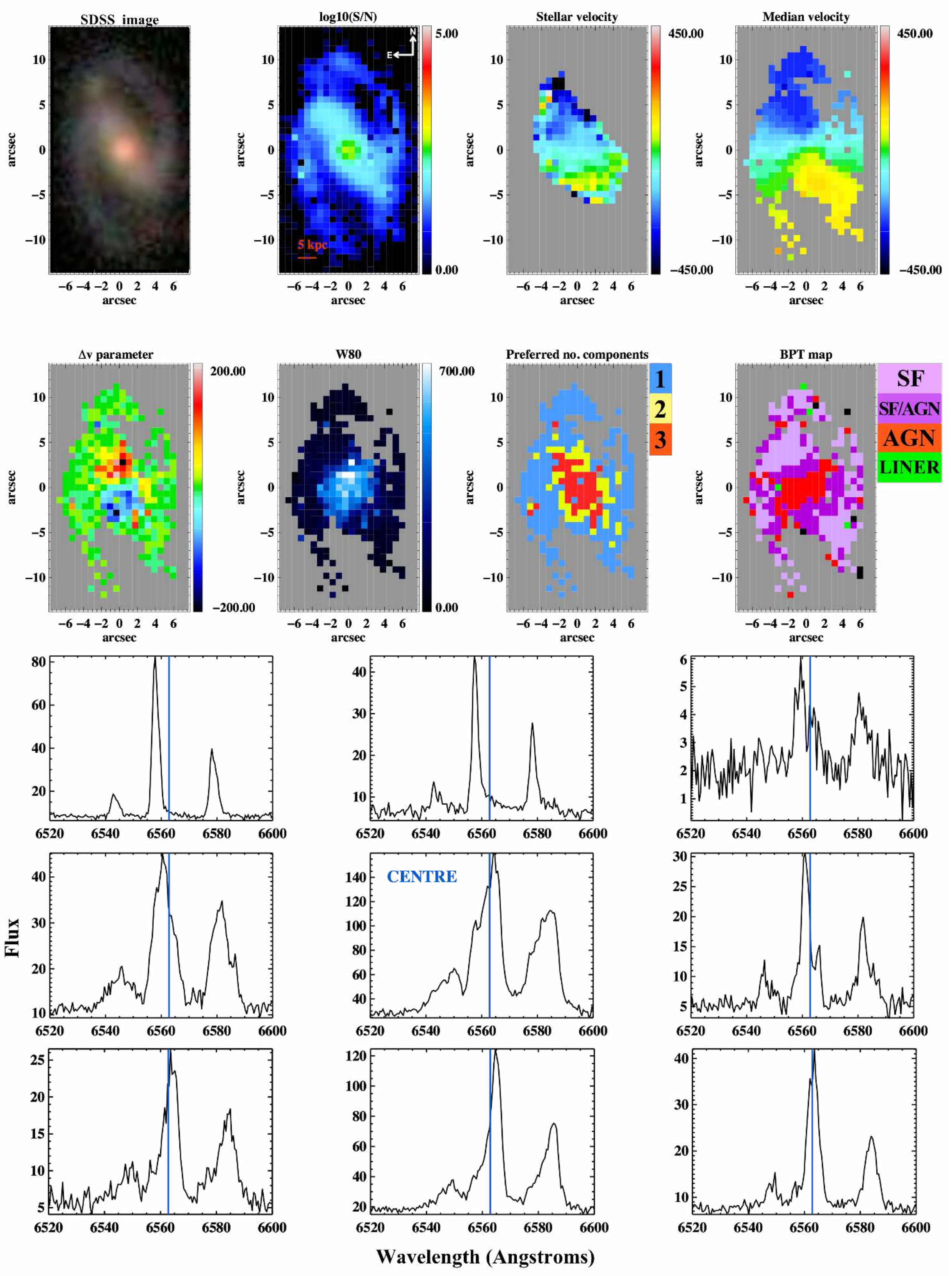}
\caption{J142237}
\label{J142237}
\end{figure*}

\begin{figure*}
\centering
 \includegraphics[width=0.8\textwidth]{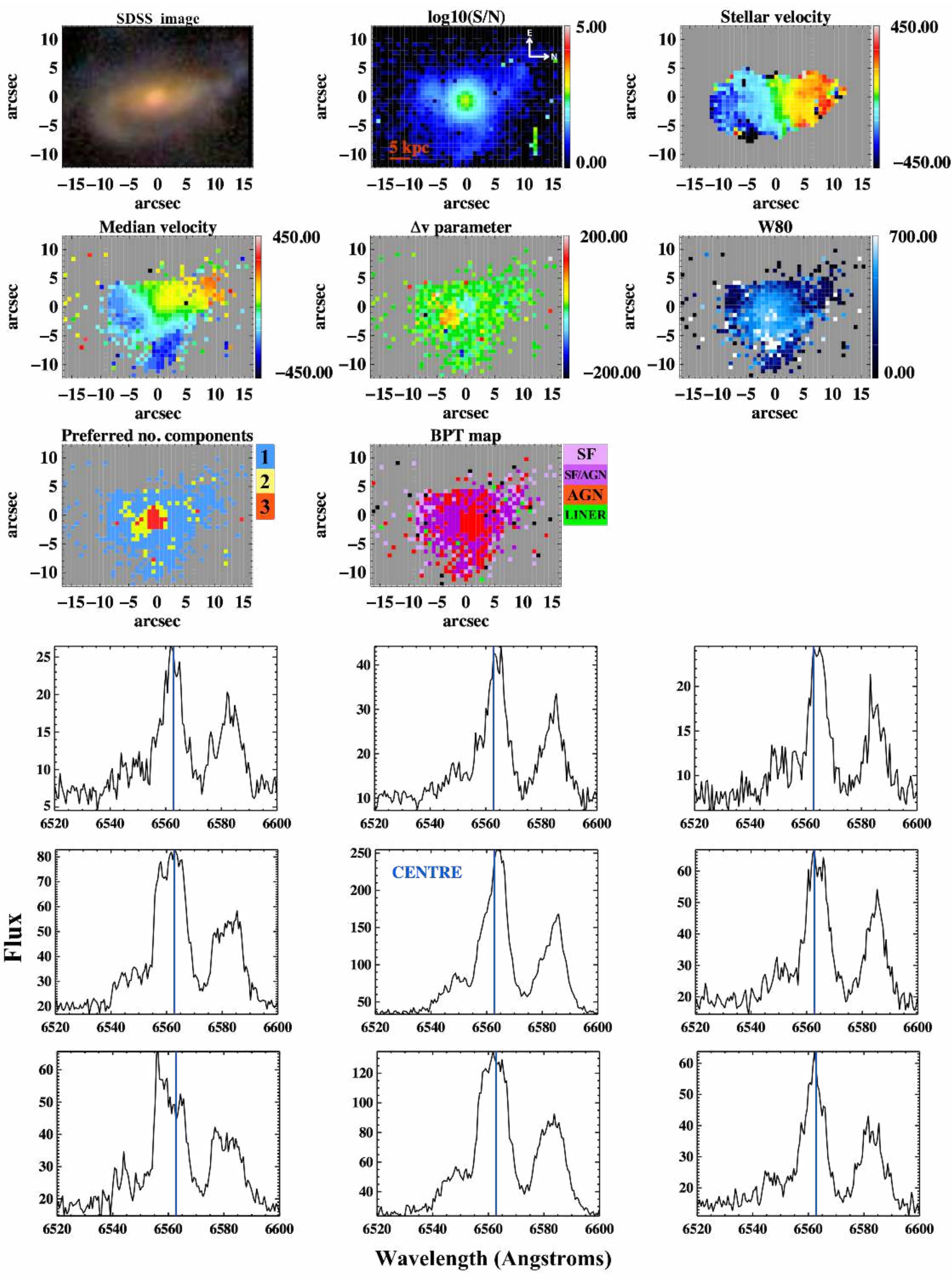}
\caption{J143046}
\label{J143046}
\end{figure*}

\begin{figure*}
\centering
 \includegraphics[width=0.8\textwidth]{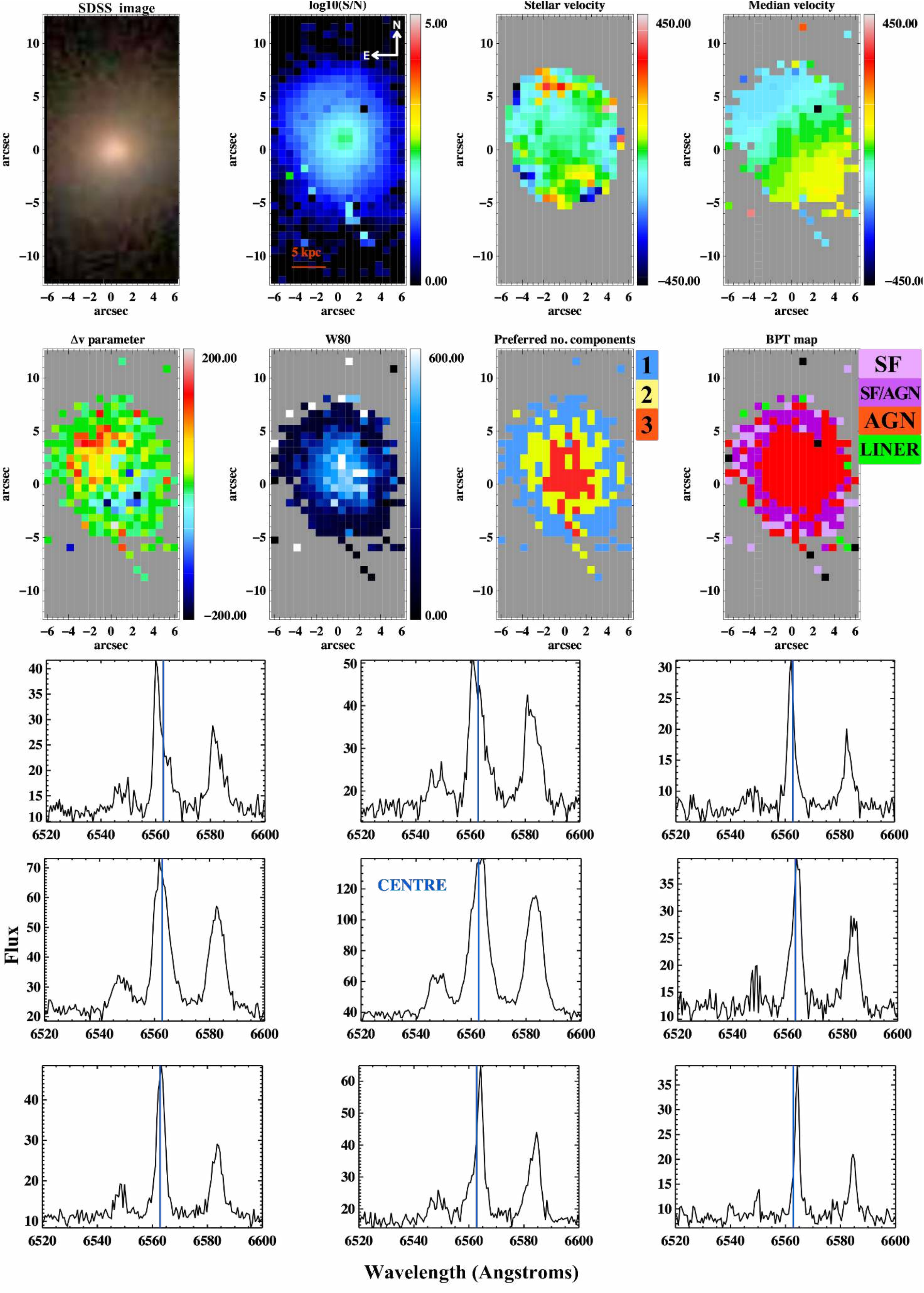}
\caption{J150754}
\label{J150754}
\end{figure*}

\begin{figure*}
\centering
 \includegraphics[width=0.8\textwidth]{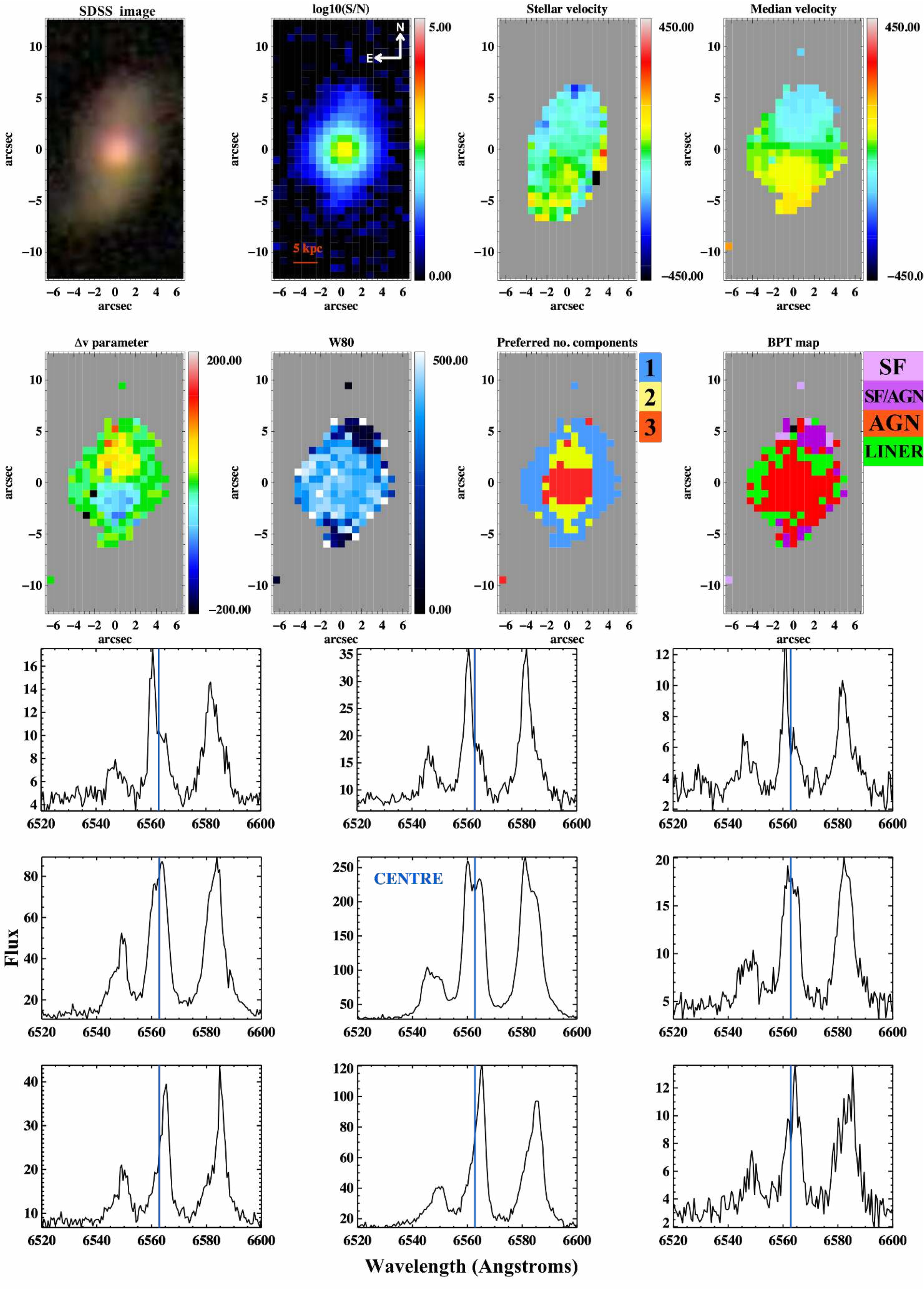}
\caption{J151147}
\label{J151147}
\end{figure*}

\begin{figure*}
\centering
 \includegraphics[width=0.8\textwidth]{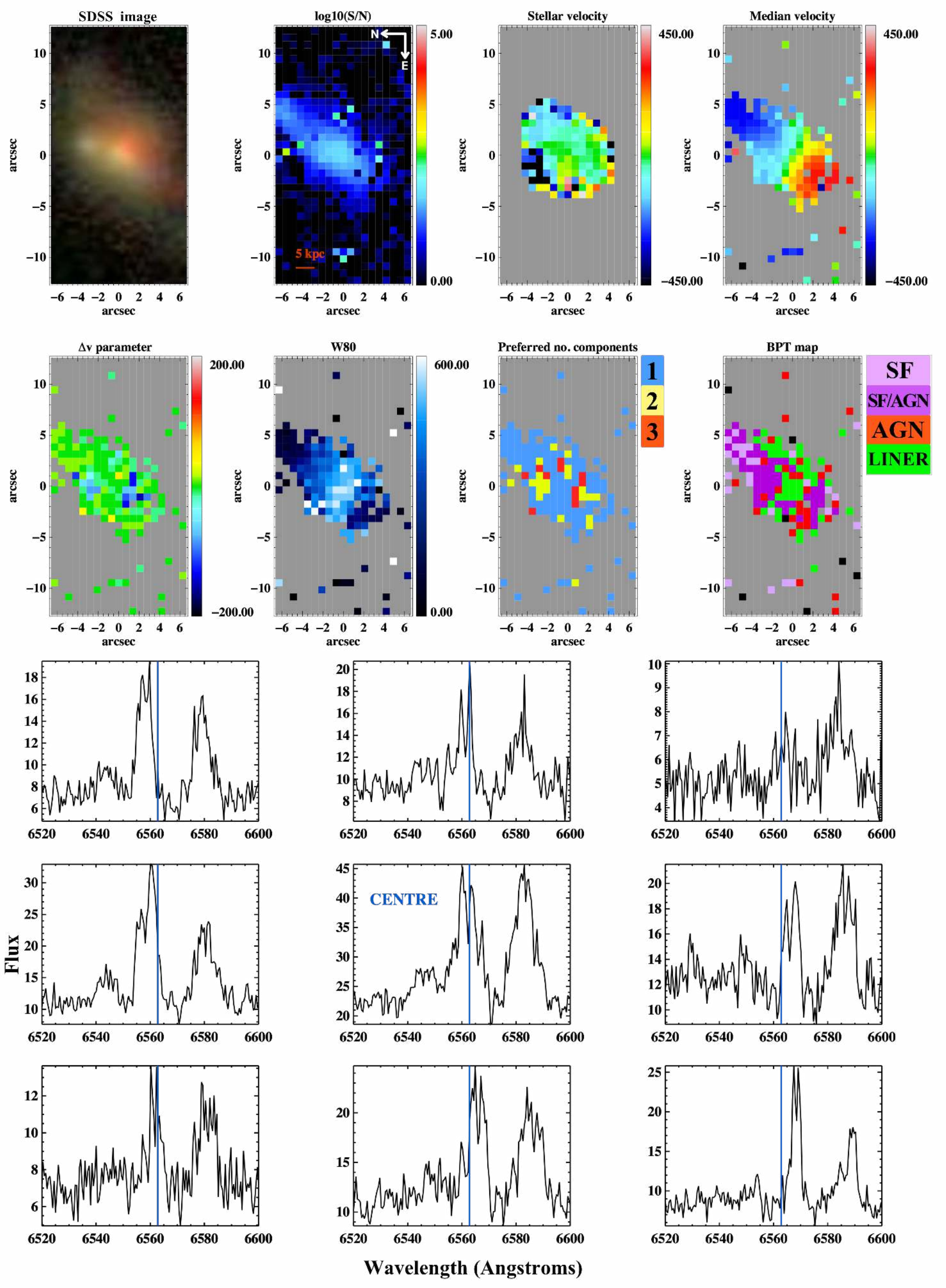}
\caption{J152133}
\label{J152133}
\end{figure*}

\begin{figure*}
\centering
 \includegraphics[width=0.8\textwidth]{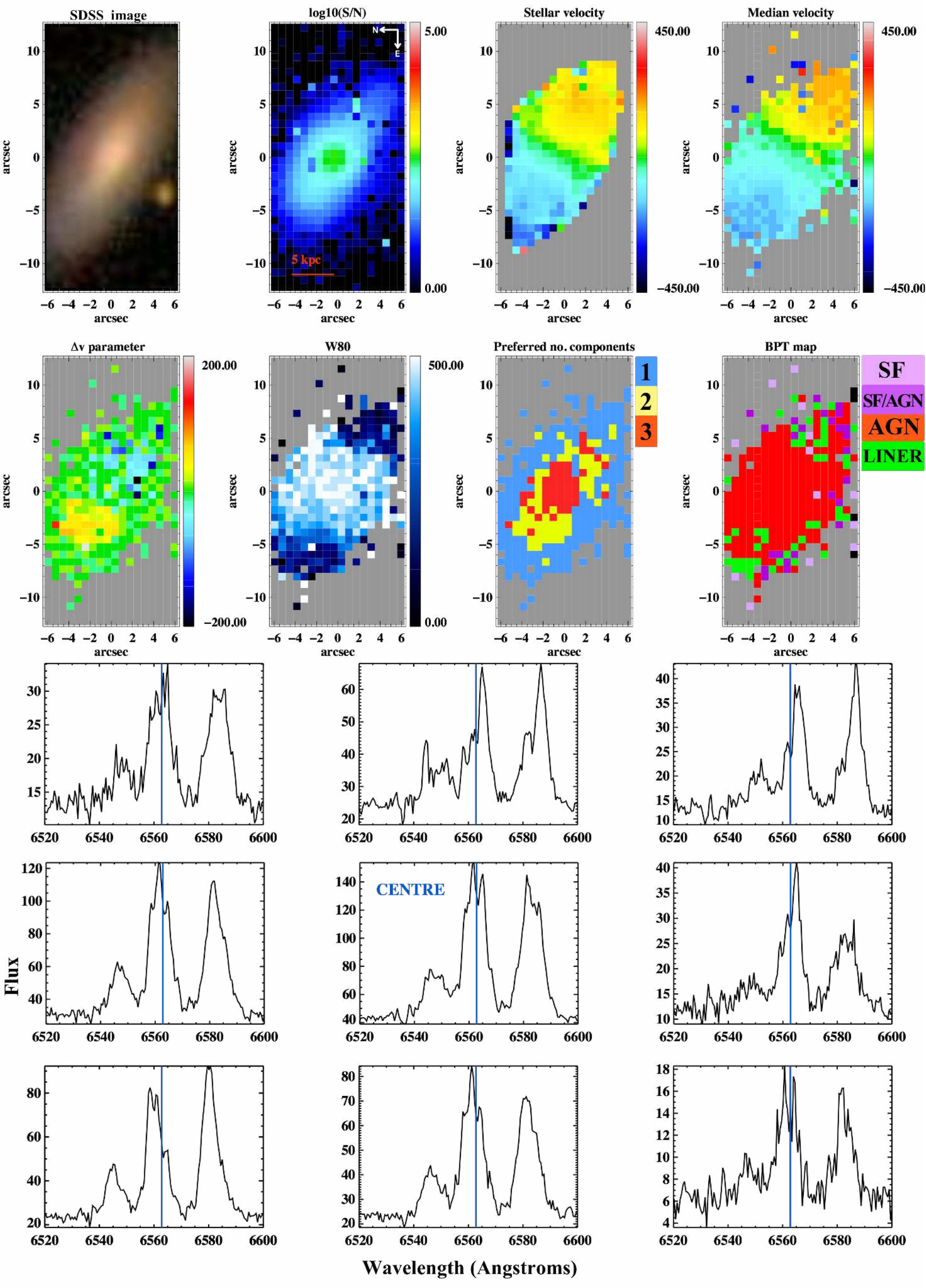}
\caption{J152637}
\label{J152637}
\end{figure*}

\bsp

\label{lastpage}

\end{document}